\documentclass[pra, 10pt, twocolumn, aps, superscriptaddress, showpacs,floatfix]{revtex4-2}
\usepackage[T1]{fontenc}
\usepackage{amssymb}
\usepackage{graphicx}
\usepackage{amsmath}
\usepackage{multirow}
\usepackage{hyperref}
\usepackage{array}
\usepackage{physics}
\usepackage{bm}
\usepackage{xcolor}
\usepackage{soul}
\usepackage{bbold}

\definecolor{lapislazuli}{rgb}{0.15, 0.38, 0.61}
\definecolor{YKblue}{rgb}{0.0, 0.18, 0.65}
\definecolor{carmine}{rgb}{0.81, 0.09, 0.13}
\definecolor{lavender}{rgb}{0.84, 0.79, 0.87}

\newcommand{\matfont}[1]{\mathsf{#1}}

\hypersetup{
	colorlinks=true,
	linkcolor=YKblue,
	linktoc=page,
	citecolor=blue,
	urlcolor=YKblue
}

\begin{document}

\title{Photon Bose-Einstein Condensation in Semiconductors: A Quantum Kinetic Theory}

\author{Jos\'e~L.~Figueiredo }
\email{jose.luis.figueiredo@tecnico.ulisboa.pt} 
\affiliation{GoLP - Instituto de Plasmas e Fus\~{a}o Nuclear, Instituto Superior T\'{e}cnico, Universidade de Lisboa, Av. Rovisco Pais 1, 1049-001 Lisboa, Portugal}
\affiliation{Department of Physics, Blackett Laboratory, Imperial College London, Prince Consort Road, London SW7 2AZ, United Kingdom}
\author{Ross C. Schofield}
\affiliation{Department of Physics, Blackett Laboratory, Imperial College London, Prince Consort Road, London SW7 2AZ, United Kingdom}
\author{Ming Fu}
\affiliation{Department of Physics, Blackett Laboratory, Imperial College London, Prince Consort Road, London SW7 2AZ, United Kingdom}
\author{Robert A. Nyman}
\affiliation{Mode Labs Ltd., 30 Upper High Street, Thame, Oxfordshire, England, OX9 3EZ}
\author{Rupert Oulton}
\affiliation{Department of Physics, Blackett Laboratory, Imperial College London, Prince Consort Road, London SW7 2AZ, United Kingdom}
\author{Hugo Ter\c{c}as }
\affiliation{Instituto Superior de Engenharia de Lisboa, Polytechnical University of Lisbon, Rua Conselheiro Emídio Navarro, 1089 Lisboa, Portugal }
\affiliation{GoLP - Instituto de Plasmas e Fus\~{a}o Nuclear, Instituto Superior T\'{e}cnico, Universidade de Lisboa, Av. Rovisco Pais 1, 1049-001 Lisboa, Portugal}
\author{Florian Mintert}
\affiliation{Department of Physics, Blackett Laboratory, Imperial College London, Prince Consort Road, London SW7 2AZ, United Kingdom}
	
\begin{abstract}

Photon condensation in semiconductor microcavities is a transformative technique for engineering quantum states of light at room temperature by tailoring strong but incoherent light-matter interactions. While continuous-wave and electrical pumping offer exceptional prospects for miniaturized quantum photonic technologies, harnessing these requires conceptual advances in understanding non-equilibrium light-matter dynamics in semiconductors. We resolve this challenge through an ab initio quantum kinetic theory capturing how Coulomb interactions of optically excited carriers and phonon scattering mediate photon thermalization and condensation in semiconductors. Our microscopic model shows that at high carrier densities, thermalization is dominated by carrier–carrier Coulomb scattering, in clear contrast to the rovibrational relaxation that governs dye-based photon condensates. The theory predicts a rich nonequilibrium phase diagram with thermal, Bose-condensed, multimode, and lasing phases, quantitatively in agreement with recent experiments. Crucially, we identify how cavity detuning controls transitions between equilibrium and gain-dominated regimes, enabling tailored design of coherent light sources. This work thus provides the foundation for semiconductor-based quantum photonic devices operating beyond conventional laser paradigms.

\end{abstract}

\maketitle

\section{Introduction}
Photon Bose--Einstein condensation (BEC) has been observed in both dye-filled~\cite{Klaers2010_2} and semiconductor microcavities~\cite{Schofield2024}, offering a versatile platform to explore macroscopic quantum states of light under nonequilibrium conditions. Unlike atoms, photons do not interact with each other, so photon condensation requires some thermalization channel able to drive them into the lowest cavity mode, in a way that mimics equilibrium Bose statistics~\cite{PhysRevA.93.013829, PhysRevLett.108.160403}. In driven-dissipative systems, such as externally excited semiconductor microcavities, photon condensation is achieved through a balance between external pumping, losses, and interactions with carriers, which collectively drive the system towards equilibrium \cite{Klaers2010,PhysRevLett.130.033602}. The presence of gain and dissipation distinguishes these condensates from equilibrium BECs and leads to effects including modified critical behavior and phase transitions shaped by the nonequilibrium balance of gain and loss~\cite{PhysRevA.104.063709,PhysRevA.91.033826}, something that is not observed in traditional matter BECs. Other platforms for condensed light include exciton-polariton~\cite{kasprzak2006bose, PhysRevLett.118.016602} and surface-plasmon–polariton condensates~\cite{PhysRevLett.127.255301, hakala2018bose}, although in both cases the condensed particles are hybrid light-matter excitations rather than purely photonic.

The first experimental realizations of photon BECs were achieved in dye-filled microcavities. Beyond the demonstration that Bose-condensation is possible at room temperature, those experiments also enabled analysis of coherence properties of condensed light \cite{PhysRevLett.112.030401}, condensation in non-Gaussian modes~\cite{PhysRevA.91.033813}, observation of non-Hermitian phase transitions~\cite{kirton2018superradiant} and condensation in the few-particle regime~\cite{rodrigues2021learning}.
Dye-based setups, however, also suffer from limitations such as the need of pulsed pumping~\cite{damm2017first} and negligibly weak effective photon-photon interactions~\cite{marelic2016phase}.
These limitations can be overcome with the help of semiconductor microcavities.

Considering a semiconductor environment, the interaction between cavity photons and charge carriers gives rise to effective photon--photon interactions, mediated by the electron--hole plasma~\cite{RevModPhys.71.1591} that are likely to result in collective phenomena such as spatial pattern formation~\cite{amo2009collective}, soliton dynamics~\cite{egorov2009bright} and excitation spectra~\cite{utsunomiya2008observation}, analogous to Bogoliubov modes in atomic condensates. Given the effective nature of these interactions, they are expected to be tuneable through external control parameters such as carrier density, cavity geometry, and detuning, making semiconductor microcavities a flexible testbed for non-equilibrium quantum field theories. Semiconductor-based Photon BEC thus provides a fascinating platform to study quantum fluid phenomena~\cite{PhysRevLett.121.183604,PhysRevA.92.043802,alyatkin2021quantum}, coherence build-up, and the emergence of long-range order in driven-dissipative conditions~\cite{RevModPhys.85.299} with a promising prospect for both classical and quantum photonic technologies.

In dye-based systems, photon thermalization proceeds via absorption and emission involving rovibrational molecular states, and the dynamics is well captured by the Kirton--Keeling model~\cite{PhysRevA.91.033826,PhysRevLett.111.100404}, which treats the photons and dye molecules as an open, driven-dissipative quantum gas.
In contrast, semiconductor microcavities rely on a qualitatively different thermalization mechanism, since cavity photons interact with a delocalized electron--hole plasma rather than localized (classical) molecules. Therefore, Coulomb scattering is also expected to mediate energy redistribution, something that is absent in dye environments. 
Existing models for photon-Bose-Einstein-condensation, such as those derived for dye-based setups~\cite{PhysRevLett.111.100404,PhysRevA.91.033826}, or models derived from equilibrium thermodynamics~\cite{loirette2023photon,Pieczarka2024,Sobyanin2012}, necessarily fail to capture the features of light BECs in semiconductor microcavities.

While proof-of-principle experiments can be conceived without a detailed theoretical model, their interpretation—and the translation into functional technologies—requires a quantitative framework. The design of practical photonic devices hinges on predictive modeling capable of identifying the system parameters and control protocols that optimize underlying physical mechanisms. In particular, understanding how cavity geometry and external pumping shape absorption and emission spectra is essential for engineering future semiconductor-based platforms. Under genuinely open and dissipative conditions, such predictions must arise from a self-consistent theory derived from a collective Hamiltonian, coupled to its environment and accounting for the quantum nature of the carrier, lattice and light fields involved.

In this work, we construct a quantum kinetic model for photon condensation in semiconductor quantum wells.
Our framework leads to coupled nonlinear equations for the photon correlation matrix and carrier distributions, incorporating light absorption and emission, Coulomb collisions, phonon interactions, environmental losses and continuous-wave pumping on equal footing. Numerical simulations based on this model reproduce key features observed in recent experiments \cite{Schofield2024} and reveal a rich nonequilibrium phase diagram, including transitions between thermal, single-mode BEC, multimode condesation, and lasing states. In the strong-pumping limit at room temperature, we identify Coulomb scattering as the dominant thermalization mechanism, fundamentally distinguishing this platform from molecular photon BECs. These results provide the first microscopic foundation for semiconductor photon condensation and a predictive tool for engineering non-equilibrium optical phases in solid-state systems. \par 
    
This paper is organized as follows: in Sec.~\ref{TSPBEC}, we present the theoretical model describing the joint quantum dynamics in a semiconductor microcavity. In Sec.~\ref{SQC}, we characterize carrier equilibration and the impact of Coulomb interactions on open-semiconductor systems. Section~\ref{DPC} analyzes the conditions for photon condensation and the transition to a laser phase, while Section~\ref{sec_phase_diagram} is devoted to a discussion of the resulting phase diagram as well as the phase transitions therein. Finally, we summarize our findings and discuss open questions in Sec.~\ref{Sec_conclusions}.

\section{Theory of semiconductor PBEC}\label{TSPBEC}

The central goal of this work is to develop a microscopic framework capable of predicting the dynamics of photon Bose--Einstein condensation in semiconductor microcavities. Unlike dye-based platforms---where photon thermalization occurs via rovibrational transitions and can be modeled by effective reservoir theories---semiconductors involve carrier--carrier scattering, phonon-mediated processes, and photon exchange, all occurring on comparable timescales. A comprehensive model must therefore resolve the coupled evolution of photonic and electronic degrees of freedom, account for transient carrier dynamics, and treat photon absorption and emission beyond equilibrium approximations. In this section, we derive a quantum kinetic description that captures these effects systematically, starting from a second-quantized Hamiltonian and an open quantum system approach.

\begin{figure}
\centering 
\includegraphics[scale=0.42]{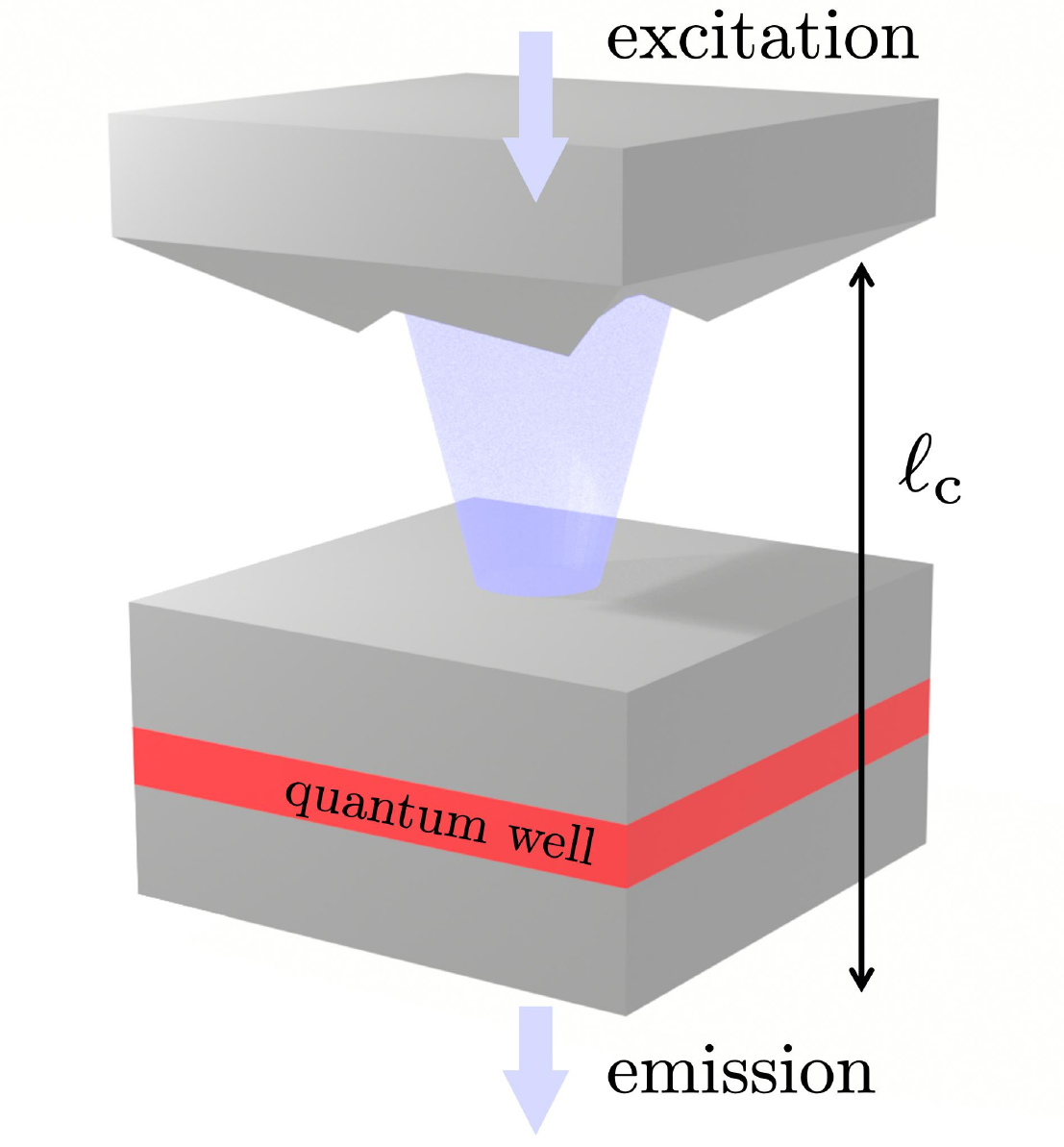}
\caption{(color online). Schematic illustration of the semiconductor quantum-well microcavity. The longitudinal distance $\ell_\text{c}$ is sufficiently small to isolate manifolds of longitudinal cavity modes. Due to energetic restrictions to a single longitudinal mode, cavity photons behave as an effective 2D photon gas.}
\label{fig_cavity}
\end{figure}

We consider a semiconductor quantum well embedded in a planar optical resonator formed by one curved and one flat high-reflectivity mirror, as illustrated in Fig.~\ref{fig_cavity}, similar to those used in Ref.~\cite{Schofield2024}. The cavity is designed with a micrometer-scale longitudinal length and centimeter-scale transverse dimensions, such that only a single longitudinal photonic mode lies near resonance with the electronic transition energy. This configuration ensures that the photonic dynamics is effectively two-dimensional, with discrete transverse modes labeled by quantum numbers $m = (m_x, m_y)$. The cavity spectrum takes the harmonic form 
$\omega_m = \omega_0 + (m_x + m_y)\, \delta\omega$,
where the energy spacing $\delta\omega$ and ground-state energy $\omega_0$ are set by the mirror curvature and cavity length~\cite{Klaers2010, Klaers2010_2}. The resulting photonic density of states supports equilibrium-like condensation in the presence of efficient thermalization channels.

The relevant microscopic degrees of freedom excited inside the described heterostructure include cavity photons, conduction-band electrons, valence-band holes, and phonons. The semiconductor can absorb or emit photons through creation or annihilation of an electron in the conduction band and a hole in the valence band.
The bosonic enhancement of the photon-carrier interaction allows highly occupied modes to couple more strongly to the electron-hole environment than weakly occupied modes.
This favors the increase in occupation of highly occupied modes and thus condensation.
Moreover, electrons and holes interact among themselves and with each other via Coulomb interaction, and with the phonon field via phonon absorption and emission.
These interactions result in equilibration and thermalization of the electron-hole environment, which is required to realize thermal states of cavity light.

Before developing a quantitative framework for these processes in Secs.~\ref{subsec:hamiltonian_model}--\ref{subsec:photonrateapproximation},
the complete model is laid out in Sec.~\ref{subsec:comprehensive} to help readers who would like to skip the derivation of the model.

\subsection{A comprehensive view on the Quantum Kinetic Theory}
\label{subsec:comprehensive}

The cavity field can be characterized by the photonic correlation matrix $\matfont{N}$, with elements $\matfont{N}_{m,n} = \langle \hat{a}_m^\dagger \hat{a}_{n} \rangle$ defined in terms of creation and annihilation operators of the cavity light field. This matrix encodes mode occupations and coherence and evolves according to the equation
\begin{equation}
\frac{\partial}{\partial t} \matfont{N} =
i[\matfont{H},\matfont{N}]-\kappa \matfont{N}+v(\matfont{N})\ ,
\label{eq1}
\end{equation}
where the first term involves the diagonal matrix $\matfont{H}$ with elements $\matfont{H}_{m,n} = \delta_{m,n}\omega_m$
and describes regular free phase evolution,
the second term describes cavity loss rate $\kappa$, and the last term $v(\matfont{N})$ denotes interactions with the optically excited medium. 
Its explicit form reads
\begin{equation}
v(\matfont{N})= \matfont{W}_\uparrow ( \matfont{N}+\mathbb{1}) \matfont{G} - \matfont{W}_\downarrow \matfont{N} \ \matfont{G} + \text{h.c.}, 
\end{equation}
with the matrix $\matfont{G}$ [Eq.~\eqref{eq:G}] comprised of the elements $\matfont{G}_{m,n}=g_mg_n$ in terms of light-matter coupling constants $g_m$ defined in Eq.~\eqref{eq:coupling}.

The matrices $\matfont{W}_\uparrow$ and $\matfont{W}_\downarrow$ describe effective rates of photon emission and absorption by the electron--hole plasma. They are given by
\begin{align}
\matfont{W}_\uparrow &= \sum_{\mathbf{k}} \matfont{O}_{\mathbf{k}} f_{e,\mathbf{k}} f_{h,-\mathbf{k}}, \label{W_u} \\
\matfont{W}_\downarrow &= \sum_{\mathbf{k}} \matfont{O}_{\mathbf{k}} (1 - f_{e,\mathbf{k}})(1 - f_{h,-\mathbf{k}}), \label{W_d}
\end{align}
where $f_{\nu,\mathbf{k}} = \langle \hat{c}_{\nu,\mathbf{k}}^\dagger \hat{c}_{\nu,\mathbf{k}} \rangle$ denotes the carrier distribution for electrons ($\nu=e$) and 
holes ($\nu=h$), and $\matfont{O}_{\mathbf{k}}$ is a diagonal matrix encoding transition strength and environmental response [see Eq.~\eqref{O_matrix}].

For an isolated electron-hole plasma in thermal equilibrium, $f_{\nu,\mathbf{k}}$ become the Fermi-Dirac probability distributions. More generically, when interactions with the remaining quasiparticles and with the environment are considered, each carrier distribution reads as a solution to
\begin{equation}
\frac{\partial}{\partial t} f_{\nu, \mathbf{k}} = \Omega_{\uparrow,\nu,\mathbf{k}} (1 - f_{\nu,\mathbf{k}}) - \Omega_{\downarrow,\nu,\mathbf{k}} f_{\nu,\mathbf{k}}, \label{eq2}
\end{equation}
with effective excitation and relaxation rates
\begin{align}
\Omega_{\uparrow,\nu,\mathbf{k}} &= (\Gamma_{\uparrow,\mathbf{k}} + \Lambda_{\uparrow,\mathbf{k}})(1 - f_{\overline{\nu}, -\mathbf{k}}) + \mu_{\uparrow,\nu,\mathbf{k}} + \eta_{\uparrow,\nu,\mathbf{k}}\ , \\
\Omega_{\downarrow,\nu,\mathbf{k}} &= (\Gamma_{\downarrow,\mathbf{k}} + \Lambda_{\downarrow,\mathbf{k}}) f_{\overline{\nu}, -\mathbf{k}} + \mu_{\downarrow,\nu,\mathbf{k}} + \eta_{\downarrow,\nu,\mathbf{k}}\ ,
\label{eq:rates}
\end{align}
containing scalar functions $\Gamma_{\circ,\mathbf{k}}$,
$\mu_{\circ,\nu,\mathbf{k}}$,
$\eta_{\circ,\nu,\mathbf{k}}$ ($\circ=\uparrow,\downarrow$)
that account for
external pumping ($\Gamma_{\uparrow,\mathbf{k}}$) and carrier-recombination rate ($\Gamma_{\downarrow,\mathbf{k}}$),
Coulomb scattering [Eqs.~\eqref{mu_up} and \eqref{mu_down}] and phonon scattering [Eqs.~\eqref{phonon_rate_up} and \eqref{phonon_rate_down}], respectively. The symbol $\overline{\nu}$ denotes the charge carrier opposite to $\nu$, {\it i.e.} $\overline{e}=h$, and $\overline{h}=e$. The terms
\begin{align}
\Lambda_{\uparrow,\mathbf{k}} &= \text{tr}\left[ \matfont{O}_{\mathbf{k}}\, \matfont{N} \,\matfont{G} \right] + \text{c.c.}, \label{L_u} \\
\Lambda_{\downarrow,\mathbf{k}} &= \text{tr}\left[ \matfont{O}_{\mathbf{k}}\, (\mathbb{1} + \matfont{N})\, \matfont{G} \right] + \text{c.c.}, \label{L_d}
\end{align}
that depend explicitly on the photon correlation matrix $\matfont{N}$, can be understood as photon-induced transition rates. 

Equations~\eqref{eq1} and~\eqref{eq2} together define a coupled set of nonlinear equations governing the dynamics of the photon and carrier subsystems. Their structure captures both the feedback between light and matter and the many-body interactions responsible for thermalization. The remainder of this section presents a detailed derivation of this dynamical system from the microscopic Hamiltonian.

\subsection{Hamiltonian model}
\label{subsec:hamiltonian_model}
	
Our starting point is the total Hamiltonian $\hat H =\hat H_\gamma+\hat H_{sc}+\hat H_I$, comprised of non-interacting parts $\hat H_\gamma$ and $\hat H_{sc}$ denoting, respectively, the bare photonic modes of the cavity and the semiconductor kinetic terms, and interactions $\hat H_I$.

Given a symmetric cavity and a homogeneous semiconductor element, the interaction between photons of odd cavity modes and charge carrier in the semiconductor is negligible, as discussed in more detail below in Appendix~\ref{Ap_interactions}.
In this case, the subsequent discussion can be restricted to photons of even cavity modes with creation and annihilation operators
$\hat{a}_m^\dagger$ and $\hat{a}_m$ and corresponding eigenfrequencies $\omega_m$ in the non-interacting part
\begin{align}
\hat H_\gamma=\sum_{m}\hbar \omega_{m}\hat a^\dagger_{m} \hat a_{m}
\end{align}
of the total Hamiltonian.

Relevant vibrational dynamics takes place only for longitudinal optical (LO) phonons within a single band.
The creation and annihilation operators for phonons are denoted by $\hat{b}_{\mathbf{k}}^\dagger$ and $\hat{b}_{\mathbf{k}}$,
labeled with the in-plane wavevector $\mathbf{k}$, and their eigenfrequency $\omega_\text{LO}$ is independent of $\mathbf{k}$ \cite{PhysRevB.77.024306}. 
Corresponding operators 
for carriers are given by $\hat{c}^\dagger_{\nu,\mathbf{k}}$ and $\hat{c}_{\nu,\mathbf{k}}$. The non-interacting component $\hat H_{sc}$ reads
\begin{align}
 \hat H_{sc} &= \sum_{\mathbf k} \left(\hbar \omega_\text{LO} \hat   b_{\mathbf k}^\dagger \hat  b_{\mathbf k}  + E_{e,\mathbf k} \hat  c^\dagger_{e,\mathbf k} \hat  c_{e,\mathbf k}
    + E_{h,\mathbf k}\hat   c^\dagger_{h,\mathbf k}\hat   c_{h,\mathbf k} \right),
\end{align}
with dispersions $E_{\nu,\mathbf{k}}$ assumed parabolic and separated by a bandgap $E_g$, {\it i.e.},  $E_{e,\mathbf k} = E_g +  \mathbf \hbar^2k^2/(2m_e)$ and $E_{h,\mathbf k} = \hbar^2\mathbf k^2/(2m_h) $, although this specific dispersion is not essential for the following derivation. 

The interacting part $\hat H_I=\hat H_{c-c}+\hat H_{c-\gamma}+\hat H_{c-p}$ is comprised of the carrier-carrier Coulomb interaction $\hat H_{c-c}$, the interaction between cavity photons and carriers in the semiconductor $\hat H_{c-\gamma}$, and the carrier-phonon interaction $\hat H_{c-p}$. These terms read
\begin{align}
   \hat  H_{c-c}  &= 
   \frac{1}{2}
   \sum_{\nu,\nu',\mathbf q}
   V_{\mathbf q}^{\nu,\nu'}
   \sum_{\mathbf k,\mathbf k'}  \hat   c^\dagger_{\nu,\mathbf k + \mathbf q}\hat  c^\dagger_{\nu',\mathbf k' - \mathbf q}\hat   c_{\nu',\mathbf k'}\hat   c_{\nu,\mathbf k} , \label{H_c_c_final}\\
    \hat   H_{c-\gamma} &= \sum_{\mathbf k,m} g_{m}\Big( \hat   a_m \hat  c^\dagger_{e,\mathbf k}\hat  c^\dagger_{h,- \mathbf k} + \hat  a^\dagger_m \hat  c_{h,-\mathbf k}\hat  c_{e,\mathbf k} \Big), \label{H_c_gamma_final}\\
   \hat    H_{c-p} &= -\sum_{\nu,\mathbf k,\mathbf q}  s_\nu \lambda_{\mathbf q} (\hat  b_{\mathbf q} -\hat   b_{-\mathbf q}^\dagger) \hat  c_{\nu,\mathbf k + \mathbf q}^\dagger \hat  c_{\nu,\mathbf k}, \label{H_c_p_final}
\end{align}
with interaction constants $V_{{\bf q}}^{\nu\nu^\prime}$, $g_m$ and $\lambda_{\bf q}$ defined and discussed in Appendix~\ref{Ap_interactions}~[see, respectively, Eqs.~\eqref{Coulomb_coupling},~\eqref{eq:coupling} and~\eqref{coupling_phonons}].

\subsection{Open quantum dynamics}

Our goal is to model the evolution of relevant observables of the quantized fields due to the effect of the interactions Eqs.~\eqref{H_c_c_final}-\eqref{H_c_p_final}, which can be achieved with the Lindblad master equation \cite{breuer2002theory,gardiner1991quantum}
\begin{equation}
\frac{\partial}{\partial t} \hat   \rho =  \frac{i}{\hbar}[\hat  \rho, \hat  H] +\mathcal L(\hat  \rho)\ ,  \label{masterEq}
\end{equation}
for the density matrix $\hat  \rho(t)$ of the joint system.
The first term in Eq.~\eqref{masterEq} represents unitary evolution induced by the system Hamiltonian $\hat H$, and all incoherent, dissipative processes are encapsulated in the Lindbladian ${\mathcal L}$. 

The equation of motion of expectation values $\langle \hat  O \rangle(t) = \text{Tr}[\hat  \rho(t)\hat  O]$  of any operator $\hat  O$ is determined from Eq.~\eqref{masterEq} and can be written as
\begin{equation}
    \frac{\partial}{\partial t} \langle \hat  O \rangle = \frac{i}{\hbar}\langle [\hat  H,\hat  O] \rangle + \langle {\mathcal L}^\dagger(\hat O)\rangle\ .   \label{observableDynamics}
\end{equation}
For a generic Lindbladian ${\mathcal L}$ of the form ${\mathcal L}(\hat\rho)=\sum_jC_j\big\langle \hat   A_j \hat\rho \hat  A_j^\dagger- \frac{1}{2}\{\hat\rho, \hat  A^\dagger_j\hat  A_j\} \big\rangle)$ with rates $C_j$ and jump operators $\hat A_j$, the central contribution to the latter term reads
\begin{equation}
    {\mathcal L}^\dagger(\hat O)=
\sum_jC_j\left( \hat   A_j^\dagger \hat  O \hat  A_j- \frac{1}{2}\{\hat  O, \hat  A^\dagger_j\hat  A_j\} \right)\ .
\end{equation}
For the semiconductor microcavity considered here, the dominant contributions to the Lindbladian are given in terms of the jump operators $\hat a_m$ with rate $\kappa$ for cavity loss, $\hat\sigma_{\mathbf k}^+=\hat  c_{e,\mathbf k}^\dagger \hat  c_{h,-\mathbf k}^\dagger$ with rate $\Gamma_{\uparrow,\mathbf k}$ for incoherent pumping,
and $\hat\sigma_{\mathbf k}^-= \hat  c_{h,-\mathbf k}\hat  c_{e,\mathbf k}$ with rate $\Gamma_{\downarrow,\mathbf k}$ 
for quantum-well fluorescence into noncavity modes.

\subsection{Derivation of the rates}\label{DOTR}

The system Hamiltonian and Lindbladian described above in Sec.~\ref{subsec:hamiltonian_model} provide the microscopic foundation for the model to-be-derived in this section, consisting of Eqs.~\eqref{eq1}--\eqref{L_d}.

Properties of the cavity light field will be characterized in terms of expectation values of products of creation operators $a^\dagger_m$ and annihilation operators $a_m$ of the light field.
Given the incoherent nature of the external pump, the light field is expected to be incoherent below condensations threshold;
{\it i.e.} the expectation values $\langle \hat  a_m\rangle$ and
$\langle \hat  a_m^\dagger\rangle$ vanish.
The correlation matrix $\matfont{N}$ with elements $\matfont{N}_{m,n}=\langle a_m^\dagger a_{n}\rangle$, on the other hand, is typically finite, and contains information about the most relevant experimentally accessible observables such as the light intensity. 

The derivation assumes an homogeneous semiconductor resulting in the diagonal form \begin{equation}
    \langle \hat  c_{\nu,\mathbf k}^\dagger \hat  c_{\nu',\mathbf k'}\rangle  = \delta_{\nu,\nu'}\delta_{\mathbf k,\mathbf k'}f_{\nu,\mathbf k} (t)
\end{equation}
of the carrier correlation function, whose diagonal elements define the carrier distribution functions $f_{\nu,\mathbf k}(t)$.
Consistently with the expectation that an incoherent pump does not create coherent light fields, also the phonon modes in the semiconductor can be assumed to be incoherent, $\langle \hat  b_{\mathbf k} \rangle=\langle \hat  b_{\mathbf k}^\dagger\rangle=0$. 

We start by evaluating the contribution of $\hat H_{c-\gamma}$ and 
Lindbladian terms to the equation of motion for $\matfont{N}$ and $f_{\nu,\mathbf k}$ following Eq.~\eqref{observableDynamics}, which we show in Sec.~\ref{sub_photon_carrier} with detail. Then, the remaining contributions -- stemming from $\hat H_{c-c}$ and $\hat H_{c-p}$ -- are briefly sketched in Secs.~\ref{sub_carrier_carrier} and~\ref{sub_carrier_phonon}, respectively, and some details of the calculations are left to Appendices~\ref{Ap_Coulomb_coll} and~\ref{Ap_Phonon}. 

\subsubsection{Carrier-photon contribution}\label{sub_photon_carrier}

Given these assumptions, the equation of motion [Eq.~\eqref{observableDynamics}] for photon-photon correlation matrix $\matfont{N}$ reads
\begin{align}
\frac{\partial}{\partial t} \matfont{N}_{m,n} &= (i\omega_m-i\omega_{n}-\kappa)\matfont{N}_{m,n} \nonumber\\
& + \frac{i}{\hbar} \sum_{\mathbf k} \Big( g_m h_{\mathbf k,n} - g_{n} h_{\mathbf k,m} \Big) \ ,
\label{d_tn}
\end{align}
with  
 \begin{align}
h_{\mathbf k,m} &= \langle \hat  c_{e,\mathbf k}^\dagger \hat  c_{h,-\mathbf k}^\dagger \hat  a_m \rangle, \label{carrierphoton}
\end{align}
the correlation function of cavity photons and semiconductor carriers. 

Excluding Coulomb and phonon interactions, the equations of motion for the carrier distribution functions $f_{\nu,\mathbf k}$ are of the form 
\begin{align}
 & \frac{\partial}{\partial t} f_{e,\mathbf k}  =    \frac{\partial}{\partial t} f_{h, -\mathbf k} = (\Gamma_{\uparrow,\mathbf k} - \Gamma_{\downarrow,\mathbf k}) u_{\mathbf k}  \nonumber \\
    & + \Gamma_{\uparrow,\mathbf k}(1 - f_{h,-\mathbf k} -  f_{e,\mathbf k}) + \frac{2}{\hbar} \sum_m g_m \,  \text{Im} \big(h_{\mathbf k,m}\big) \label{d_tf}
\end{align}
with `$\text{Im}$' denoting the imaginary part, and the correlation function 
\begin{align}
 u_{\mathbf k} &=  \langle \hat  c_{e,\mathbf k}^\dagger \hat   c_{h,-\mathbf k}^\dagger \hat   c_{h, - \mathbf k} \hat  c_{e,\mathbf k} \rangle , \label{u_k}
 \end{align}
translating carrier-carrier correlations. Additional contributions to the right-hand side of Eq.~\eqref{d_tf} stemming from the remaining interactions are easily introduced later [see Secs~\ref{sub_carrier_carrier} and~\ref{sub_carrier_phonon}]. 

According to Eqs.~\eqref{d_tn} and \eqref{d_tf}, two-point correlation functions depend on the three-point correlation function $h_{\mathbf k,m}$ and the four-point correlation function $u_{\mathbf k}$. Similarly, 
equations of motion for the correlations functions $u_{\mathbf k}$ and $h_{\mathbf k,m}$ include higher order correlation functions, and so on. In order to avoid an infinite hierarchy of coupled equations of motion, it is thus necessary to approximate the dynamics of $u_{\mathbf k}$ and $h_{\mathbf k,m}$ through the so-called \textit{cluster expansion}~\cite{fricke1996transport}. It consists of truncating the higher-order correlators into all possible combinations of lower-order terms plus quantum fluctuations, and then including only the relevant fluctuations depending on the strength of the couplings under consideration. This procedure permits us to derive a closed system involving a finite number of variables, as shown in the following. For a complete review on the cluster-expansion method, see Ref.~\cite{2006PQE} and references therein.

In the spirit of the cluster expansion, we start by rewriting $u_{\mathbf k}$ as
\begin{equation}
   u_{\mathbf k} =  |p_{\mathbf k }|^2 + f_{e,\mathbf k} f_{h,- \mathbf k} + \delta u_{\mathbf k},\label{cDcDcc}
\end{equation}
where $p_{\mathbf k}(t) = \langle  \hat  c_{h,-\mathbf k}\hat  c_{e,\mathbf k} \rangle$ denotes the (interband) optical polarization and $\delta u_{\mathbf k}$ accounts for quantum fluctuations. We stress that the expansion above is exact, as long as the fluctuations are correctly calculated~\footnote{Equation \eqref{cDcDcc} defines the  fluctuations through $\delta u_{\mathbf k} \equiv u_{\mathbf k} - |p_{\mathbf k }|^2 - f_{e,\mathbf k} f_{h,- \mathbf k}$, which can be interpreted as the departure of $u_{\mathbf k}$ from its classical value.}. It can be shown that $\delta u_{\mathbf k}$ contributes with terms which are at least quartic in the coupling constants. Since we are interested in the low-coupling limit, we take $\delta u_{\mathbf k}  \simeq 0$ and retain only the single-particle contributions in Eq.~\eqref{cDcDcc}. \par 
The optical polarization evolves in time according to 
\begin{align}
    \frac{\partial}{\partial t} p_{\mathbf k} &= -i\Big(  \frac{E_{e,\mathbf k}+E_{h, \mathbf k}}{\hbar} - \frac{i}{2}\Gamma_{\uparrow,\mathbf k} - \frac{i}{2}\Gamma_{\downarrow,\mathbf k}\Big)  p_{\mathbf k} \nonumber \\
    &+  \frac{i}{\hbar}\sum_{m} g_m\Big( w_{e,\mathbf k ,m} + w_{h,-\mathbf k ,m} \Big) ,
\end{align}
where $w_{\nu,\mathbf k ,m} =  \langle \hat  a_m \hat  c_{\nu,\mathbf k}^\dagger \hat  c_{\nu,\mathbf k}\rangle$ defines a new correlator associated with intraband transitions mediated by photon exchange. A finite polarization signals the presence of an excitonic component in the semiconductor. However, for room temperature conditions, this contribution is usually small due to the small exciton binding energy, and therefore $p_{\mathbf k} \simeq 0$. Thus, to the desired order, the cluster expansion provides $ u_{\mathbf k} \simeq  f_{e,\mathbf k} f_{h,- \mathbf k}$. \par

Turning our attention to the remaining correlator $h_{\mathbf k,m}$, it is clear that neglecting quantum fluctuations as before leads to $h_{\mathbf k,m} \sim \langle \hat a_m\rangle = 0$ when an incoherent photon gas is considered. We must thus fully take into account the quantum fluctuations and perform the truncation only at a higher order.

From Eq.~\eqref{observableDynamics} we get
\begin{align}
\frac{\partial}{\partial t} h_{\mathbf k,m} &= -i \varphi_{\mathbf k,m}h_{\mathbf k,m} + Y_{\mathbf k,m} , \label{photon-assisted}
\end{align}
where the source term is given by
\begin{align}
    Y_{\mathbf k,m} &= i \sum_{m'} g_{m'}\matfont{N}_{m',m} - i g_m \sum_{\mathbf k'} u_{\mathbf k,\mathbf k'} \nonumber \\
   & - i \sum_{m'} g_{m'} \Big( l_{e, \mathbf k ,m',m} +  l_{h,- \mathbf k ,m',m} \Big)  
     . \label{Ykm}
\end{align}
Additionally, $ \varphi_{\mathbf k,m} \equiv  \Delta_{\mathbf k,m} -i \alpha_{\mathbf k,m}$ denotes a complex frequency of oscillation with real and imaginary parts given by
\begin{align}
     \Delta_{\mathbf k,m} &= \frac{E_{e,\mathbf k} +  E_{h,  \mathbf k}}{\hbar}-\omega_m   \label{detunig} \\
\text{and} \ \     \alpha_{\mathbf k,m} &= \frac{1}{2}(\Gamma_{\uparrow,\mathbf k} + \Gamma_{\downarrow,\mathbf k} + \kappa_m) .\label{alpha}
\end{align}
The real part of the frequency plays the role of the detuning for each event converting an electronic excitation with momentum $\mathbf k$ into an a cavity photon in mode $m$, and vice-versa. We also defined the higher-order correlators 
\begin{align}
 l_{\nu,\mathbf k,m,m'} &= \langle \hat  c_{\nu,\mathbf k}^\dagger \hat  c_{\nu,\mathbf k}\hat  a_m^\dagger \hat  a_{m'} \rangle , \label{photonphoton}\\
 u_{\mathbf k,\mathbf k'}  &= \langle \hat  c_{e,\mathbf k}^\dagger \hat  c_{h,-\mathbf k}^\dagger \hat  c_{h, - \mathbf k'} \hat  c_{e,\mathbf k'} \rangle,
\end{align}
contributing to the source term.

To proceed, Eq.~\eqref{photon-assisted} must be solved and its solution replaced back into Eqs.~\eqref{d_tf} and \eqref{d_tn}. Since this solution still features contributions from the higher-order correlators contained in $Y_{\mathbf k,m}$, it is convenient to cluster-expand these terms so to arrive at a solution containing only carrier and photon distributions. Hence, to second-order in the interactions and neglecting a small contribution from the polarization, we may write $ l_{\nu,\mathbf k,m,m'} = \matfont{N}_{m,m'} f_{\nu,\mathbf k}$ and $u_{\mathbf k,\mathbf k'} = \delta_{\mathbf k,\mathbf k'}f_{e,\mathbf k} f_{h,- \mathbf k}$. These terms provide nonlinear contributions to the time evolution of distribution functions, which mediate the allowed photon absorption and emission events. \par 
After applying the cluster expansion, and assuming that $h_{\mathbf k,m}$ vanishes for $t\rightarrow -\infty$, a solution to Eq.~\eqref{photon-assisted} reads
\begin{equation}
   h_{\mathbf k,m}(t) = \int_{-\infty}^0 d\tau \ e^{i \varphi_{\mathbf k,m} \tau} Y_{\mathbf k,m}(t+\tau) \label{sol1},
\end{equation}
which contains non-Markovian contributions by taking into account all instants of the past. Non-Markovian contributions included in $h_{\mathbf k,m}$ are a direct consequence of removing the latter from the chain. This feature is common to  the cluster-expansion technique, and is discussed at length in the quantum-kinetic literature (see, e.g., Refs.~\cite{Kira_Koch_2011,vasko2005quantum}).  

Locality in time can  be recovered by applying a Markovian approximation to Eq.~\eqref{sol1}. This consists in assuming that $Y_{\mathbf k,m}(t)$ is slow-varying in time when compared to $\text{Re}\,\varphi_{\mathbf k,m}$, $Y_{\mathbf k,m}(t+\tau) \simeq Y_{\mathbf k,m}(t)$, allowing the time integration to be performed. Then, with the help of the cluster expansion discussed above, Eq.~\eqref{d_tf} reduces to 
\begin{align}
   \frac{\partial}{\partial t} f_{e,\mathbf k} &=   \frac{\partial}{\partial t} f_{h, -\mathbf k} =  \mathcal J_{\mathbf k} + \mathcal A_{\mathbf k}, \label{d_tf_final}
\end{align}
where 
\begin{align}
\mathcal J_{\mathbf k} &= - \Gamma_{\downarrow,\mathbf k} f_{e,\mathbf k} f_{h,-\mathbf k} +  \Gamma_{\uparrow,\mathbf k} (1-f_{e,\mathbf k})(1- f_{h,-\mathbf k})\label{carrierLind}
\end{align}
is the Lindbladian contribution and
\begin{align}
\mathcal A_{\mathbf k}  &=  -\Lambda_{\downarrow,\mathbf k} f_{e,\mathbf k}  f_{h,-\mathbf k} + \Lambda_{\uparrow,\mathbf k} (1- f_{e,\mathbf k} )(1- f_{h,-\mathbf k} ) \label{carrierAbsEm}
\end{align}
represents electron-hole excitation and recombination via photon absorption and emission. In particular, $\Lambda_{\downarrow,\mathbf k}$ denotes the recombination rate of electron-hole pairs out of mode $\mathbf k$ due to spontaneous and stimulated photon emission,
\begin{equation}
\Lambda_{\downarrow,\mathbf k} = \frac{\pi}{\hbar^2} \sum_{m,m'} g_mg_{m'} o_{\mathbf k,m} n_{m',m} + \text{c.c.},
\label{eq:Gamma_down}
\end{equation}
while 
\begin{equation}
\Lambda_{\uparrow,\mathbf k}  =  \frac{\pi}{\hbar^2} \sum_{m,m'} g_mg_{m'} o_{\mathbf k,m} (\delta_{m',m} + n_{m',m} ) + \text{c.c.}, \label{L_uk}
\end{equation}
is the rate of carrier excitation due to photon absorption.
In Eq.~\eqref{L_uk} we defined the complex Lorentzian factor
\begin{equation}
    o_{\mathbf k,m} = \frac{1}{\pi}\frac{\alpha_{\mathbf k,m}+i\Delta_{\mathbf k,m}}{\Delta_{\mathbf k,m}^2 + \alpha_{\mathbf k,m}^2} ,
    \label{eq:ok}
\end{equation} 
with $\Delta_{\mathbf k, m}$ and $\alpha_{\mathbf k,m}$ being, respectively, the external and detuning rates defined in Eqs.~\eqref{detunig} and \eqref{alpha}. Applying the same approximations to Eq.~\eqref{d_tn} leads to
 \begin{align}
   \frac{\partial}{\partial t} \matfont{N}_{m,m'} &= -  (\kappa + i\omega_{m'}-i\omega_m)\matfont{N}_{m,m'} +  \matfont{B}_{m,m'} , \label{d_tn_final}
\end{align}
with 
\begin{align}
   & \matfont{ B}_{m,m'}  = \frac{1}{\hbar^2}\sum_{m''} g_{m''} \Big[ - g_{m'} W_{\downarrow,m} \matfont{N}_{m,m''} - g_{m}W_{\downarrow,m'} \matfont{N}_{m'',m'} \nonumber \\ 
   &g_{m'} W_{\uparrow,m} (\delta_{m,m''} + \matfont{N}_{m,m''})  + g_{m} W_{\uparrow,m'}^\ast (\delta_{m'',m'} + \matfont{N}_{m'',m'}) \Big]\label{photonAbsEm}
\end{align}
and 
\begin{align}
    W_{\uparrow,m} &= \pi\sum_{\mathbf k}  o_{\mathbf k,m} f_{e,\mathbf k} f_{h,-\mathbf k},\\
        W_{\downarrow,m} &= \pi\sum_{\mathbf k}  o_{\mathbf k,m} (1-f_{e,\mathbf k})(1- f_{h,-\mathbf k}).
\end{align}
Equation \eqref{photonAbsEm} should be interpreted as the photonic counterpart to $\mathcal A_{\mathbf k}$.   \par 
By defining the matrices 
\begin{align}
\big[\, \matfont{O}_{\mathbf k} \, \big]_{m,m'} &= \delta_{m,m'} o_{\mathbf k,m} \label{O_matrix} \\
\matfont{G}_{m,m'} &= \frac{1}{\hbar^2} g_m g_{m'}\label{eq:G} ,\\
\matfont{W}_\uparrow &= \pi\sum_{\mathbf k}  \matfont{O}_{\mathbf k} f_{e,\mathbf k} f_{h,-\mathbf k}, \\
\matfont{W}_\downarrow &= \pi\sum_{\mathbf k}  \matfont{O}_{\mathbf k} (1-f_{e,\mathbf k})(1- f_{h,-\mathbf k}) , 
\end{align}
the absorption and emission rates can be written in matrix form as 
\begin{align}
    \Lambda_{\uparrow, \mathbf k}  &= \text{tr}\Big[ \matfont{O}_{\mathbf k} \,\matfont{N} \, \matfont{G}\Big] + \text{c.c.}, \label{em}\\
      \Lambda_{\downarrow, \mathbf k}  &= \text{tr}\Big[ \matfont{O}_{\mathbf k} \,(\mathbb{1} + \matfont{N}) \, \matfont{G}\Big] + \text{c.c.}, \label{ab}
\end{align}
where 'tr' denotes the trace over photonic indices, $\text{tr} \,\matfont{M} \equiv \sum_m \matfont{M}_{m,m}$. Similarly, by defining the matrix $\matfont{B}$ with elements $\matfont{ B}_{m,m'}$, we may also rewrite Eq.~\eqref{photonAbsEm} in matrix form as
\begin{equation}
    \matfont{B} = \matfont{G} \,(\mathbb{1}+\matfont{N})\,  \matfont{W_\uparrow} - \matfont{G} \, \matfont{N} \,  \matfont{W_\downarrow} + \text{h.c.},
\end{equation}
such that Eq.~\eqref{d_tn_final} reduces to the anticipated Eq.~\eqref{eq1}.

\subsubsection{Carrier-carrier contribution}\label{sub_carrier_carrier}

The effect of carrier-carrier interactions to the rate equation~\eqref{d_tf_final} constitutes a fundamental mechanism responsible for the relaxation of optically excited carrier distributions, which plays a decisive role in the large excitation regime. This effect is determined by considering the contribution of the Coulomb Hamiltonian $\hat  H_{c-c}$ on the unitary evolution of carrier distributions as dictated by Eq.~\eqref{observableDynamics}. We denote this contribution by $\mathcal C_{\nu,\mathbf k}$, where the dependence on $\nu$ arises due to the mass difference between the bands. The form of $\mathcal C_{\nu,\mathbf k}$ can be derived with cluster expansion techniques, which are discussed in detail in Appendix~\ref{Ap_Coulomb_coll}. The final result is 
\begin{align}
\mathcal C_{\nu,\mathbf k}  =  \mu_{\uparrow,\nu, \mathbf k} (1-f_{\nu,\mathbf k} )-  \mu_{\downarrow,\nu, \mathbf k}f_{\nu,\mathbf k}  ,  \label{C_nu}
\end{align}
with Coulomb scattering rates given by
\begin{align}
\mu_{\uparrow,\nu, \mathbf k}&=  \sum_{\nu,\mathbf k',\mathbf q}  P^{\nu,\nu'}_{\mathbf k,\mathbf k'}(\mathbf q)  f_{\nu,\mathbf k + \mathbf q} f_{\nu',\mathbf k' - \mathbf q}(1-f_{\nu',\mathbf k'} ) \label{mu_up},\\
\mu_{\downarrow,\nu, \mathbf k}&=  \sum_{\nu,\mathbf k',\mathbf q} P^{\nu,\nu'}_{\mathbf k,\mathbf k'}(\mathbf q) f_{\nu',\mathbf k'}(1-    f_{\nu,\mathbf k + \mathbf q} )(1-f_{\nu',\mathbf k' - \mathbf q}) \label{mu_down}.
\end{align}
The factor $P^{\nu,\nu'}_{\mathbf k,\mathbf k'}(\mathbf q)$ is determined from 
\begin{align}
    P^{\nu,\nu'}_{\mathbf k,\mathbf k'}(\mathbf q) &= \frac{2\pi}{\hbar} |V^{\nu,\nu'}_\mathbf{q}|^2 \nonumber \\
    &\times \delta(E_{\nu,\mathbf k} + E_{\nu',\mathbf k'} - E_{\nu,\mathbf k + \mathbf q} - E_{\nu',\mathbf k' - \mathbf q}). \label{ProbScatt}
\end{align}
and corresponds to the probability per unit time for a Coulomb scattering event of the type
\begin{equation}
	(\nu,\mathbf k)+(\nu',\mathbf k') \longrightarrow (\nu,\mathbf k + \mathbf q)+(\nu',\mathbf k'-\mathbf q),
\end{equation}
with $\mathbf q$ the transferred momentum. In the high-density limit (high fermion degeneracy), the Coulomb interaction is the main responsible for redistributing the carriers in each band towards equilibrium, which is an essential mechanism for photon condensation to occur.  

\subsubsection{Carrier-phonon contribution}\label{sub_carrier_phonon}

While carriers and photons are inherently part of the quantum system of interest, the phonons, being delocalized over the entire heterostructure, constitute a large thermal reservoir that is maintained at room temperature due to the strong coupling to its surroundings. For this reason, it is convenient to remove the phonon distributions from the dynamical state functions in our description, under the assumption of a large thermal reservoir. That can be achieved by applying the Born-Markov approximation to the carrier-phonon interaction and assuming that the correlations with phonon degrees of freedom are dynamically eliminated on a time scale $\tau_c \sim 100\,$fs~\cite{10.1063/1.3117236}, much smaller than all the other time scales of the system.\par

In Appendix~\ref{Ap_Phonon} we show how to transform the carrier-phonon coherent evolution onto Lindbladian incoherent terms, resulting in contributions similar to those of the external pumping and losses featuring Eq.~\eqref{masterEq}. The result is a phonon-scattering term on the right-hand side of \eqref{d_tf_final} denoted by $ \mathcal Q_{\mathbf \nu,\mathbf k}$, which accounts for phonon absorption and emission. It reads 
\begin{equation}
	 \mathcal  Q_{\mathbf \nu,\mathbf k}= \eta_{\uparrow,\nu,\mathbf k} (1- f_{\nu, \mathbf k} )  -\eta_{\downarrow,\nu,\mathbf k}  f_{\nu, \mathbf k} ,
\end{equation}
with $\eta_{\downarrow,\nu,\mathbf k}$ and $\eta_{\uparrow,\nu,\mathbf k}$ being, respectively, the outcoming and incoming scattering rates due to carrier-phonon events. According to the discussion in Appendix~\ref{Ap_Phonon}, the phonon system is maintained in thermal equilibrium, with $\langle \hat b_{\mathbf q}^\dagger \hat b_{\mathbf q} \rangle  = \mathcal N( \hbar \omega_{LO}) $ valid for all times. Here $\mathcal N(x)$ denotes the Bose-Einstein distribution at room temperature. With this in mind, the above rates read
\begin{align}
	 \eta_{\uparrow,\nu,\mathbf k} &= \frac{2\pi}{\hbar}\sum_{\mathbf q}|\lambda_{\mathbf q}|^2 f_{\nu,\mathbf k-\mathbf q} \Big[ \mathcal N \delta(E_{\nu,\mathbf k} - E_{\nu, \mathbf k - \mathbf q} - \hbar\omega_{LO}) \nonumber \\ 
	& +   (1+ \mathcal N)\delta(E_{\nu,\mathbf k} - E_{\nu, \mathbf k - \mathbf q} + \hbar\omega_{LO}) \Big]\label{phonon_rate_up} ,\\ 
	\eta_{\downarrow,\nu,\mathbf k} &= \frac{2\pi}{\hbar} \sum_{\mathbf q}|\lambda_{\mathbf q}|^2 f_{\nu,\mathbf k-\mathbf q} \Big[ \mathcal N \delta(E_{\nu,\mathbf k} - E_{\nu, \mathbf k - \mathbf q} -\hbar\omega_{LO}) \nonumber \\ 
	& +   (1+ \mathcal N)\delta(E_{\nu,\mathbf k} - E_{\nu, \mathbf k - \mathbf q} + \hbar\omega_{LO}) \Big]\label{phonon_rate_down}.
\end{align}
To the expressions above contribute both phonon absorption ($\sim \mathcal N$) and spontaneous/stimulated emission ($\sim 1+\mathcal N$). In particular, the first and second terms in Eq.~\eqref{phonon_rate_up} correspond to the scattering of carriers from mode $\mathbf k+\mathbf q$ to mode $\mathbf k$ through the absorption and spontaneous plus stimulated emission of a phonon in mode $\mathbf q$, respectively. A similar interpretation applies to Eq.~\eqref{phonon_rate_down}. 

\par 

After joining the results derived in Secs.~\ref{sub_photon_carrier}--\ref{sub_carrier_phonon}, we are immediately led to the anticipated Eq.~\eqref{eq2}. A list summarizing the different scattering rates is given in Tab.~\ref{tab0}.

\subsection{Diagonal approximation}
\label{subsec:photonrateapproximation}

Equation ~\eqref{eq1} defines an equation of motion for the photon correlator $\matfont{N}_{m,m'}=\langle \hat a^\dagger_m \hat a_{m'}\rangle$.
The diagonal terms $m=m'$ therein represent the average occupation of each cavity mode, while off-diagonal terms are associated with mode correlations. Both diagonal and off-diagonal terms contribute to electromagnetic observables, such as the light intensity, defined by
\begin{equation}
	I(\mathbf r,t) = \sum_{m,m'} \mathcal E_{m}^\ast(\mathbf r)\mathcal E_{m'}(\mathbf r) \matfont{N}_{m,m'}(t), \label{intensity}
\end{equation}
or photon statistics $g^{(1)}$ and $g^{(2)}$. Therefore, in order to accurately reproduce intensity profiles for generic states of light, both occupations and correlations must be included as dynamical variables. The number of independent photonic variables scales as $\sim M^2$ with the total number of cavity modes $M$, which rapidly becomes untreatable. Thus, it is convenient to include only the elements associated to the biggest contributions to Eq.~\eqref{intensity}. \par 
Solutions of Eq.~\eqref{eq1} take the form $\matfont{N}_{m,m'} = e^{-\kappa t} e^{i(\omega_m - \omega_{m'})t} F_{m,m'}(t)$, with $F_{m,m'}(t)$ some function of time. When $F_{m,m'}(t)$ varies slowly as compared to the energy differences $\omega_m - \omega_{m'}$, the phase factor $e^{i(\omega_m - \omega_{m'})t}$ results in a large separation of time scales between occupations and off-diagonal correlations. Assuming that this condition is verified, we can integrate Eq.~\eqref{d_tn_final} in time with respect to the fast time scales in order to approximately eliminate the highly oscillatory correlations.
The result is the photon-rate equation
\begin{equation}
    \frac{\partial}{\partial t} n_m = -(\kappa + \gamma_{\downarrow,m}^\text{(diag.)}) n_m + \gamma_{\uparrow,m}^\text{(diag.)} (1+n_m), \label{photon_rate}
\end{equation}
for the occupations $\matfont{N}_{m,m} \equiv n_m$,
where the photon rates 
\begin{align}
\gamma_{\uparrow,m}^\text{(diag.)}  &=\frac{2\pi}{\hbar} |g_m|^2 \sum_{\mathbf k}  L_{\mathbf k,m}f_{e,\mathbf k}  f_{h,-\mathbf k}, \label{g_u}\\  \gamma_{\downarrow,m}^\text{(diag.)} &=\frac{2\pi}{\hbar} |g_m|^2 \sum_{\mathbf k} L_{\mathbf k,m} (1- f_{e,\mathbf k} )(1- f_{h,-\mathbf k} ). \label{g_d}
\end{align}
are scalars resulting from the matrices $\matfont{W}_\uparrow$ and $\matfont{W}_\downarrow$ in Eq.~\eqref{W_u} and \eqref{W_d}. The factors
\begin{equation}
   L_{\mathbf k,m} \equiv \Re(o_{\mathbf k,m}) = \frac{1}{\pi} \frac{\alpha_{\mathbf k,m}}{\Delta_{\mathbf k,m}^2 + \alpha_{\mathbf k,m}^2 } \label{Lorentzian}
\end{equation}
have a Lorentzian form, with center at $\Delta_{\mathbf k,m} = 0$ and width $\alpha_{\mathbf k,m}$ [see Eq.~\eqref{alpha}]. 

Given the diagonal approximation for the correlation matrix $\matfont{N}$, the equation of motion for carriers [Eq.~\eqref{eq2}] reduces to
\begin{align}
&\frac{\partial}{\partial t} f_{\nu, \mathbf{k}} =  - \Big[(\Gamma_{\downarrow,\mathbf{k}} + \Lambda_{\downarrow,\mathbf{k}}^\text{(diag.)}) f_{\overline{\nu}, -\mathbf{k}} + \mu_{\downarrow,\nu,\mathbf{k}} + \eta_{\downarrow,\nu,\mathbf{k}}\Big]f_{\nu,\mathbf{k}} \nonumber \\
&+\Big[ (\Gamma_{\uparrow,\mathbf{k}} + \Lambda^\text{(diag.)}_{\uparrow,\mathbf{k}})(1 - f_{\overline{\nu}, -\mathbf{k}}) + \mu_{\uparrow,\nu,\mathbf{k}} + \eta_{\uparrow,\nu,\mathbf{k}}\Big] (1 - f_{\nu,\mathbf{k}}) \label{carrier_rate}
\end{align}
with scalar effective rates
\begin{align}   
\Lambda_{\uparrow,\mathbf k}^\text{(diag.)} &= \frac{2\pi}{\hbar} \sum_m |g_m|^2 L_{\mathbf k,m} n_m
,
\label{lambda_up_diag} \\
\Lambda_{\downarrow,\mathbf k}^\text{(diag.)} &=\frac{2\pi}{\hbar} \sum_m |g_m|^2 L_{\mathbf k,m} (1+n_m)\ ,
\label{lambda_down_diag}
\end{align}
which are now independent of off-diagonal elements of $\matfont{N}$. Since the rates $\mu_{\circ,\nu,\mathbf{k}}$ and $\eta_{\circ,\nu,\mathbf{k}}$ for carrier-carrier interactions and carrier phonon interactions do not depend on the state of the light in the cavity, Eqs.~\eqref{mu_up} and \eqref{mu_down} are left unchanged.

Equations~\eqref{photon_rate} and~\eqref{carrier_rate} define the full nonequilibrium transport dynamics in a semiconductor microcavity, resolving both photon and carrier distributions with explicit inclusion of environmental effects. The latter essentially replace energy-conserving delta functions by Lorentzian functions, which results in an energy broadening for the allowed light-matter scattering events. In other words, inelastic carrier-photon scattering becomes possible because the energy difference can be exchanged with the environment. When $\alpha_{\mathbf k,m}\rightarrow 0$, which corresponds to the limit of an isolated cavity, we recover energy conservation, $ L_{\mathbf k,m} \rightarrow \delta(\Delta_{\mathbf k,m})$, such that only elastic light-matter transitions are allowed. In this limit, the light-matter rates defined above verify Fermi's golden rule. 

\begin{table*}[t]
  \caption{Summary of scattering rates.}
  \label{tab0}
  \centering
  \renewcommand{\arraystretch}{1.2}
  \begin{ruledtabular}
    \begin{tabular}{lll}
      \textbf{Species} & \textbf{Rate} & \textbf{Description} \\
      \hline
      \textit{Carriers} & $\Gamma_{\uparrow,\mathbf{k}}$   & Pumping rate \\
                        & $\Gamma_{\downarrow,\mathbf{k}}$ & Electron--hole recombination rate (nonradiative) \\
                        & $\Lambda_{\uparrow,\mathbf{k}}$  & Electron--hole generation rate (radiative) \\
                        & $\Lambda_{\downarrow,\mathbf{k}}$& Electron--hole recombination rate (radiative) \\
                        & $\mu_{\nu,\uparrow,\mathbf{k}}$  & Coulomb ingoing scattering rate \\
                        & $\mu_{\nu,\downarrow,\mathbf{k}}$& Coulomb outgoing scattering rate \\
                        & $\eta_{\uparrow,\nu,\mathbf{k}}$ & Ingoing scattering rate due to phonon processes \\
                        & $\eta_{\downarrow,\nu,\mathbf{k}}$& Outgoing scattering rate due to phonon processes \\
      \hline
      \textit{Photons} & $\kappa$              & Cavity-loss rate \\
                       & $\gamma_{\uparrow,m}$ & Photon emission rate \\
                       & $\gamma_{\downarrow,m}$ & Photon absorption rate \\
    \end{tabular}
  \end{ruledtabular}
\end{table*}

\subsection{A comparison between dye-based and semiconductor-based quantum kinetic theories}

A central motivation to realize photon BECs with semiconductor environments are the differences in the thermalization processes as compared to the well-established case of dye-based systems. The differences in the thermalization properties necessarily lead to differences between the respective quantum-kinetic models.

The goal of this section is thus to identify the central differences between the model of Eqs.~\eqref{photon_rate},~\eqref{carrier_rate} and the quantum kinetic theory~\cite{PhysRevLett.111.100404} of photon BEC in dye-filled microcavities.

\subsubsection{A brief reviw of the quantum kinetic theory of dye-based photon condensation}

The Kirton--Keeling (KK) model treats $N$ dye molecules as two-level systems in near-thermal equilibrium with a solvent bath. The KK rate equations couple the photon occupation $n_m = \langle a_m^\dagger a_m \rangle$ in mode $m$ to a single excitation fraction $f(\mathbf r)$ of dye molecules located at position $\mathbf r$. Specifically, the photon number follows the equation of motion
\begin{equation}
	\frac{\partial}{\partial t}  n_m =  - \big[\kappa + (1-f)\gamma^\text{(KK)}(\delta_m) \big] n_m + f\gamma^\text{(KK)}(-\delta_m) (1+n_m),  \label{KK_1}
\end{equation}
while the excitation fraction obeys
\begin{align}
&\frac{\partial}{\partial t} f(\mathbf r) = -\Big[  \Gamma_{\downarrow} + \sum_m |\psi_m(\mathbf r) |^2\, \gamma^\text{(KK)}(-\delta_m) (1+n_m)\Big]f \nonumber \\
	&  + \Big[ \Gamma_{\uparrow}  + \sum_m |\psi_m(\mathbf r)|^2 \, \gamma^\text{(KK)}(\delta_m)n_m\Big] (1-f) ,   \label{KK_2}
\end{align} 
with $\psi_m(\mathbf r)$ the Hermite-Gauss mode functions and rates $\gamma^\text{(KK)} (\delta_m)$ that depend on the detuning $\delta_m$ between the cavity frequency $\omega_m$ and the zero-point rovibrational frequency, as well as on the temperature of the dye system. 

The rovibrational states are assumed to be in thermal equilibrium at temperature $T$, which motivates the use of a Kennard-Stepanov (KS) relation for the ratio
$\gamma^\text{(KK)}(\delta) = e^{\hbar \delta/k_BT}\gamma^\text{(KK)}(-\delta)$
of excitation and deexitation rates of the dye molecules~\cite{PhysRevA.93.013829}.
This detailed balance between absorption and emission enforced by the solvent via Kennard--Stepanov relations leads to a Bose--Einstein-like photon distribution at an effective temperature set by the dye bath.
The KK model successfully explains photon condensation in dyes by assuming an ideal thermal reservoir of dye molecules that rapidly thermalizes through vibrational collisions.
However, it neglects any back-action of the photon gas on the (molecular) vibrational distribution by assuming that the dye remains at fixed temperature throughout. \par 

\subsubsection{Main differences}

Contrarily to the KK model, where the dye acts as an infinite reservoir that can absorb or supply quanta without being appreciably perturbed from equilibrium, the corresponding photon-carrier dynamics for the semiconductor is fully energy-conserving.

While the single excitation fraction $f$ in Eq.~\eqref{KK_2} is a spatially varying quantity, the analogue quantity in the semiconductor model are the carrier distributions $f_{e,{\bf k}}$ (for electrons) and $f_{h,{\bf k}}$ (for holes) in Eq.~\eqref{carrier_rate} that have a momentum dependence, but no spatial dependence. 
The spatial dependence of $f$ in the dye model results in increased competition for excitations among cavity modes with mode functions of overlapping maxima, whereas there is substantially reduced competition between modes  with wave functions with non-overlapping maxima.
This competition governs, to a large extent, the transitions between the distinct phases of the dye system~\cite{PhysRevLett.120.040601}.

According to Eq.~\eqref{KK_2}, the rates for excitation and deexcitation of a dye molecule at position $\mathbf r$ are given by
\begin{equation}
    \sum_m |\psi_m(\mathbf r)|^2 \, \gamma^\text{(KK)}_{\downarrow}(\delta_m)n_m \label{KK_ex_rate}
\end{equation}
and 
\begin{equation}
    \sum_m |\psi_m(\mathbf r) |^2\, \gamma^\text{(KK)}_{\uparrow}(\delta_m) (1+n_m),\label{KK_deex_rate}
\end{equation}
respectively, with $\gamma^\text{(KK)}_{\uparrow}(\delta_m)$ a function of the (constant) temperature of the dye mixture. The analogue rates of excitation and deexcitation of $\nu-$carriers in the semiconductor model are, according to Eqs.~\eqref{lambda_up_diag} and~\eqref{lambda_down_diag}, given by
\begin{equation}
     \frac{2\pi}{\hbar} (1 - f_{\overline{\nu}, -\mathbf{k}})\sum_m |g_m|^2 L_{\mathbf k,m} n_m  \label{S_ex_rate}
\end{equation}
and 
\begin{equation}
    \frac{2\pi}{\hbar} f_{\overline{\nu}, -\mathbf{k}} \sum_m |g_m|^2 L_{\mathbf k,m} (1+n_m) ,\label{S_deex_rate}
\end{equation}
respectively. These semiconductor rates include energy-dependent Lorentzian factors $L_{\mathbf{k},m}$ [Eq.~\eqref{Lorentzian}] that take into account the pump rate and hence play an important role in establishing the steady-state carrier distribution. Contrarily to the dye system, which is maintained in equilibrium at room temperature, the thermalization of semiconductor carrier distributions is a dynamical process that is fully described by Eq.~\eqref{carrier_rate}. Additionally, semiconductor rates of Eqs.~\eqref{S_ex_rate} and~\eqref{S_deex_rate} depend explicitly on the carrier distributions $f_{e,{\bf k}}$ and $f_{h,{\bf k}}$ themselves, while Eqs.~\eqref{KK_ex_rate} and~\eqref{KK_deex_rate} are independent of $f$. Such nonlinear feedback between the carrier distributions and the scattering rates is responsible for several distinctive features of semiconductor platforms, including pump-induced shifts of the condensation threshold and nontrivial spectral redistribution of carriers under strong excitation, both of which have been observed in experiments~\cite{rozas2018temperature,schneider2013electrically}. 

The semiconductor emission rate into photon mode $m$ is obtained by summing over all electron-hole pairs that can recombine into that mode. According to Eq.~\eqref{g_u}, the quantity
\begin{equation*}
  f \gamma_{\uparrow}^\text{(KK)}(\delta_m)  
\end{equation*}
in the dye model should be compared with the semiconductor emission rate
\begin{align*}
 \frac{2\pi}{\hbar} |g_m|^2 \sum_{\mathbf k} L_{\mathbf k, m} f_{e,\mathbf k}f_{h,-\mathbf k} .
\end{align*}
Correspondingly, the absorption rate [Eq.~\eqref{g_d}] in the semiconductor model includes the double Pauli blocking factors for the electron and hole distributions,
\begin{equation*}
  \frac{2\pi}{\hbar} |g_m|^2 \sum_{\mathbf{k}} L_{\mathbf{k}, m} \big(1 - f_{e,\mathbf{k}}\big)\big(1 - f_{h,-\mathbf{k}}\big),
\end{equation*}
which plays a role analogous to the dye model's rate 
\begin{equation*}
(1 - f)\gamma_{\uparrow}^\text{(KK)}(\delta_m).
\end{equation*}

In the limit of weak pumping and fast thermalization, the semiconductor model [Eqs.~\eqref{photon_rate} and \eqref{carrier_rate}] predicts that the cavity’s absorption-to-emission ratio obeys the van Roosbroeck–Shockley (vRS) relation [see Fig.~\ref{fig_vRS_profiles}], representing the quantum analogue of the Kennard–Stepanov law~\cite{10.1063/1.4721495} in dye systems.
This ensures that thermally excited carriers provide an effective bath for the photons. However, the semiconductor model is not restricted to this limit, and can also capture the opposite regime where pumping is so strong that the timescale of photon-condensation becomes comparable to the timescale of carrier thermalization.
In such cases, the assumption of a static thermal reservoir breaks down, and a transient, partial equilibrium develops between the photon gas and the electron-hole plasma~\cite{PhysRevLett.131.146201}.
\par

Beyond these formal generalizations, the semiconductor model introduces other key physical differences from the KK dye model. Most notably, the light-matter interactions are now governed by pair statistics, {\it i.e.}, photon can only be absorbed if an electron is available in the valence band and, simultaneously, an empty state is available in the conduction band. This requirement appears as the product $(1-f_{e,\mathbf{k}})(1-f_{h,-\mathbf{k}})$ in Eq.~\eqref{g_d}. Likewise, stimulated emission into mode $m$ requires an occupied electron state and an occupied hole state of momentum $\mathbf{k}$, as reflected by the $f_{e,\mathbf{k}}f_{h,-\mathbf{k}}$ factor in Eq.~\eqref{g_u}. These bilinear terms imply a saturable gain and absorption due to the coupling between both bands, which differs from the KK model where saturability enters in a simpler way through the finite fraction $f$ of excited dye molecules (stimulated processes scale linearly in $f$ or $1-f$).  \par 

Another crucial difference is that thermal equilibrium is not enforced externally in the semiconductor model.
While the KK theory assumes the dye's rovibrational state remains thermally distributed at all times as a result of fast vibronic relaxation in the solvent, in our model no such assumption is made, and the carrier populations thermalize via Coulomb collisions and phonon scattering (with rates denoted by $\mu$ and $\eta$).
This means that under strong driving, the carrier plasma can depart from equilibrium (e.g. developing a non-thermal energy tail or a population inversion), such that the photons interact with a time-dependent electron-hole mixture that is not thermal  -- a regime that the KK model cannot describe.

Although Eq.~\eqref{KK_2} includes direct pumping of the cavity and decay from the cavity in terms of constant rates $\Gamma_{\uparrow}$ and $\Gamma_{\downarrow}$, the analogue relation of the semiconductor model contains additional terms associated with carrier-carrier and carrier-phonon events. Comparing Eqs.~\eqref{carrier_rate} and~\eqref{KK_2} allows to establish the following correspondence between dye and semiconductor models,
\begin{align*}
	\Gamma_{\uparrow} &\longrightarrow \Gamma_{\uparrow}(1 - f_{\overline{\nu},-\mathbf k}) + \mu_{\uparrow,\nu,\mathbf k} + \eta_{\uparrow, \nu,\mathbf k},\\
	\Gamma_{\downarrow} &\longrightarrow \Gamma_{\downarrow} f_{\overline{\nu},-\mathbf k} + \mu_{\downarrow,\nu,\mathbf k} + \eta_{\downarrow, \nu,\mathbf k}.
\end{align*}
with Coulomb ($\mu_{\uparrow/\downarrow,\nu,\mathbf k}$) and phonon ($\eta_{\uparrow/\downarrow,\nu,\mathbf k}$) terms given in Eqs.~\eqref{mu_up}--\eqref{mu_down} and \eqref{phonon_rate_up}--\eqref{phonon_rate_down}, respectively. In terms of the carrier distributions, the Coulomb collision rate scales as $\mu \sim f^3$ while the phonon rate $\eta \sim f$. Thus, for small excitation number, thermalization of matter is promoted by a competition between photon and phonon processes, while for large excitation ({\it i.e.}, large pumping), thermalization is mainly a result of carrier--carrier collisions. A summary of the main differences between Kirton-Keeling model and the present theory is enclosed in Table \ref{tab:comparison}.

\begin{table*}[t]
  \caption{Comparative summary of the main features of the theory of dye-based and semiconductor photon BEC systems.}
  \label{tab:comparison}
  \centering
  \renewcommand{\arraystretch}{1.3}
  \begin{ruledtabular}
    \begin{tabular}{lll}
      \textbf{Feature} &
      \textbf{Dye Systems (Kirton--Keeling \cite{PhysRevLett.111.100404})} &
      \textbf{Semiconductor (This Work)} \\
      \hline
      Thermalization mechanism &
      Fixed solvent temperature maintains thermal equilibrium &
      Emergent temperature from pumping + carrier interactions \\
      Key thermalization process &
      Vibrational coupling to molecular rovibrational states &
      Coulomb scattering (high densities) or phonon scattering (low densities) \\
      Critical limitation &
      No photon back-action on molecular reservoir &
      Self-consistent carrier dynamics modified by photon gas \\
    \end{tabular}
  \end{ruledtabular}
\end{table*}

\section{Semiconductor quantum-well characterization}\label{SQC}

Before addressing the emergence of condensation, it is essential to characterize the underlying relaxation dynamics of the semiconductor quantum well. In this section, we analyze how each interaction channel—carrier–carrier, carrier–photon, and environment coupling—shapes the approach to equilibrium. Understanding the relative strength and timescales of these processes is crucial for identifying the regimes in which thermalization becomes efficient, and thus for establishing the physical conditions that enable condensation. The interplay between competing interactions is governed by the ratios of the corresponding rates listed in Table~\ref{tab0}, which must be determined self-consistently from the evolving photon and carrier distributions.

Since closed solutions to the rate equations are unavailable, we employ a fourth-order Runge-Kutta scheme to propagate the system in time. At each time step, the scattering rates derived in Sec.~\ref{TSPBEC} are evaluated self-consistently using the current state of the system. For computational efficiency, all simulations are performed assuming isotropic carrier distributions, {\it i.e.}, $f_{\nu, \mathbf k} \equiv f_{\nu, k}$ with $k \equiv |\mathbf k|$. \par
Motivated by experimental implementations, we assume optical pumping is provided by a continuous-wave laser with frequency $\omega_L$ and spectral linewidth $ \delta\omega_L $, verifying $\hbar \omega_L > E_g $. Given that $ \omega_L $ lies in the optical regime, the corresponding wavelength is much larger than the lattice constant. Consequently, the laser field can be considered spatially homogeneous over the crystal unit cell, and its coupling to electron-hole states is approximately independent of momentum $ k $. This justifies modeling the $ k $-dependent pumping rate using a Lorentzian profile:
\begin{equation}
    \Gamma_{\uparrow, k} \sim  \frac{ \Gamma_{\uparrow}}{\pi}\frac{\delta\omega_L}{(\frac{E_{e, k} + E_{h, k}}{\hbar} - \omega_L)^2 + \delta\omega_L^2}, 
\end{equation}
corresponding to the laser spectral density evaluated at the energy detuning of each electron-hole pair transition relative to the laser frequency. Above, $\Gamma_{\uparrow}$ denotes the overall pumping scale. Additionally, we restrict our analysis to the one-dimensional cavity spectrum and assume that the carrier decay rate is a constant for all states, $\Gamma_{\downarrow,  k} \equiv \Gamma_{\downarrow}$.  A list summarizing the relevant parameters is given in Table~\ref{tab1}, together with reference values used in the simulations. 

\begin{table*}[t]
  \caption{List of simulation parameters and reference values for InGaAs quantum wells.}
  \label{tab1}
  \centering
  \renewcommand{\arraystretch}{1.2}
  \begin{ruledtabular}
    \begin{tabular}{lll}
      \textbf{Category} & \textbf{Parameter} & \textbf{Description, Value} \\
      \hline
      \textit{Laser} & $\Gamma_\uparrow$   & Pumping rate, $1$--$10^3$~THz \\
                     & $\omega_L$          & Laser frequency, $380$~THz \\
                     & $\delta\omega_L$    & Laser linewidth, $100$~MHz \\
      \hline
      \textit{Material / Geometric} & $\kappa$            & Cavity loss rate, $0.01$--$0.1$~THz \\
                     & $\Gamma_\downarrow$ & Carrier recombination rate, $10^{-3}$--$10^{-2}$~THz \\
                     & $\omega_0$          & Cavity ground-state energy, $315$--$330$~THz \\
                     & $\delta\omega$      & Mode spacing, $0.03$~THz \\
                     & $E_g$               & Band gap, $1.3$~eV \\
                     & $a$                 & Lattice constant, $0.59$~nm \\
                     & $m_e$               & Electron mass, $0.041\,M_e$ \\
                     & $m_h$               & Hole mass, $0.5\,M_e$ \\
                     & $A$                 & Quantum-well area, $1~\text{cm}^2$ \\
    \end{tabular}
  \end{ruledtabular}
\end{table*}

\subsection{Coulomb dynamics and equilibration}\label{CDE}
In what follows, we start by characterizing the evolution of electronic distributions driven solely by Coulomb interactions. To do so, we start by neglecting interactions with light and the lattice by setting $g_m=\lambda_{\mathbf q} = 0$, which allow us to focus on the effects of carrier-carrier collisions. Later, after introducing the light emission and absorption processes, we will show how the carrier evolution is modified by the presence of a nonzero $g_m$. Nevertheless, the main features of electronic equilibration in high-excitation regimes can be understood by taking into account carrier-carrier collisions only.\par
 
Let us assume that the system is initially in its ground state, which corresponds to a fully occupied valence band and empty conduction band, {\it i.e.}, $f_{\nu,  k}(0) = 0$ for both electrons and holes. At $t=0^+$ the pumping is switched on, leading to the generation of electron-hole pairs. These excitations accumulate around a peak centered at the characteristic momentum $k_p$ defined by $k_p =  K(\omega_L)$, where the function 
\begin{equation}
   K(\omega)= \frac{1}{\hbar}\sqrt{2\overline{m}( \hbar\omega - E_g)} \label{K_Omega}
\end{equation}
satisfies the resonance condition for parabolic bands $E_{e,K(\omega)} + E_{h,K(\omega)} -  \hbar\omega = 0$, with $\overline{m}$ being the reduced exciton mass, which assumes negligible photon momentum exchange with carriers. The resulting peak in the carrier distribution subsequently broadens due to Coulomb collisions as the electron-hole plasma relaxes towards equilibrium.\par 

\begin{figure}
\centering 
\hspace{-0.3cm}
\includegraphics[scale=0.42]{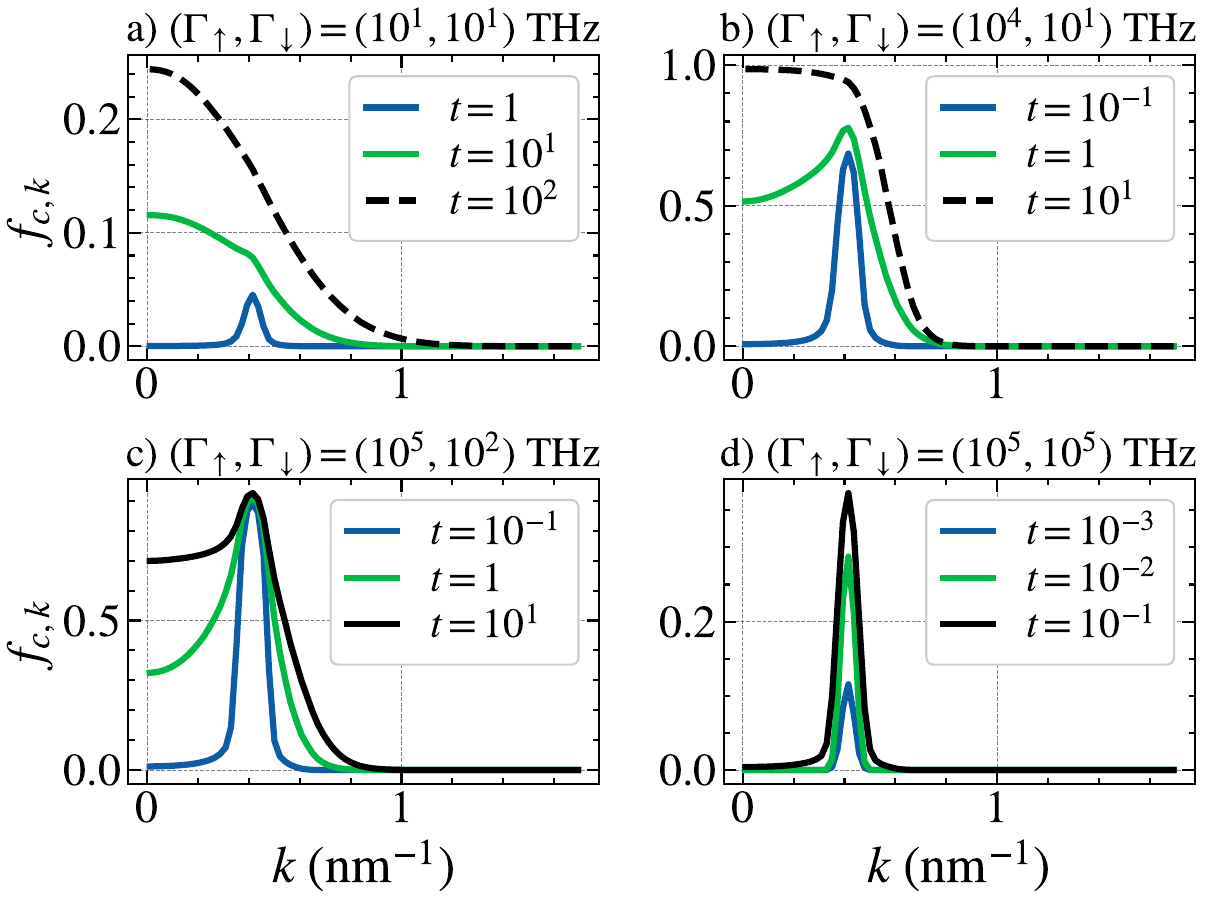}
\caption{(color online). Time evolution of the electron distribution function for distinct values of $\Gamma_{\uparrow}$ and $\Gamma_{\downarrow}$ indicated in the titles. Black curves correspond to the steady-state distribution for each case. The dashed curves of the top panels are a result of fitting the (steady-state) data to a Fermi-Dirac distribution, indicating that thermalization was reached. Contrarily, the steady-state distributions of the bottom panels do not reach a Fermi-Dirac shape. Legends display the time coordinate in units of $10^{-15}\,$s. Other simulation parameters are $\omega_L = 2 \pi \times 380\,$THz, $\delta\omega_L= 100\,$MHz, $E_g = 1.3\,$eV, $m_e = 0.04 M_e$, $m_h = 0.5 M_e$, with $M_e$ the bare electron mass.}
\label{fig_electron_profiles}
\end{figure}

For sufficiently low pumping and non-radiative carrier-loss rates, Coulomb collisions are effective in thermalizing the electronic distributions. Under these conditions, the steady-state carrier populations converge toward a Fermi--Dirac distribution with a common temperature, indicating the onset of thermal equilibrium. Conversely, when $\Gamma_{\uparrow}$ and $\Gamma_{\downarrow}$ exceed a critical threshold, excitation and relaxation via the environment outpace Coulomb scattering, and the system fails to thermalize. Representative scenarios are illustrated in Fig.~\ref{fig_electron_profiles}, where we show the electron distribution function for increasing times and different values of $(\Gamma_{\uparrow}, \Gamma_{\downarrow})$.\par 

The top panels of Fig.~\ref{fig_electron_profiles} illustrate steady-state carrier distributions that have reached thermal equilibrium (dashed lines). In panel a), the system is in the non-degenerate regime, and the resulting distribution is well described by a Maxwell--Boltzmann tail. In contrast, panel b) corresponds to a degenerate case, where the conduction-band occupation reaches $f_{e,k} \sim 1$ for low-energy states, resulting in a Fermi--Dirac-like profile. The degree of degeneracy is controlled by the pumping strength, which sets the carrier density. Higher densities lead to faster thermalization due to increased scattering rates, as evidenced by the shorter equilibration time in the degenerate regime. \par 

For stronger coupling to the environment -- either by enhancing the pumping rate or the non-radiative carrier decay --  Coulomb collisions become less effective in driving the system toward thermal equilibrium. This behavior is illustrated in the bottom panels of Fig.~\ref{fig_electron_profiles}. In panel c), partial thermalization is observed:  although a portion of the carrier population relaxes toward a Fermi-Dirac distribution, a distinct non-thermal peak remains near $k=k_p$ indicating the residual carrier generation at the pump frequency. This non-equilibrated fraction persists even at long times. In panel d), the effect becomes more pronounced, and the entire carrier distribution remains peaked around $k=k_p$, signaling the breakdown of thermalization. The latter case represents the strongly driven regime in which exchange with the environment dominates over carrier-carrier scattering. \par 

\begin{figure}
\centering 
\hspace{0.6cm}
\includegraphics[scale=0.55]{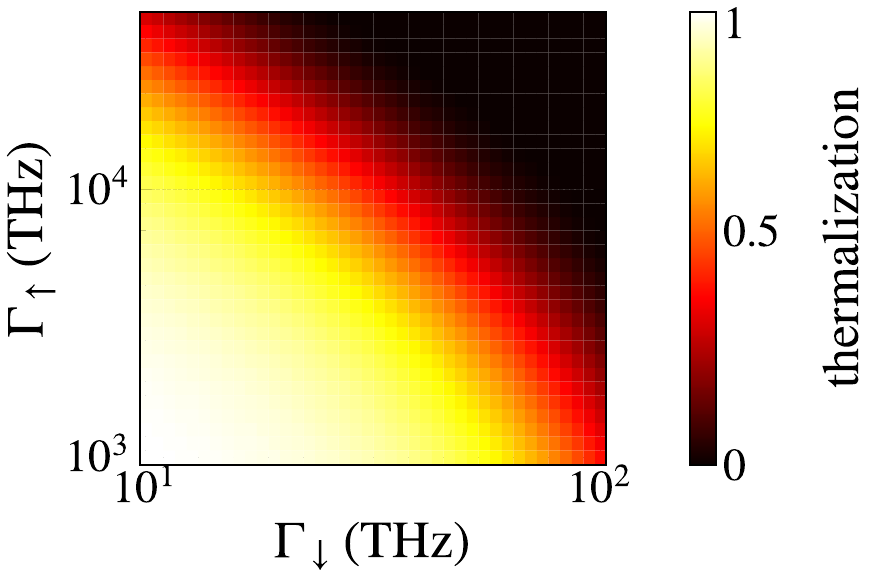}
\caption{(color online). Steady-state thermalization degree as a function of the pumping and (nonradiative) carrier-recombination rates. Unit thermalization degree is attained when the steady-state distribution is well described by a Fermi-Dirac function [Fig.~\ref{fig_electron_profiles} a) and b)], while a value of zero refers to a fully unthermalized equilibrium [Fig.~\ref{fig_electron_profiles} d)].
The thermalization parameter was computed directly from the normalized deviations between the measured distribution and the expected distribution in thermalized conditions.
To do so, estimated temperatures and chemical potentials were determined by fitting the tail of each steady-state distribution to a Fermi-Dirac function. The fitting parameters were then used to estimate the value of the distribution around the pumping region centered at $k = k_p$. Other simulations parameters are those of Fig.~\eqref{fig_electron_profiles}.}
\label{fig_Coulomb_thermalization}
\end{figure}

To summarize these findings, Fig.~\ref{fig_Coulomb_thermalization} presents a thermalization map of the steady-state electronic distributions as a function of pumping and carrier-decay rates. This diagram delineates the transition between thermalized and non-thermalized regimes. A thermalization degree of one corresponds to fully equilibrated Fermi-Dirac distributions, as seen in panels a) and b) of Fig.~\ref{fig_electron_profiles}. In contrast, a value of zero indicates complete thermalization failure, as exemplified in panel d). Intermediate values reflect partial equilibration, where residual non-thermal features remain.

\subsection{Including light absorption and emission processes}\label{ILAEP}

Having characterized carrier equilibration in the absence of light, we now incorporate the dynamics of photon absorption and emission.
As established in Sec.~\ref{TSPBEC}, the emission and absorption rates of cavity photons with energy $\hbar \omega_m $, denoted $\gamma_{\uparrow,m}^\text{(diag.)}$ and $ \gamma_{\downarrow,m}^\text{(diag.)}$ respectively, are determined by Eqs.~\eqref{g_u} and \eqref{g_d}, and depend explicitly on the instantaneous carrier distributions.
These rates are inherently time-dependent, evolving in response to the redistribution of electrons and holes.
As such, their steady-state values depend not only on the microscopic interactions but also on external parameters -- namely the pumping and recombination rates $ \Gamma_{\uparrow} $, $ \Gamma_{\downarrow} $, and the cavity loss rate $ \kappa $. \par 

\begin{figure}
\centering 
\includegraphics[scale=0.5]{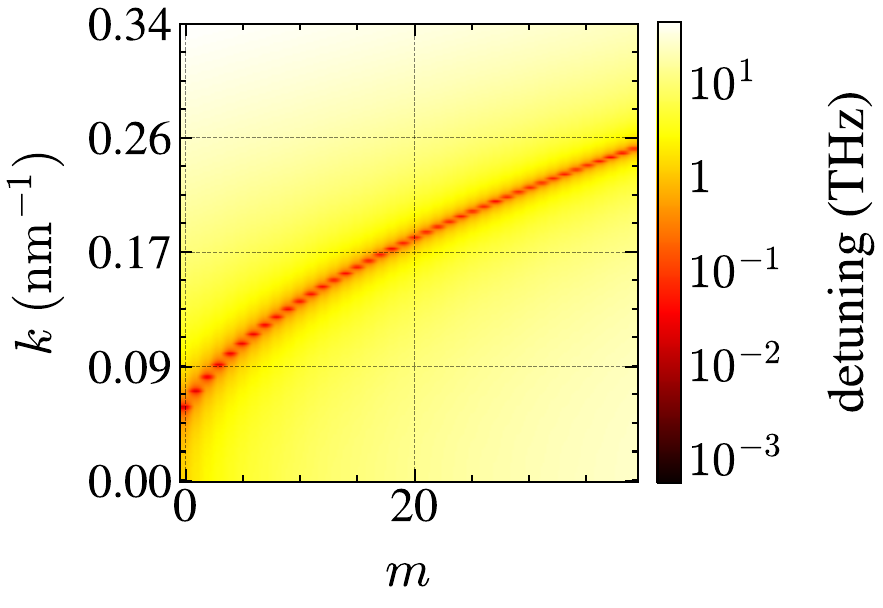}
\caption{(color online). Detuning of each light-matter process involving an electron-hole pair with carrier momentum $k$ and a photon in mode $m$. The darker curve represents the transitions for which the detuning vanishes, defined by $k \equiv K(\omega_m)$ with $K(\omega)$ given in Eq.~\eqref{K_Omega}. Numerical values $ \hbar\omega_0 = 1.32\,$eV, $\hbar\delta\omega = 0.1\,$meV, $E_g = 1.3\,$eV, $m_c = 0.04 m_e$ and $m_h = 0.5m_e $ were used.}
\label{fig_detuning}
\end{figure}

Microscopically, the probability density for a light--matter process involving a photon in mode $ m $ and an electron--hole pair with momentum $ k $ is proportional to $ |g_m|^2 L_{k,m} $, where $ L_{k,m} $ is the Lorentzian defined in Eq.~\eqref{Lorentzian}. The linewidth $ \alpha_{k,m} = (\Gamma_{\uparrow,k} + \Gamma_{\downarrow} + \kappa)/2 $ quantifies the coupling to the environment, while the detuning $ \Delta_{k,m} = (E_{e,k} + E_{h,k})/\hbar- \omega_m $ indicates the energy mismatch between the photon and the electron--hole transition. Fig.~\ref{fig_detuning} shows the detuning profile $ \Delta_{k,m} $ for representative parameters used in the simulations. \par 

The ratio between the detuning $ \Delta_{k,m} $ and the linewidth $ \alpha_{k,m} $ sets the effective coupling to the environment. In the limit $ \alpha_{k,m} / \Delta_{k,m} \ll 1 $, spectral broadening is negligible, and only resonant transitions ($ \Delta_{k,m} \approx 0 $) significantly contribute -- this corresponds to a closed system with vanishing losses. In contrast, when the ratio becomes comparable to the threshold value $ \alpha_{k,m} / \Delta_{k,m} \sim 1 $, environmental coupling is strong enough to permit off-resonant transitions, effectively rendering the system open. 

\subsubsection{Closed system}

When $\alpha_{k,m}\rightarrow 0$, the factor $L_{k,m}$ becomes a $\delta$-function of the form $L_{k,m} = \delta(\Delta_{k,m})$, which permits the emission and absorption rates of Eqs.~\eqref{g_u} and \eqref{g_d} to be calculated exactly, providing 
\begin{align}  
  \gamma_{\uparrow,m}^\text{(diag.)}  &=  \frac{|g_m|^2 }{\hbar} \overline{m}A \widetilde{f}_{e,m}\widetilde{f}_{h,m}. \label{g_u_2} , \\ 
  \gamma_{\downarrow,m}^\text{(diag.)} &= \frac{|g_m|^2 }{\hbar} \overline{m}A(1-\widetilde{f}_{e,m})(1-\widetilde{f}_{h,m}), \label{g_d_2}
\end{align}
Here $\widetilde{f}_{\nu,m} \equiv f_{\nu,K(\omega_m)}$ denotes the value of $f_{\nu,k}$ evaluated at $k = K(\omega_m)$, thus satisfying energy conservation. 

Combining Eqs.~\eqref{g_u_2} and \eqref{g_d_2} we arrive at the following expression for the ratio between (effective) absorption and emission rates of an isolated cavity,
$R_m \equiv R(\omega_m)$, 
\begin{align}
    R_m &\equiv \frac{\gamma_{\downarrow,m}^\text{(diag.)}n_m}{\gamma_{\uparrow,m}^\text{(diag.)}(1+n_m)}\nonumber \\
&=\left(\widetilde{f}_{e,m}^{-1}-1\right)\left(\widetilde{f}_{h,m}^{-1}-1\right)\frac{n_m}{n_m + 1}. \label{R_m}
\end{align}
This result generalizes the van Roosbroeck–Shockley (vRS) relation to the non-equilibrium regime and takes into account both spontaneous and stimulated emission of light \cite{PhysRev.94.1558}. In the special case where all species are in thermal equilibrium at temperature $T$, Eq.~\eqref{R_m} becomes $R_m = 1$, independent of the photon energy.

\subsubsection{Open system}
In real experiments, where the coupling to the environment cannot be neglected, Eq.~\eqref{R_m} no longer holds and the inclusion of a finite-width Lorentzian $L_{ k,m}$ requires analytical evaluation of the summations over $k$ to determine emission and absorption patterns. Moreover, as demonstrated in Sec.~\ref{SQC}, the degree of thermalization itself depends sensitively on the strength of this environmental coupling, which affects both carrier and photon dynamics. \par

To assess the validity of the vRS relation for open systems, we numerically solve the rate equations using thermal initial distributions for the carriers,
\begin{equation}
   f_{\nu,k}(0) = \frac{1}{e^{(E_{\nu,k} - U_\nu)/k_B T} + 1 }, \label{Fermi_D}
\end{equation}
with $T$ the temperature and $U_e = E_g$, $U_h = 0$ the chemical potential of each band. We also assume a vanishing occupation of photon modes at the initial time, $n_m(0) = 0$. To isolate the impact of environmental coupling, we focus solely on the photon-loss rate $ \kappa $, setting $ \Gamma_\uparrow = \Gamma_\downarrow = 0 $. \par
\begin{figure}
\centering 
\hspace{-0.7cm}
\includegraphics[scale=0.51]{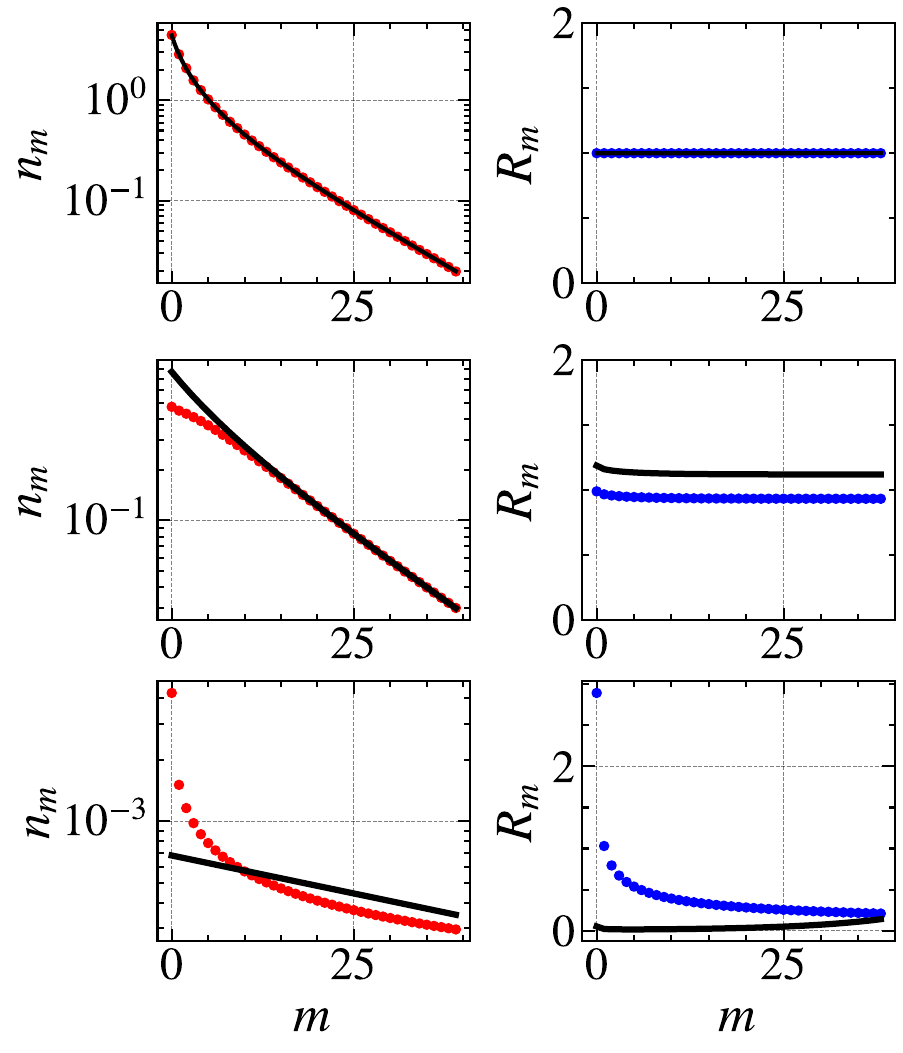}
\caption{(color online). Steady-state photon distribution (left panels, red data) and corresponding photon absorption and emission ratio (right panels, blue data) as a function of the mode number. Different values for the photon-loss rate were used: $\kappa = 10^{-2}\,$THz (top row), $\kappa = 1\,$THz (middle row) and $\kappa = 10^2\,$THz (bottom row). The black line depicted on the left panels corresponds to the result of fitting the tail of each spectrum to a Bose-Einstein distribution. The vRS relation of Eq.~\eqref{R_m} corresponds to the black curves on the right panels. Other simulation parameters are those of Fig.~\ref{fig_detuning}.}
\label{fig_vRS_profiles}
\end{figure}

Numerical results of the steady-state photon spectra at room temperature for different photon-loss rates are shown in the left panels of Fig.~\ref{fig_vRS_profiles}. The right panels show the numerical ratio between steady-state absorption and emission rates (blue points) alongside with the corresponding prediction given by the vRS relation of Eq.~\eqref{R_m} (black curves). The top row represents the case of a closed system, where the effect of photon losses to the environment is negligible. In this case, the photon gas is able to fully thermalize to a Bose-Einstein distribution due to multiple absorption and emission events. Since the system is approximately closed, the vRS relation is able to predict the absorption-to-emission ratio with good accuracy. According to Eq.~\eqref{R_m}, the latter is constant equal to unity because $R_m =1$ when carriers and light are in thermal equilibrium.  When the photon-loss rate is increased (middle and bottoms rows), the photon spectra are no longer described by the Bose-Einstein distribution. In these cases, the vRS relation breaks down leading to significant deviations of $n_m$ from the Bose-Einstein distribution, as evidenced by the discrepancies observed in the right panels. The bottom panels, in particular, highlight a strongly dissipative regime dominated by photon losses.

\section{Dynamics of photon condensation}\label{DPC}
Having established the conditions under which carrier thermalization is effective, we now turn our attention to the emergence of photon condensation. By analyzing the steady-state behavior of photon and carrier distributions under varying pump and loss conditions, we identify key features of the condensation transition. The evolution of the photon population is tracked to reveal condensation thresholds and spectral reshaping characteristic of Bose-Einstein condensation in driven-dissipative systems.

We start by noting that the steady-state solution of Eq.~\eqref{photon_rate} can be written as
\begin{equation}
    n_m^\text{s.s.} = \frac{1}{\Psi_m^\text{s.s.} -1}, \label{n_m_ss}
\end{equation}
with 
\begin{equation}
	\Psi_m(t) = \frac{\kappa + \gamma^\text{(diag.)}_{\downarrow,m}(t)}{\gamma^\text{(diag.)}_{\uparrow,m}(t)} \label{Psi_m}
\end{equation}
denoting the ratio between outcoming and incoming rates and $\Psi_m^\text{s.s.} \equiv \Psi_m(t \rightarrow\infty)$ being its steady-state value.
Equation~\eqref{n_m_ss} establishes that the steady-state photon spectrum takes the form of a Bose-Einstein distribution when $\Psi_m^\text{s.s.} \sim e^{(\hbar\omega_m-\mu)k_BT}$, where $\mu$ is the photon chemical potential. This requires the cavity losses to satisfy $\kappa\ll \gamma_{\downarrow,m}$ for all modes.
When $\kappa$ becomes comparable to $\gamma_{\downarrow,m}$, Eq.~\eqref{n_m_ss} predicts deviations from a Bose-Einstein function. 

Condensation of mode $m$ is defined by the divergence of its steady state population, {\it i.e.} by $n_m^\text{s.s.}\rightarrow\infty$, which, according to Eq.~\eqref{n_m_ss}, can only occur if $\Psi_m^\text{s.s.}$ approaches unity. Since the dynamical evolution of $\Psi_m(t)$ depends on the carrier distributions through Eqs.~\eqref{g_u} and \eqref{g_d}, there must be a threshold carrier density above which the steady state amplitude $\Psi_m^\text{s.s.}$ of a given mode $m$ approaches the critical value of $1$. The density threshold defines a critical pumping rate $\Gamma_c^{(m)}$ above which mode $m$ becomes macroscopical occupied.\par

In what follows, we denote a steady-state photon spectrum with a condensed ground state as a \textit{Bose-Einstein condensate (BEC)}.
A steady state in which a set of excited modes is condensed but not the ground state defines the \textit{laser} phase \cite{PhysRevLett.120.040601}, while
spectra where both the ground state and one or more excited states are condensed, are referred to as \textit{multimode condensates}. Conversely, if no macroscopic occupation emerges, the photon gas is said to be in the \textit{thermal} phase when thermal equilibrium is reached, and in the \textit{uncondensed} phase otherwise. Transitions between phase are promoted by a competition between the cavity losses and the (time-dependent) absorption and emission profiles governed by the carrier distributions. \par

As shown in the Sec.~\ref{ILAEP}, multiple photon absorption and emission events resulting from excited carrier distributions can lead to thermalized photon spectra if the photon-loss rate is not too large. Importantly, it must be noted that light thermalization requires all cavity modes to be optically accessible through electron-hole absorption and emission. For the latter to be verified, the band gap must be smaller than the cavity ground-state energy. This situation was analyzed in Sec.~\ref{SQC}, where we used $E_g = 1.3\,$eV and $\hbar\omega_0= 1.32\,$eV (see Fig.~\ref{fig_detuning}). On the contrary, when $E_g>\hbar\omega_0$, emission and absorption into the ground state is significantly reduced, as the maximum emission is expected for some higher-order cavity mode. In these conditions, the small cut-off prevents a thermal photon distribution to form. \par

With this in mind, this section is devoted to an analysis of the steady-state photon spectra resulting from changing the pumping rate and the cavity cut-off frequency. The latter can be easily varied experimentally by controlling the cavity length, which modifies the Lorentzian factor $L_{k,m}$ and consequently the absorption and emission profiles. Moreover, we restrict our analysis to values of $\Gamma_\uparrow$ and $\Gamma_\downarrow$ that are lower than Coulomb-collision rates, allowing the carriers to thermalize while interacting with the photon gas. This restriction is necessary in order for photon condensation to occur. According to Fig.~\ref{fig_Coulomb_thermalization}, carrier thermalization is possible for $\Gamma_\downarrow < 10^2\,$THz and $\Gamma_\uparrow < 10^4\,$THz. Additionally, the photon-loss rate should be fixed at a value which allows the photon gas to thermalize before being lost to the environment. This can be achieved with $\kappa <100\,$GHz (see Fig.~\ref{fig_vRS_profiles}), corresponding to $r_t r_b\sim0.9995$
with $r_{t}$ and $r_{b}$ being the top and bottom mirror reflectivities, respectively. Such high-quality mirrors are available experimentally \cite{Schofield2024}.  

In Sec.~\ref{BTTT}, we examine cavity configurations in which the ground-state energy $\hbar\omega_0$ exceeds the material band gap $E_g$. When this condition is verified, the ground-state cavity mode becomes optically accessible, allowing the photon gas to thermalize. Our results show that, when the degree of carrier excitation surpasses a critical value, the emission into the ground-state cavity mode becomes macroscopical and a phase transition takes place, leading to the formation of a photon BEC. Next, the case of $\hbar\omega_0<E_g $ is analyzed in Sec.~\ref{NSHP}. Here, the suppression of absorption and emission into the ground-state mode leads to the formation of steady-state photon spectra which are not in thermal equilibrium. As a result, macroscopic occupations might be established for excited cavity modes, typical of laser operation. In this case, increasing the pumping rate results in a phase transition into a laser or multimode-condensate phase. 

\subsection{Thermal--BEC transition }\label{BTTT}

Let us consider a cavity geometry corresponding to a ground-state photon energy $\hbar\omega_0 = 1.32\,$eV, slightly above the semiconductor band gap $E_g = 1.3\,$eV. Throughout the discussion, we characterize the cavity by its ground-state wavelength $\lambda_0 = 2\pi c/\omega_0$ rather than by $\omega_0$, yielding $\lambda_0 = 950\,$nm for the parameters considered. This choice allows a more direct comparison with experimental results, which are typically reported in terms of wavelength rather than energy.

The condition $\hbar\omega_0 > E_g$ leads to a detuning profile of the type shown in Fig.~\ref{fig_detuning}. In this case, the curve of light–matter resonances with vanishing detuning intersects the $m=0$ axis at a finite value of $k$. Consequently, for every photon mode there exists an electron–hole excitation with the same energy $\hbar \omega_m$. This ensures that radiative recombination can occur into all cavity modes, thereby enabling full photon thermalization.

\begin{figure}
\centering 
\hspace{-0.5cm}
\includegraphics[scale=0.5]{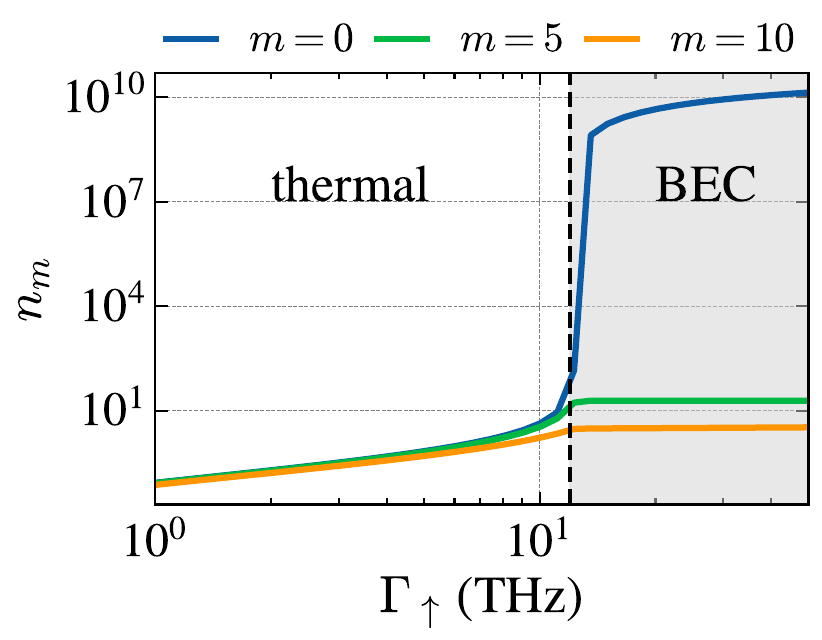}
\caption{(color online). Steady-state photon occupation as a function of pumping rate for the ground state and two excited modes. For $\Gamma_\uparrow \geq \Gamma_c^{(0)} = 12\,$THz, the occupation of the ground state becomes macroscopic while excited modes remain thermally distributed. The abrupt increase in the ground-state occupation defines the onset of photon condensation. Other simulation parameters are $\Gamma_{\downarrow} = 100\,$ GHz, $\kappa = 100\,$ GHz, $A = 10^4\, \mu\text{m}^2$}
\label{fig_occupations1}
\end{figure}

In Fig.~\ref{fig_occupations1} we show steady-state photon occupations as a function of pumping rate for the ground state and two excited modes. The onset of the BEC phase is signaled by a large increase in the ground-state occupation after a small variation of the pumping rate close to the critical value $\Gamma_c^{(0)} \simeq 12\,$THz, while excited modes remain thermally occupied.
The growth of $n_0$ above threshold is super-linear, reflecting that additional pump power is almost entirely funneled into the condensate mode.
Meanwhile, the population of excited modes saturates above threshold. This is the analogue of gain clamping in lasers, here arising from the equilibrium condition that the excited-state photon occupations cannot increase further without violating the Boltzmann distribution at the fixed reservoir temperature.

\begin{figure}
\centering 
\hspace{-0.5cm}
\includegraphics[scale=0.42]{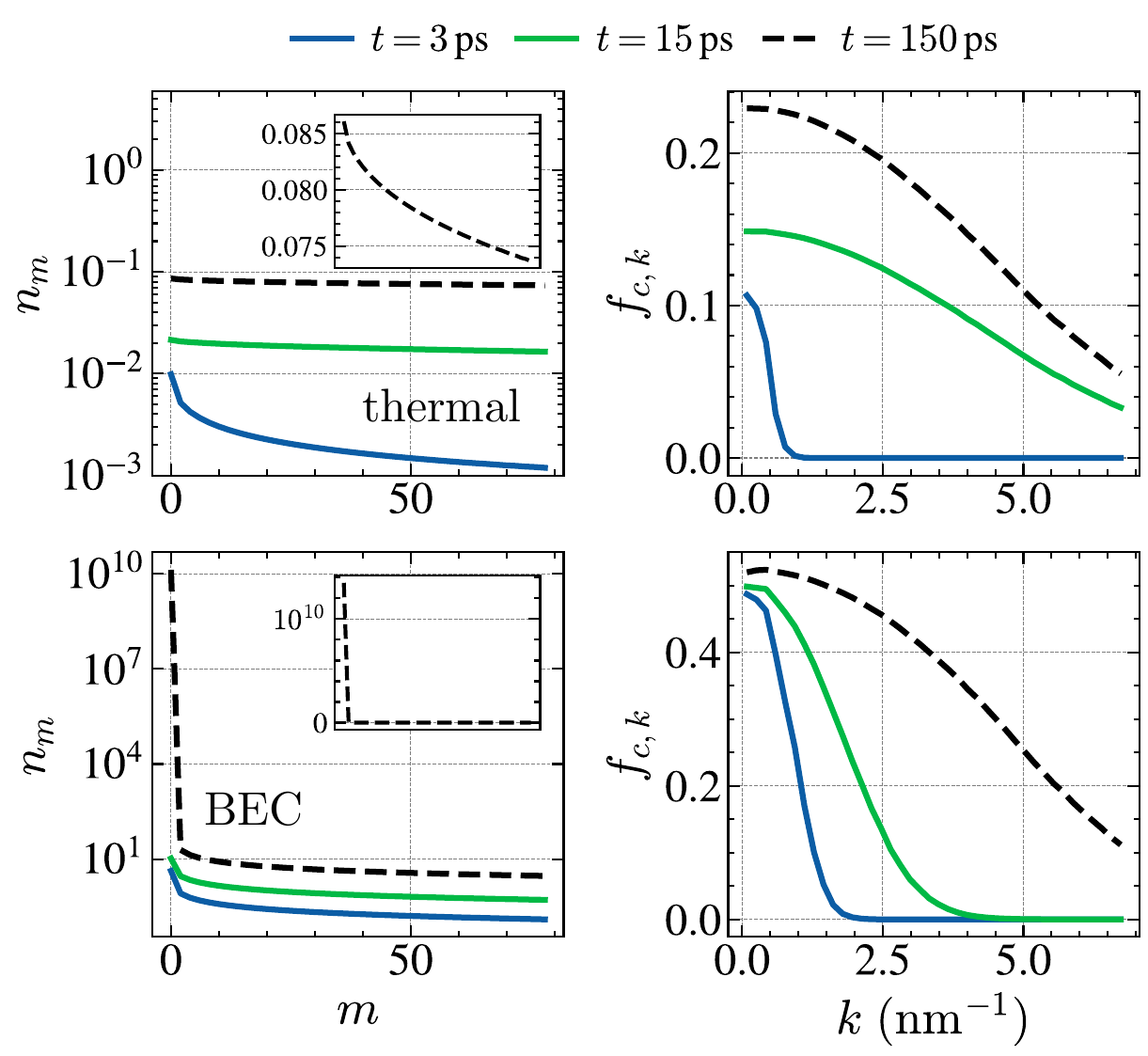}
\caption{(color online). Photon and electron distributions for different time instants below (top panels: $\Gamma_\uparrow < \Gamma_c^{(0)}$) and above (bottom panels: $\Gamma_\uparrow > \Gamma_c^{(0)}$) threshold. The dashed lines represent the steady-state data. Inset: steady-state photon spectra in linear scale. Other simulation parameters are those of Fig.~\ref{fig_occupations1}}
\label{fig_BEC_to_thermal_profiles}
\end{figure}

 Figure~\ref{fig_BEC_to_thermal_profiles} depicts the photon and electron distribution functions for various times and pumping rates located below (top panels) and above (bottom panels) the threshold for ground-state condensation. Dashed lines represent steady-state solutions.
 In the first time instants, the photon spectra looks similar both below and above threshold, corresponding to a photon gas partially equilibrated with the electron-hole plasma at low temperature.
 As time increases and more electron-hole excitations become available, the effective rates of light emission and absorption increase and approach the steady-state value. Below threshold, the photon population decays approximately exponentially with mode energy, indicating a thermal distribution.
 This thermal Boltzmann-like profile is a signature of photon thermalization via repeated absorption and re-emission by electron-hole pairs: higher-frequency modes (larger $m$) have lower occupancy, consistent with a Bose--Einstein distribution $n_m = [\exp((\hbar\omega_m-\mu)/k_B T)-1]^{-1}$ with a negative effective chemical potential $\mu$ relative to the cut-off. Hence absorption and emission processes are in detailed balance under weak pumping, and the carriers serve as an effective thermal reservoir that imprints a well-defined temperature on the photon gas. Below the critical pump rate, all cavity modes including the lowest-energy mode ($m=0$) exhibit a thermal (Boltzmann) occupancy spectrum. Higher-order modes strictly follow the expected exponential decay, while the ground state initially conforms to this distribution. As the pump power approaches the critical value, $\Gamma_c^{(0)}$, the occupancy of the $m=0$ mode begins to rise above the thermal trend. Precisely at $\Gamma_c^{(0)}$, $n_0$ diverges (in practice limited only by finite system size), marking the onset of a non‐equilibrium phase transition. Immediately above threshold, the $m=0$ mode hosts a macroscopic population, thereby constituting a photon condensate. This dramatic redistribution of photons into the ground state is the defining signature of condensation. The transition is driven by the upward shift of the photon chemical potential, $\mu$, toward the cavity ground‐state energy $\hbar\omega_0$. Increasing carrier excitation injects photons into the cavity, raising $\mu$ until it meets $\hbar\omega_0$ at $\Gamma_c^{(0)}$ at the onset of condensation.

\begin{figure}
\centering 
\hspace{-0.5cm}
\includegraphics[scale=0.45]{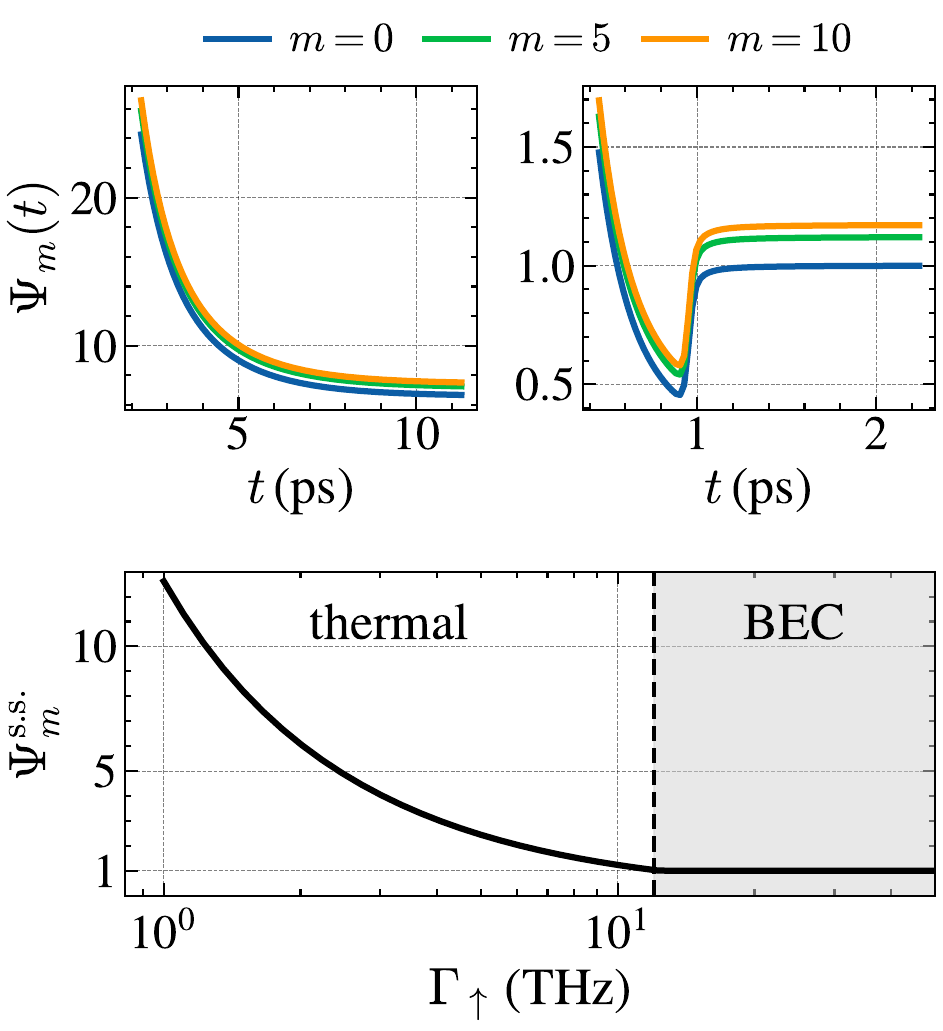}
\caption{(color online). The role of the ratio $\Psi_{m}$ in determining the phase. Top panels: time evolution of $\Psi_{m}$ below (left panel) and above (right panel) threshold for the ground state and two excited modes. Bottom panel: steady-state ratio $\Psi_{m}^\text{s.s.}$ for the ground state ($m=0$) as a function of pumping rate. }
\label{fig_ratio_m}
\end{figure}

The criteria for condensation to occur can be understood as follows: as the pumping is turned on and carrier distributions start to get filled, multiple photon start being emitted due to electron-hole recombination.
The photon spectrum then thermalizes with the electron-hole gas, approaching Eq.~\eqref{n_m_ss} for later times. While the electron-hole density is still increasing, the  value of $\Psi_m(t)$ as defined in Eq.~\eqref{Psi_m} varies in time until steady-state is reached. According to Eq.~\eqref{n_m_ss}, if the steady-state value of $\Psi_0(t)$ approaches the critical value of $1$, the ground-state occupation $n_0(t\rightarrow \infty)$ diverges, and a BEC is formed. In Fig.~\ref{fig_ratio_m} we provide representative curves of the ratio $\Psi_m$. Top panels display the time evolution of $\Psi_m(t)$ for several modes below (right panel) and above (left panel) threshold until a steady state value is reached.
The latter happens when all carrier distribution functions become stationary. The thermal phase is characterized by steady-state values of $\Psi_m$ that are larger than unity for all modes, which, according to Eq.~\eqref{n_m_ss}, results in finite ({\it i.e.}, non-macroscopical) photon occupations in steady state. Contrarily, the BEC phase requires $\Psi_m$ to approach unity for the ground state, as shown in the top right panel of Fig.~\ref{fig_ratio_m}. Moreover, the variation of the steady-state ratio for $m=0$ as a function of pumping power is shown in the bottom panel therein, providing an additional order parameter related to the thermal--BEC phase transition. In other words, close to the critical point (marked with a dashed vertical line), the steady-state ratio for $m=0$ attains nonanalytic behavior as a function of $\Gamma_\uparrow$, becoming unity in the BEC phase. On the contrary, this curve is always analytic for all the excited modes.\par

\begin{figure}
\centering 
\hspace{-1cm}
    \includegraphics[scale=0.45]{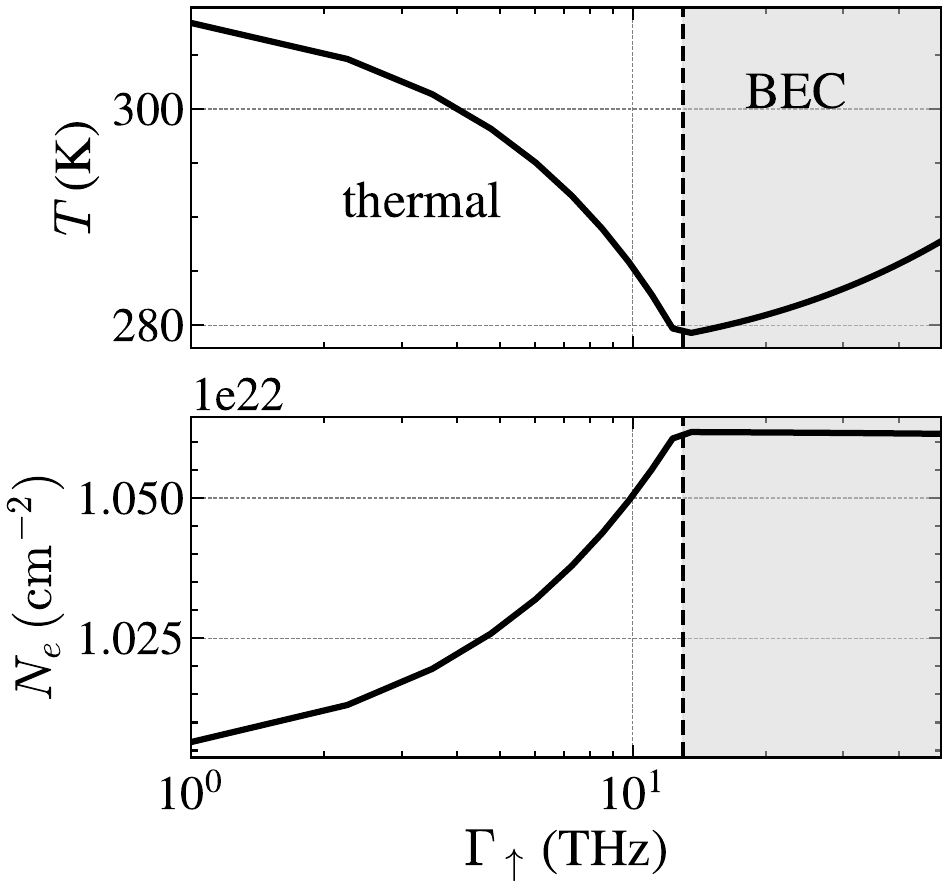}
\caption{(color online). Top panel: Photon temperature in steady-state as a function of pumping rate, determined from fitting the steady-state photon distribution to the Bose--Einstein function. Bottom panel: steady-state carrier density as a function of pumping rate.}
\label{fig_el_temp_dens}
\end{figure}

Near the critical point, the sharp increase in ground-state photon occupancy leads to modifications in the absorption and emission rates. As a result, sharp variations pinpointing the phase transition are also visible in other macroscopic quantities of the system. In Fig.~\ref{fig_el_temp_dens} we plot the steady-state temperature of the photon gas (top panel) and final density of the electron-hole system (bottom panel) for varying pumping strength, where the same nonanalytic trend is observed close to the critical value. In the thermal phase, the decrease in the temperature for increasing pump power reflects the increasing efficiency of reabsorption processes, leading to a more sharply peaked electron distribution. This behavior persists until the critical point is reached, at which stimulated emission into the ground mode dominates over reabsorption. Beyond this threshold, the carriers no longer thermalize efficiently with the photons, and the temperature saturates. Simultaneously, the bottom panel shows that the steady-state carrier density also exhibits a characteristic saturation behavior. It increases linearly with pump in the thermal regime, reflecting accumulation of carriers, but becomes clamped just above the threshold due to rapid depletion by stimulated processes into the condensate. These results are consistent with the kinetic picture underlying the phase transition: below threshold, carrier dynamics are governed by detailed balance and thermal relaxation, while above threshold, the photon condensate modifies the recombination dynamics, imposing a constraint on the carrier population and locking the system into a new steady state. Together, the evolution of these thermodynamic quantities across the transition reinforces the interpretation of photon condensation as a non-equilibrium, yet thermodynamically driven, phase transition in a coupled light--matter system.

\subsection{Non-thermal spectra and higher-order phases}\label{NSHP}
When the cavity cutoff frequency is lowered so that the lowest cavity mode energy $\hbar\omega_0$ drops below the semiconductor band gap $E_g$, the system enters a qualitatively new regime. In contrast to Section~\ref{BTTT}, here the ground-state mode becomes optically inaccessible due to energy conservation constraints. In simple terms, if $\hbar\omega_0 < E_g$, an electron-hole pair recombination cannot directly produce a photon in the cavity ground-state mode because the minimum emitted energy is $E_g$. As a result, the photon gas can no longer achieve a global thermal equilibrium distribution. Instead, the steady-state spectrum departs from a Bose--Einstein form, and the system undergoes non-thermal phase transitions into higher-order mode condensates (or laser-like states) depending on parameters such as the cavity cutoff and pump power.

\begin{figure}
\centering 
\hspace{0cm}
\includegraphics[scale=0.45]{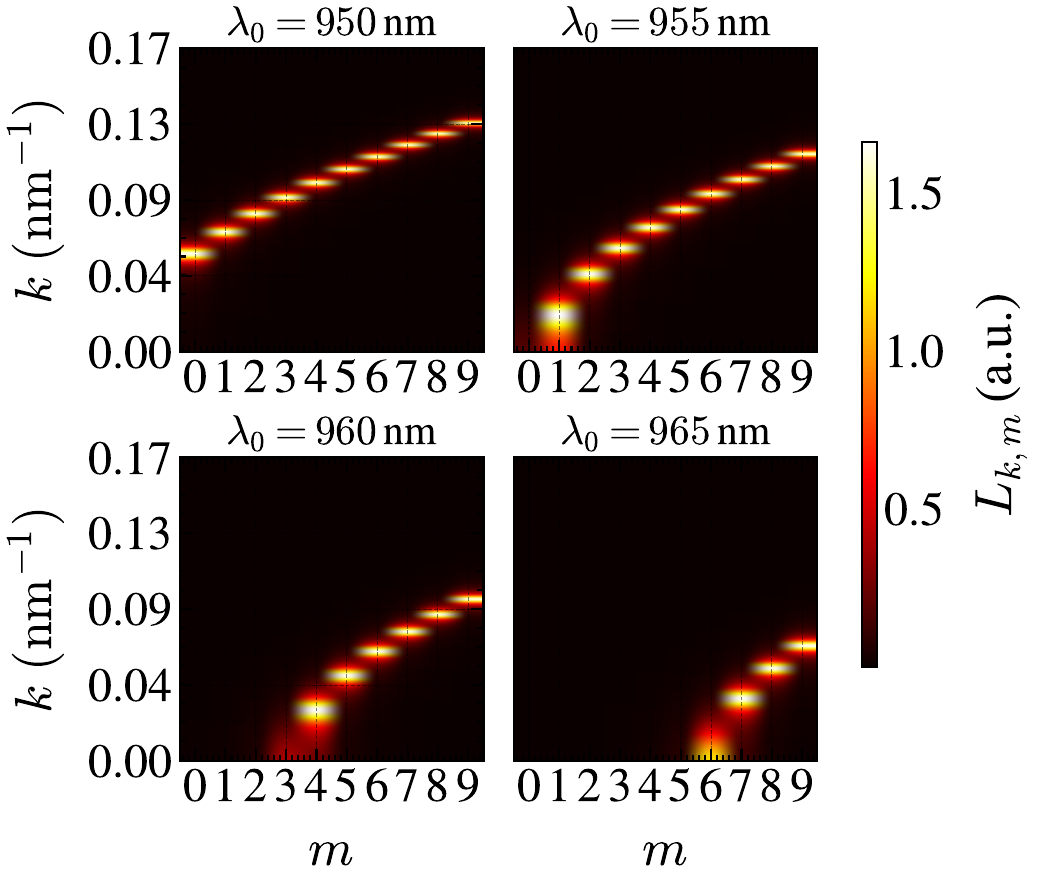}
\caption{(color online). Normalized Lorentzian profiles for different cavity geometries. When $\lambda_0 = 950\,$nm, all modes are optically accessible through electron-hole excitations and recombination, while for the remaining panels the ground-state becomes progressively unaccessible. Each profile is centered at the curve defined by $\Delta_{k,m}=0$ where $\Delta_{k,m}$ is the detuning. The width of the nonzero region is essentially $(\Gamma_{\downarrow} + \kappa)/2$. Here we used $\Gamma_{\downarrow} = 100\,$ GHz, $\kappa = 100\,$GHz. }
\label{fig_lorentzian_profiles}
\end{figure}

Figure~\ref{fig_lorentzian_profiles} illustrates the spectral accessibility of photon modes for different cavities by plotting the normalized Lorentzian line-shape $L_{k,m}$ that governs light--matter coupling as a function of detuning. For the smaller cavity with $\lambda_0 = 950~\mathrm{nm}$, $\omega_0$ lies above or very close to $E_g$, and consequently all photon modes (including $m=0$) can be resonantly populated via electron-hole recombination. As $\lambda_0$ increases, however, $\omega_0$ decreases, and beyond a critical length $\lambda_\mathrm{crit}$ (for which $\hbar\omega_0 = E_g$), the resonance condition $\Delta_{k,m}=0$ no longer occurs close to $m=0$. In such cases, the ground mode lies outside the peak of the gain spectrum and receives only the tail of the broadened emission line. \par 

\begin{figure}
\centering 
\hspace{-0.5cm}
\includegraphics[scale=0.45]{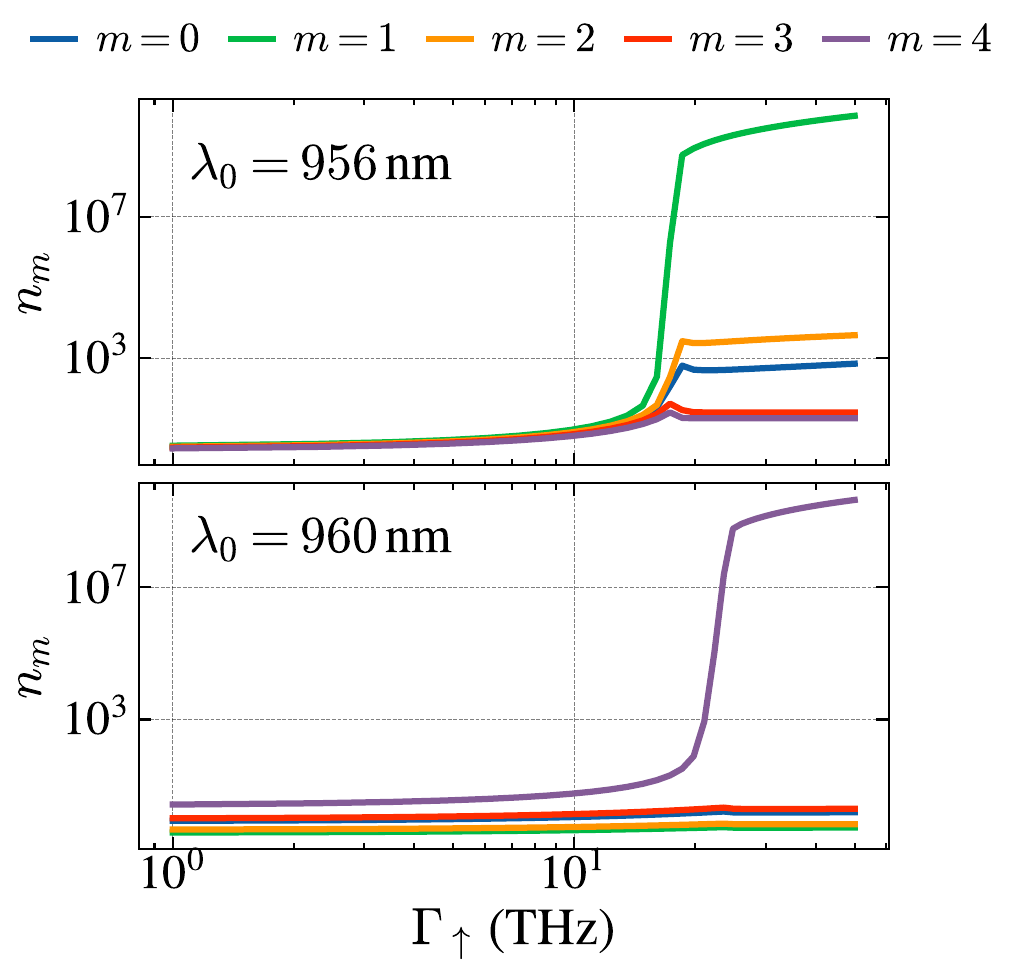}
\caption{(color online). Steady-state photon occupation as a function of pumping rate for the first cavity modes and different cavity distances. The top panel shows a transition to a multimode--condensate phase, while the bottom panel shows a transition into a laser phase. Other simulation parameters are $\Gamma_{\downarrow} = 100\,$ GHz, $\kappa = 100\,$ GHz, $A = 10^4\, \mu\text{m}^2$.}
\label{fig_occupations2}
\end{figure}

Under these conditions, the steady-state reached under continuous pumping is markedly non-thermal. To probe the nature of the phase transition in this regime, we examine the steady-state occupations $n_m$ of the cavity modes as a function of pump power $\Gamma_\uparrow$ for various cavity cutoffs. Figure~\ref{fig_occupations2} summarizes the results for two representative cavity cut-offs beyond $\lambda_\mathrm{crit}$. In the top panel, corresponding to a moderately increased cavity length, the system undergoes a transition to a multimode condensate as the pump power exceeds a critical threshold. Here, the first excited mode ($m=1$) is the primary mode that reaches macroscopic occupation, but the ground ($m=0$) and second excited ($m=2$) modes also acquire significant occupations once the threshold is crossed, which is in stark contrast to the single-mode condensation observed in the thermal case. The evolution of steady-state photon occupation for $m=0,1,2$, however, shares similarities with that of the ground state in the large cut-off limit treated in Sec.~\ref{BTTT}, {\it i.e.}, it is a monotonically increasing function of the pump power, with super-linear increase close to the critical point. For the remaining non-condensed modes, the occupation either saturates or decreases after crossing the critical point. The emergence of multiple condensed modes is a consequence of the broad gain profile overlapping several nearby cavity modes. As shown in Figure~\ref{fig_lorentzian_profiles}, for $\lambda_0 = 956~\mathrm{nm}$ the light--matter coupling is centered near $m=1$ but still appreciable for $m=0$ and $m=2$. As a result, the gain is sufficient to drive macroscopic occupation in all three modes once the pump exceeds threshold. The behavior of these modes resembles a generalized Bose condensation, with stimulated emission into each mode reinforcing its growth. \par

In contrast, the bottom panel of Figure~\ref{fig_occupations2} demonstrates a laser phase for a larger cavity with $\lambda_0 = 960~\mathrm{nm}$. Here, only the $m=4$ mode lies near the peak of the gain spectrum and receives sufficient gain to overcome losses. As the pump is increased, the occupation $n_4$ rises sharply, while lower modes remain suppressed despite their energetically favorable positions, highlighting the breakdown of thermal ordering in this regime. Moreover, and contrarily to standard laser models, here lasing is achieved without electronic inversion. This is possible because absorption and emission profiles are not symmetric due to the environmental broadening and cavity geometry, allowing for net amplification to occur even in the absence of excited matter distributions. \par 

\begin{figure}
\centering 
\hspace{-1cm}
\includegraphics[scale=0.55]{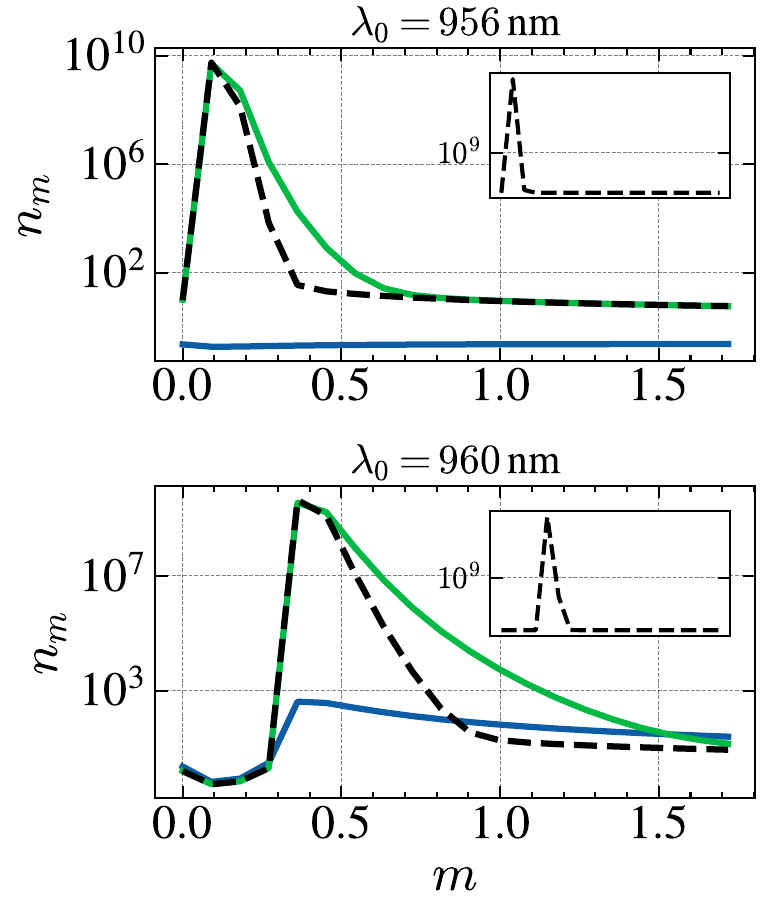}
\caption{(color online). Photon distribution for increasing times leading to condensation of excited modes in unthermalized spectra. The top panel shows the multimode BEC phase, while the bottom panel displays the laser phase. The legend is the same as in Fig.~\ref{fig_BEC_to_thermal_profiles}. Inset show steady-state spectra in linear scale.}
\label{fig_multimode_profiles}
\end{figure}

Figure~\ref{fig_multimode_profiles} provides a time-resolved view of these transitions. Initially, the photon spectrum resembles a thermal distribution, seeded by spontaneous emission. As stimulated emission sets in, certain modes grow exponentially, forming a non-thermal peak. In the multimode case, multiple modes including the ground state exhibit synchronized growth. In the laser case, a set of excited modes dominates, while others are suppressed. The steady-state spectrum deviates strongly from a Bose--Einstein distribution, reflecting kinetic selection governed by gain, losses, and spectral overlap.\par 
The higher-order condensate phases described here contrast sharply with the thermal ground-state condensation discussed in Sec.~\ref{BTTT}. The photon gas undergoes an out-of-equilibrium phase transition into a gain-selected driven-dissipative steady state. These results demonstrate how cavity engineering and spectral detuning allow access to distinct nonequilibrium condensation regimes, connecting the physics of photon BEC with that of traditional semiconductor lasers. The accessibility and control of these regimes in semiconductor platforms mark a decisive step toward engineered quantum fluids of light beyond equilibrium.

\section{Phase diagram of the photon gas}\label{sec_phase_diagram}
The nonequilibrium phase behavior of the photon gas in a semiconductor microcavity reflects a rich interplay between the cavity spectral structure, carrier-mediated gain, and the efficacy of thermalization. Transitions between the different phases, which are associated with the condensation or decondensation of one or more photon modes, can be induced either by modifying the cavity spectral properties through its geometry, or by changing the degree of carrier excitation through the external pump, {\it i.e.} by operating directly on the optical medium. 

\begin{figure*}
\centering 
\includegraphics[scale=0.55]{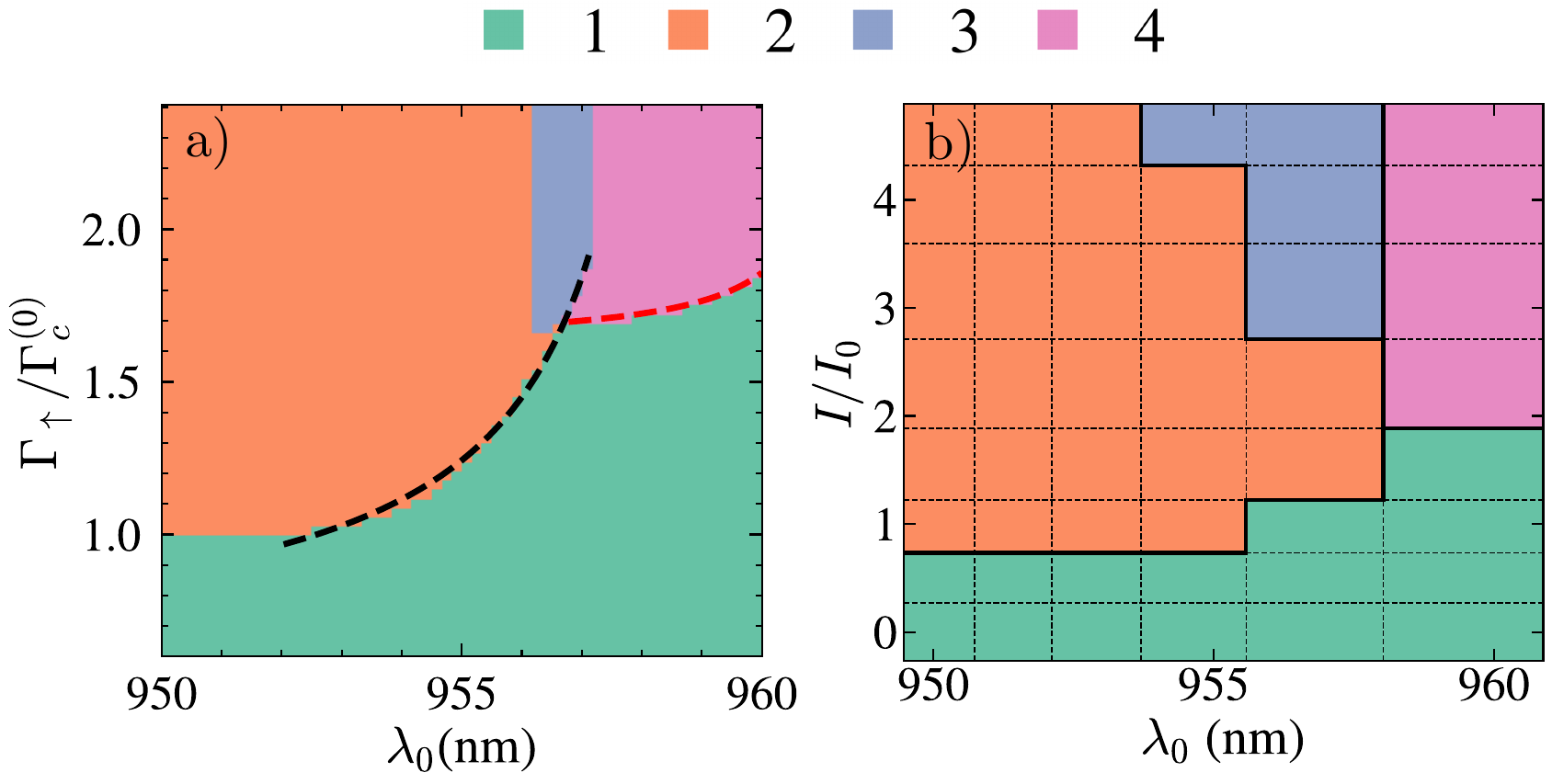}
\caption{(color online). a) Simulated phase diagram of the photon gas as a function of cavity cut-off wavelength $\lambda_0$ and pump rate $\Gamma_\uparrow$ normalized to the critical pumping for $\lambda_0 = 950\,$nm. The dashed black and red lines  represent the analytical estimations for the $1\rightarrow 2$ and $1\rightarrow 4$ phase boundaries, respectively, found from a fit to Eq.~\eqref{final}. Other simulation parameters are $\Gamma_{\downarrow} = 100\,$ GHz, $\kappa = 100\,$ GHz, $A = 10^4\, \mu\text{m}^2$, $\ell_c = 1\mu$m. b) Experimental phase diagram reproduced from Ref.~\cite{Schofield2024}. Due to nonlinear and saturation processes taking place at the semiconductor heterostructure, further measurements would be required to find the exact relation between pump rate and pump intensity. For this reason, the $y-$axis given in terms of the normalized pump intensity for a better comparison with panel a). The $y-$axis Color code: 1 -- thermal cloud, 2 -- BEC , 3 -- Multimode BEC , 4 -- Laser. }
\label{fig_phase_diagram}
\end{figure*}

In Sec.~\ref{DPC} we focused on the role of the pump rate in driving a thermal photon gas into either a single Bose-Einstein condensate, a multimode condensate, or a laser, depending on the cavity cut-off frequency [see, e.g., Figs.~\ref{fig_occupations1} and~\ref{fig_occupations2}]. In order to gain quantitative insights on the role of the cavity cut-off in determining the phase, in Fig.~\ref{fig_phase_diagram}(a) we show the full phase diagram depicting the steady-state phase of the photon gas, built from extensive numerical simulations of the kinetic model of Eqs.~\eqref{photon_rate} and~\eqref{carrier_rate} for a representative region of the parameters $\lambda_0$ and $\Gamma_\uparrow$. The phase boundaries were determined numerically by examining the evolution of the steady-state ratio $\Psi_{m}$ along the parameter space. According to Eqs.~\eqref{n_m_ss} and~\eqref{Psi_m}, the steady-state value of $\Psi_{m}$ approaching unity indicates the threshold for condensation of mode $m$ [see Fig.~\eqref{fig_ratio_m} for representative trends]. Moreover, Fig.~\ref{fig_phase_diagram}(b) displays the phase diagram measured in a III–V semiconductor microcavity experiment, reproduced from Ref.~\cite{Schofield2024}, under experimental conditions that were replicated in our simulations. Remarkably, the numerical diagram not only captures the correct topology of the phases but also quantitatively agrees with the experimental boundary positions. The close correspondence between theoretical predictions and experimental observations provides strong validation of the model and its underlying approximations, thereby substantiating the predictive capability of our framework.

The phase diagram of Fig.~\ref{fig_phase_diagram}(a) provides additional insights into the nature of the boundaries by allowing cuts of fixed pump power to be examined as a function of the cut-off. For constant $\Gamma_\uparrow$ above BEC threshold, increasing the cavity eventually pushes the system out of the condensate phase. In particular, as $\lambda_{0}$ approaches a critical value $\lambda_{\rm crit}$, the required $\Gamma_{\uparrow,c}^{(0)}$ for BEC diverges. Physically, when the cavity photon energy falls below the band gap, $\lambda_{0} > \ell_{\rm crit} \approx 957~\text{nm}$ for $E_{g}=1.3~\text{eV}$, the ground-state occupation is rapidly suppressed, which indicates a transition between multimode BEC (region 3) and laser (region 4). As a consequence, for $\lambda_{0} > \lambda_{\rm crit}$ the system never enters the BEC phase, remaining thermally distributed up to the laser threshold. The onset of the laser phase requires an excited cavity mode to experience the highest net gain from the carrier population, indicating poor photon thermalization.

Due to the convoluted kinetic model governing the many-body evolution of the system, exact analytical results for the phase boundaries are not available. Formally, we know that the onset of condensation of a given mode $m$ is defined by $\Psi_m(t\rightarrow\infty) = 1$, which can be used to identify
all phase boundaries implicitly. Hence, assuming $\kappa \ll \gamma^\text{(diag.)}_{\downarrow,m}$ (which is confirmed by simulations), the condensation onset for mode $m$ can be written as
\begin{equation}
 \frac{\gamma^\text{(diag.)}_{\downarrow,m}}{\gamma^\text{(diag.)}_{\uparrow,m}} = 1,\label{gamma_down_up_diag}
\end{equation}
with $\gamma^\text{(diag.)}_{\downarrow,m}$ and $\gamma^\text{(diag.)}_{\uparrow,m}$ given by Eqs.~\eqref{g_u}
    and~\eqref{g_d} evaluated at $t\rightarrow \infty$. The latter require a sum over $k$ involving the steady-state distributions. Since each term in the sum is weighted by a Lorentzian, we approximate the result by its dominant resonant contribution, yielding
\begin{align}
    \gamma^\text{(diag.)}_{\downarrow,m} &\simeq D f_{e} f_h,\label{g_ress_d} 
    \\ 
    \gamma^\text{(diag.)}_{\uparrow,m} &\simeq D (1-f_e)(1- f_h),\label{g_ress_u}
 \end{align}
where $D$ is a constant, and $f_e$ and $f_h$ are the carrier distribution function evaluated at resonance, {\it i.e.}, for the value of $k$ that verifies $E_{e,k} + E_{h,k} = \hbar \omega_m$.

In steady state, we have from Eq.~\eqref{carrier_rate}
\begin{equation}
    \frac{ (\Gamma_{\uparrow,k} + \Lambda^\text{(diag.)}_{\uparrow,k})(1 - f_{\overline{\nu}, k}) + \mu_{\uparrow,\nu,k} + \eta_{\uparrow,\nu,k}}{(\Gamma_{\downarrow,k} + \Lambda_{\downarrow,k}^\text{(diag.)}) f_{\overline{\nu}, k} + \mu_{\downarrow,\nu,k} + \eta_{\downarrow,\nu,k}} = \frac{f_{\nu,k}}{1-f_{\nu,k}}. \label{eq_e}
\end{equation}
If we assume that the steady state phonon and Coulomb rates vanish identically, then joining Eqs.~\eqref{gamma_down_up_diag}--\eqref{eq_e} leads to the following approximate relation for the condensation threshold for mode $m$:
\begin{equation}
     \Gamma_{\uparrow,c}^{(m)} \simeq \Gamma_{\downarrow} + \Lambda_{\downarrow,k}^\text{(diag.)} - \Lambda^\text{(diag.)}_{\uparrow,k} \label{Gamma_up_c_0}
\end{equation}
with $k$ denoting the electron-hole state with energy equal to $\hbar\omega_m$. Equation~\eqref{Gamma_up_c_0} is particularly useful to determine the boundary between the thermal phase and any other phase with only one condensed mode, say mode $m$, which implies that $n_{m}^\text{s.s.} \gg n_{n}^\text{s.s.}$ holds for $n \neq m$. In this case, we are allowed to retain only the condensed mode in Eqs.~\eqref{lambda_up_diag} and \eqref{lambda_down_diag}, which results in $\Lambda_{\downarrow,k}^\text{(diag.)} \simeq L_{k,m} n_m^\text{s.s.}$ and $\Lambda_{\uparrow,k}^\text{(diag.)} \simeq L_{k,m} (1+n_0^\text{s.s.})$. After some algebra we obtain
\begin{equation}
     \Gamma_{\uparrow,c}^{(m)}(\lambda_0)  = c_1 +  \frac{c_2}{1 + c_3/\lambda_0^2 + c_4/\lambda_0}  \label{final}
\end{equation}
as an estimate for the phase boundaries, where $c_j$ are constants that depend on the material parameters.

Both $1\rightarrow 2$ and $1\rightarrow 4$ transitions involve single condensed modes, which permits to apply Eq.~\eqref{final}. Indeed, in Fig.~\ref{fig_phase_diagram}(a) we include estimations for these phase-transition boundaries as a result of fitting Eq.~\eqref{final} to the numerical data. The analytic expression reproduces the numerical threshold line across a large range of $\lambda_{0}$, validating the sequence of approximations, despite the significant deviations close to $\lambda_0 = 950\,$nm, where the ground-state photon energy is maximum. Note that, to optically excite higher-energy photon modes, electron–hole pairs at higher energies are required, where the carrier distributions are expected to be more dilute and thermalization less effective. Then, the assumption that the threshold is governed solely by the resonant pair [Eqs.~\eqref{g_ress_d} and~\eqref{g_ress_u}] cease to hold, which can help to explain the poorer agreement of the black curve in this region.

\section{Discussion and conclusion}
\label{Sec_conclusions}

In conclusion, we have developed an \textit{ab initio} quantum kinetic framework that provides a microscopic foundation for photon Bose–Einstein condensation in semiconductor microcavities. This model self-consistently incorporates carrier–photon coupling, Coulomb-mediated carrier collisions, phonon scattering, continuous pumping, and losses on equal footing. By bridging the gap between simplified equilibrium models and the complex nonequilibrium dynamics of recent experiments, our theory enables quantitative predictions of condensate behavior under realistic device conditions. In particular, the model reveals that Coulomb-mediated thermalization is the dominant pathway for photon equilibration at high carrier densities -- a fundamental distinction from dye-based photon BECs -- and that the effective photon temperature is set by the pumping conditions rather than the ambient lattice temperature. These insights, combined with a rich phase diagram encompassing thermal, single-mode condensate, multimode, and lasing regimes, accurately reproduce observed phase transitions in semiconductor microcavities. The agreement between our simulations and experimental phase boundaries underscores the predictive power of the framework as a tool for interpreting experiments and designing new photonic structures. In particular, we identify how tunable parameters, such as the cavity distance or pump power, dictate the crossover between thermal regime and one of the condensed phases, thus informing future experimental endeavours. 

Leveraging these theoretical advances, we establish a predictive basis for engineering practical photonic devices that exploit photon condensation. The semiconductor platform, unlike its dye-based counterpart, supports true steady-state operation without long-lived molecular dark states. Our results confirm that a continuous-wave photonic condensate can be sustained in a III–V semiconductor at room temperature, enabling electrically driven condensate devices free of the pulsed-pumping constraints of dye systems. Crucially, because condensation occurs without the need for population inversion, the threshold for coherent light generation can be dramatically lowered. This points to the feasibility of ultra-low-threshold condensate lasers, where our model provides guidance on optimizing cavity parameters and pump conditions to minimize the threshold. Indeed, by quantitatively linking microscopic interactions to macroscopic observables like critical pump power, the theory allows device designers to predict the onset of condensation or multimode behavior under various geometries and excitation schemes. 

In combination with the inherent stability and integrability of semiconductor microcavities, these insights pave the way for on-chip, electrically injected photon-BEC devices that transcend the performance limits of conventional lasers. For example, our findings inform the development of: ~i) ultra-low threshold lasers exploiting condensation rather than inversion could enable energy-efficient optical interconnects \cite{yuhao_2022}, with recent prototypes achieving sub-$\mu$W thresholds in related polariton systems \cite{colombelli_2015, lagree_2024}, ii) reconfigurable photonic circuits could make use of the BEC phase coherence for analog quantum simulation \cite{vrenetar_2021}, and iii) hybrid quantum light sources may harness the strong photon-photon interactions to generate correlated photon pairs, complementing existing quantum dot approaches \cite{reiserer_2022}. In each case, the ab initio model presented here provides concrete design parameters to realize these applications, firmly connecting novel device concepts with predictive theoretical support.

\section{Acknowledgments}

Stimulating discussions with Rick Mukherjee and Yijun Tang are gratefully acknowledged.
J.~L.~F. and H.~T. acknowledge Funda\c{c}\~{a}o para a Ci\^{e}ncia e a Tecnologia (FCT-Portugal) through the Grants No. PD/BD/135211/2017, UI/BD/151557/2021, and through Contract No. CEECIND/00401/2018 and Project No. PTDC/FIS-OUT/3882/2020, respectively.

\appendix
\section{Interaction coupling constants in the semiconductor microcavity}\label{Ap_interactions}

The relevant interactions within the semiconductor environment are obtained via second-quantization methods by defining the field operators for the relevant quasiparticles. Then, the second-quantized version of each interaction follows from replacing the classical field by their quantum counterpart in the classical expression. For a more complete discussion, see Refs.~\cite{2006PQE,haug2004quantum}.

\subsection{Carrier--carrier interaction}
The electronic wave operator that collectively describes all carrier bands of the semiconductor quantum well can be decomposed as
\begin{equation}
    \hat  \Psi(\mathbf r) = \sum_{j,\nu,\mathbf k} \Phi_{j,\nu,\mathbf k}(\mathbf r) \hat  c_{j,\nu,\mathbf k}, \label{Psi_operator}
\end{equation}
where $j$ is an additional quantum number denoting the carrier subband for the reduced $z-$dimension. Hence $ \Phi_{j,\nu,\mathbf k}(\mathbf r) = \zeta_j(z)\phi_{\nu,\mathbf k}(x,y)$, with $\phi_{\nu,\mathbf k}(x,y)$ being a Bloch solution for the quasi-2D lattice, and $\zeta_j(z)$ the quantum-well envelope function. Additionally, let us assume that the motion of the carriers is restricted to the first subband $j=1$ and neglect inter-subband transitions. This is valid when the quantum-well thickness is sufficiently small so that it can be approximated by a two-dimensional system. As a result, the electron dynamics is confined to the $x-y$ plane, {\it i.e.} $|\zeta_{j=1}(z)|^2 \simeq \delta(z)$. 

The Coulomb interacting Hamiltonian can be written as
\begin{equation}
    \hat H_{c-c} = \frac{1}{2} \int d\mathbf r \int d\mathbf r' \,  \hat \Psi^\dagger(\mathbf r) \hat \Psi^\dagger (\mathbf r')\frac{e^2}{4\pi \epsilon |\mathbf r - \mathbf r'|}  \hat \Psi(\mathbf r')\hat \Psi(\mathbf r) \label{H_cc_1}
\end{equation}
with $\epsilon = \epsilon_0 \epsilon(\infty)$ the semiconductor permittivity and $\hat n(\mathbf r) = \hat \Psi^\dagger(\mathbf r)\hat \Psi(\mathbf r)$ the density operator.

For low-energy excitations, we restrict Eq.~\eqref{H_cc_1} to intraband transitions, which is valid for temperatures that verify $k_BT \ll E_g$, where $E_g$ denotes the band gap separating the bands. Then, replacing Eq.~\eqref{Psi_operator} into Eq.~\eqref{H_cc_1} and assuming the plane waves for the $x-y$ motion, $\phi_{\nu,\mathbf k}(\mathbf r_\perp) = \frac{1}{\sqrt{A}} e^{i \mathbf k\cdot \mathbf r_\perp}$, leads to Eq.~\eqref{H_c_c_final} of the main text:
\begin{equation}
     \hat  H_{c-c}  = 
   \frac{1}{2}
   \sum_{\nu,\nu',\mathbf q}
   V_{\mathbf q}^{\nu,\nu'}
   \sum_{\mathbf k,\mathbf k'}  \hat   c^\dagger_{\nu,\mathbf k + \mathbf q}\hat  c^\dagger_{\nu',\mathbf k' - \mathbf q}\hat   c_{\nu',\mathbf k'}\hat   c_{\nu,\mathbf k} ,
\end{equation}
with 
\begin{equation}
    V_{\mathbf q}^{\nu,\nu'} = \frac{Q_\nu Q_{\nu'}}{2 A \epsilon |\mathbf q|} \label{Coulomb_coupling}
\end{equation}
the two-dimensional Coulomb matrix element, $A$ the area of the sample, $Q_\nu = s_\nu e$ the electric charge of $\nu$ carriers and $s_\nu$ the sign. 

\subsection{Carrier--photon interaction}
The carrier-photon interaction stems from the coupling of the semiconductor dipole moment to the electric field of the cavity. The latter, denoted $\hat {\mathbf D} (\mathbf r)$, can be expanded in terms of cavity-photon operators as
\begin{equation}
    \hat {\mathbf D} (\mathbf r) = \sum_{m} \bm{\mathcal E}_{m}(\mathbf r) \hat  a_m + \text{h.c.},
\end{equation}
with expansion coefficients
\begin{align}
    \bm{\mathcal E}_{m}(x,y) & = \bm{\varepsilon}_m \sqrt{\frac{ \hbar\omega_m}{2 \epsilon_0\mathcal V }} \nonumber \\
    &\times\frac{\mathcal H_{m_x}\Big(\frac{x}{\ell_\text{HO}}\Big)\mathcal H_{m_y}\Big(\frac{y}{\ell_\text{HO}}\Big)}{\sqrt{\pi 2^{m_x+m_y} m_x! m_y!}  } e^{-(x^2+y^2)/2\ell_\text{HO}},
\end{align}
where $\mathcal H_n(x)$ are Hermite polynomials, $\mathcal V$ is the cavity volume, and $\ell_\text{HO} = c/\sqrt{\omega_{0} \delta\omega }$ the cavity-oscillator length. 

The carrier-photon interaction follows from 
\begin{equation}
     \hat  H_{c-\gamma} = -e \int  d \mathbf r \     \hat \Psi^\dagger(\mathbf r) \, \mathbf r \cdot    \hat {\mathbf D} (\mathbf r) \,    \hat  \Psi(\mathbf r)
\end{equation}

The former interactions allow for all possible events involving carrier scattering mediated by either absorption or emission of a photon. However, for practical purposes, the most important events will be electron-hole creation via photon absorption and electron-hole recombination via photon emission \cite{haug2004quantum}. Restricting the analysis to the latter transitions corresponds to the rotating-wave approximation, which leads to the simplified carrier-photon interaction given in Eq.~\eqref{H_c_gamma_final} of the main text,
\begin{equation}
  \hat   H_{c-\gamma} = \sum_{\mathbf k,m} g_{m}\Big( \hat   a_m \hat  c^\dagger_{e,\mathbf k}\hat  c^\dagger_{h,- \mathbf k} + \hat  a^\dagger_m \hat  c_{h,-\mathbf k}\hat  c_{e,\mathbf k} \Big),
\end{equation}
with couplings
\begin{equation}
    g_m = \frac{\textbf{d}}{a^2} \cdot \int d\mathbf r_\perp   \, \bm{\mathcal E}_{m}(\mathbf r_\perp ). \label{def_g_m}
\end{equation}
Here $\textbf{d} = -e\int_{cell} d\mathbf r_\perp \phi_{e,\mathbf k}^\ast(\mathbf r_\perp) \mathbf  r_\perp \phi_{h,\mathbf k}(\mathbf r_\perp)$ is the optical dipole moment (which is essentially independent of $\mathbf k$ and of order $\text{d}\sim 3\times 10^{-30}\,\text{C}\cdot\text{m}$ for InGaAs quantum wells) and $a$ is the lattice constant. Using the properties  $\mathcal H_m (-x) = (-1)^m \mathcal H_m(x)$ and $\mathcal H_{2m}(0)= (-2)^m (2m-1)!!$ we get
\begin{equation}
 g_m=\begin{cases}
     \sqrt{\frac{\hbar \omega_m d^2} {\epsilon_0 \mathcal V
}} \left(\frac{\ell_\text{H0}}{a}\right)^2\frac{(m_x-1)!! (m_y-1)!!}{ \sqrt{ m_x! m_y!  }} \quad & \text{if} \ \   m_x, m_y \in \text{even},\\
\\
\quad \quad \quad \quad  0  &\text{otherwise}.
\end{cases}  
\label{eq:coupling}
\end{equation}

The decoupling of odd cavity modes results from the variation of the optical field $\mathcal{E}_m(x,y)$ being negligible across a single lattice site, which renders the carriers locally insensitive to the electric field. In practical experiments, however, a small coupling to odd modes is expected due to imperfections in the crystal potential and nonzero momentum transfer between the photons and the electron-hole pairs. 

In what follows, we assume a perfect crystal and consider only the even modes. To ease the notation, we replace $m\rightarrow m/2$ for the even modes, such that $g_m$ is nonzero for all integers this the reduced mode basis. 

\subsection{Carrier-phonon interaction}
Akin to the carrier-photon interaction, the carrier-phonon interaction is also of electrostatic nature and results from the coupling between the carrier dipole and the lattice polarization field. The latter can be decomposed in terms of phonon operators as
\begin{equation}
    \hat {\mathbf P}(\mathbf r) =  \sum_{\mathbf q} \bm{\mathcal P}_{\mathbf q}(\mathbf r) \hat  b_\mathbf{q} + \text{h.c.}
\end{equation}
where the expansion coefficients read
\begin{equation}
     \bm{\mathcal P}_{\mathbf{q}}(\mathbf r) =  \frac{X\epsilon_0}{e}  \frac{\mathbf q}{q} e^{i\mathbf q\cdot \mathbf r},
\end{equation}
in terms of the Fr\"{o}hlich parameter for LO phonons \cite{haug2004quantum}
\begin{equation}
    X = \sqrt{\frac{ \hbar \omega_{LO}  e^2}{\ell_\text{c} A}\frac{1}{2 \epsilon_0} \Bigg(\frac{1}{\epsilon(\infty)} - \frac{1}{\epsilon(0)}\Bigg)}.
\end{equation}

Performing a rotating-wave approximation to neglect interband transitions, we are led to
\begin{equation}
      \hat    H_{c-p} = -\sum_{\nu,\mathbf k,\mathbf q}  s_\nu \lambda_{\mathbf q} (\hat  b_{\mathbf q} -\hat   b_{-\mathbf q}^\dagger) \hat  c_{\nu,\mathbf k + \mathbf q}^\dagger \hat  c_{\nu,\mathbf k},
\end{equation}
with coupling constant
\begin{equation}
       \lambda_{\mathbf q} = \frac{iX}{\sqrt{ A |\mathbf q|}}. \label{coupling_phonons}
\end{equation}

\section{Derivation of Coulomb rates}\label{Ap_Coulomb_coll} 
In this Appendix we show how to derive Eqs.~\eqref{C_nu}--\eqref{ProbScatt} of the main text. We begin with 
\begin{equation}
	\mathcal C_{\nu,\mathbf k} = \frac{i}{\hbar} \big\langle \big[ \hat  H_1, \hat c_{\nu,\mathbf k}^\dagger \hat c_{\nu,\mathbf k} \big] \big\rangle   , 
\end{equation}
where $\hat  H_1$ is the Coulomb interaction defined in Eq.~\eqref{H_c_c_final}. After a straightforward application of fermionic anticommutation relations, and making use of the properties $[ A, B] = \{ A, B\} - 2  B A$ and $[AB,C] = A[B,C] + [A,C]B$, we are led to 
\begin{align}
	\big[ \hat  H_1, \hat c_{\nu,\mathbf k}^\dagger \hat c_{\nu,\mathbf k} \big] &= \sum_{\nu' , \mathbf p,\mathbf q} V_{\mathbf q}^{\nu,\nu'} \Big( \hat c_{\nu,\mathbf k}^\dagger  \hat c_{\nu',\mathbf p}^\dagger  \hat c_{\nu,\mathbf k-\mathbf q}  \hat c_{\nu',\mathbf p+\mathbf q}  \nonumber \\
	& - \hat c_{\nu,\mathbf k+\mathbf q}^\dagger  \hat c_{\nu',\mathbf p}^\dagger  \hat c_{\nu,\mathbf k }  \hat c_{\nu',\mathbf p+\mathbf q} \Big).
\end{align}
Consequently, the Coulomb contribution becomes
\begin{equation}
	\mathcal C_{\nu,\mathbf k} =   -\frac{2}{\hbar}\sum_{\nu' , \mathbf k',\mathbf q} V_{\mathbf q}^{\nu,\nu'} \text{Im}\Big(a^{\nu,\nu'}_{\mathbf k,\mathbf k',\mathbf q}\Big), \label{C3}
\end{equation}
where $a^{\nu,\nu'}_{\mathbf k,\mathbf k',\mathbf q} = \big\langle \hat   c^\dagger_{\nu,\mathbf k + \mathbf q}\hat  c^\dagger_{\nu',\mathbf k' - \mathbf q}\hat   c_{\nu',\mathbf k'}\hat   c_{\nu,\mathbf k} \big\rangle$ is a higher-order correlator associated with two-body carrier-carrier processes. If we were to neglect all quantum fluctuations, then $a^{\nu,\nu'}_{\mathbf k,\mathbf k',\mathbf q} = \delta_{\mathbf q,\mathbf 0} f_{\nu,\mathbf k}f_{\nu',\mathbf k'} - \delta_{\mathbf k'- \mathbf k,\mathbf q}\delta_{\nu,\nu'}f_{\nu,\mathbf k}f_{\nu',\mathbf k'}$ which correspond to the classical ({\it i.e.}, mean field) plus Fock contributions. However, since both these contributions are real, they do not contribute to Eq.~\eqref{C3}, and a nontrivial result can only be found by taking the quantum fluctuations of fermionic operators into account. This is most easily done by solving the equation of motion for the correlator of interest. \par 
From the Lindblad equation we obtain
\begin{equation}
	\frac{\partial}{\partial t} a^{\nu,\nu'}_{\mathbf k,\mathbf k',\mathbf q} = -i  \chi \,  a^{\nu,\nu'}_{\mathbf k,\mathbf k',\mathbf q}  + Y^{\nu,\nu'}_{\mathbf k,\mathbf k',\mathbf q} + L^{\nu,\nu'}_{\mathbf k,\mathbf k',\mathbf q} , \label{A4}
\end{equation}
where $\chi \equiv \chi^{\nu,\nu'}_{\mathbf k,\mathbf k',\mathbf q} = (E_{\nu,\mathbf k} + E_{\nu',\mathbf k'} - E_{\nu', \mathbf k - \mathbf q} - E_{\nu,\mathbf k + \mathbf q})/\hbar$ is the natural frequency of oscillation, 
\begin{equation}
	Y^{\nu,\nu'}_{\mathbf k,\mathbf k',\mathbf q} = \frac{i}{\hbar} \big\langle\big[ \hat H_1, \hat   c^\dagger_{\nu,\mathbf k + \mathbf q}\hat  c^\dagger_{\nu',\mathbf k' - \mathbf q}\hat   c_{\nu',\mathbf k'}\hat   c_{\nu,\mathbf k}\big]\big\rangle 
\end{equation}
is the source term and $L^{\nu,\nu'}_{\mathbf k,\mathbf k',\mathbf q} $ is a small Lindblad contribution which can be safely neglected here. A solution to Eq. \eqref{A4} is found to be
\begin{equation}
	a^{\nu,\nu'}_{\mathbf k,\mathbf k',\mathbf q} (t) = e^{-i \chi t } \int_{-\infty}^t d\tau \, e^{i\chi \tau} \,  Y^{\nu,\nu'}_{\mathbf k,\mathbf k',\mathbf q} (\tau),
\end{equation}
provided that correlations vanish as $t\rightarrow -\infty$. Using the Markovian approximation as discussed in the main text, we may perform the integral and obtain 
\begin{equation}
	a^{\nu,\nu'}_{\mathbf k,\mathbf k',\mathbf q} (t) \simeq   \Bigg[ \pi \delta(\chi) - i \mathcal P \frac{1}{\chi}\Bigg]Y^{\nu,\nu'}_{\mathbf k,\mathbf k',\mathbf q} (t), \label{A6}
\end{equation}
where $\mathcal P$ denotes the principal part. \par
\begin{widetext}
Having obtained an explicit form for $a^{\nu,\nu'}_{\mathbf k,\mathbf k',\mathbf q} (t)$, we still need to relate $Y^{\nu,\nu'}_{\mathbf k,\mathbf k',\mathbf q}$ with the carrier distributions in order to close the system. Starting from its definition, and after a lengthy yet straightforward calculation, we arrive at
\begin{align}
	Y^{\nu,\nu'}_{\mathbf k,\mathbf k',\mathbf q} = \frac{i}{\hbar} &\sum_{\sigma, \mathbf p',\mathbf q'} \Big[ V_{\mathbf q'}^{\sigma, \nu'} \big\langle \hat c_{\nu,\mathbf k + \mathbf q}^\dagger \hat c_{\nu',\mathbf k' + \mathbf q}^\dagger \hat c_{\sigma,\mathbf p'}^\dagger \hat c_{\nu',\mathbf k' - \mathbf q'} \hat c_{\sigma,\mathbf p' + \mathbf q'} c_{\nu,\mathbf k}\big \rangle  - V_{\mathbf q'}^{\sigma, \nu'} \big\langle \hat c_{\nu,\mathbf k + \mathbf q}^\dagger \hat c_{\sigma,\mathbf p' + \mathbf q'}^\dagger \hat c_{\nu',\mathbf k ' - \mathbf q - \mathbf q'}^\dagger \hat c_{\sigma,\mathbf p'} \hat c_{\nu',\mathbf k'} c_{\nu,\mathbf k}\big \rangle \nonumber \\
	& + V_{\mathbf q'}^{\nu,\sigma} \big\langle \hat c_{\nu,\mathbf k + \mathbf q}^\dagger \hat c_{\nu',\mathbf k' + \mathbf q}^\dagger  \hat c_{\nu',\mathbf k'}\hat c_{\sigma,\mathbf p'}^\dagger \hat c_{\nu,\mathbf k - \mathbf q'} \hat  c_{\sigma,\mathbf p'+ \mathbf q'}\big \rangle - V_{\mathbf q'}^{\nu, \sigma} \big\langle \hat c_{\sigma,\mathbf p' + \mathbf q'}^\dagger \hat c_{\nu,\mathbf k + \mathbf q-\mathbf q'}^\dagger  \hat c_{\sigma,\mathbf p'}\hat c_{\nu',\mathbf k' - \mathbf q}^\dagger \hat c_{\nu',\mathbf k '} \hat  c_{\nu,\mathbf k}\big \rangle \Big]. \label{A7}
\end{align}
The cluster expansion technique can now be used to factorize the three-body correlators above into products of carrier distributions \cite{2006PQE}. Since both $Y^{\nu,\nu'}_{\mathbf k,\mathbf k',\mathbf q}$ and $\mathcal C_{\nu,\mathbf k}$ contain one interaction amplitude $V_{\mathbf q}^{\nu,\nu'}$ each, we obtain the corrections up to second order in $V_{\mathbf q}^{\nu,\nu'}$ by retaining the singlet terms in Eq.~\eqref{A7}. As an example, the singlet contribution to the first correlator is
\begin{align}
	&\big\langle \hat c_{\nu,\mathbf k + \mathbf q}^\dagger \hat c_{\nu',\mathbf k' + \mathbf q}^\dagger \hat c_{\sigma,\mathbf p'}^\dagger \hat c_{\nu',\mathbf k' - \mathbf q'} \hat c_{\sigma,\mathbf p' + \mathbf q'} c_{\nu,\mathbf k}\big \rangle  \simeq \delta_{\sigma, \nu'} \delta_{\mathbf q,\mathbf 0} \delta_{\mathbf p',\mathbf k'-\mathbf q'} f_{\nu',\mathbf p'} f_{\nu',\mathbf k'} f_{\nu,\mathbf k}  + \delta_{\nu,\sigma} \delta_{\mathbf q,\mathbf q'} \delta_{\mathbf p',\mathbf k} f_{\nu',\mathbf k'-\mathbf q} f_{\nu,\mathbf k + \mathbf q} f_{\nu,\mathbf k} \nonumber \\
	& +\delta_{\nu,\nu'} \delta_{\mathbf k' -\mathbf k, \mathbf q} \delta_{\mathbf q',\mathbf 0} f_{\nu,\mathbf k + \mathbf q} f_{\sigma, \mathbf p'} f_{\nu,\mathbf k}  - \delta_{\sigma,\nu'} \delta_{\sigma,\nu} \delta_{\mathbf q,\mathbf q'} \delta_{\mathbf p',\mathbf k} f_{\sigma,\mathbf p'} f_{\nu,\mathbf k + \mathbf q} f_{\nu,\mathbf k} - \delta_{\mathbf q,\mathbf 0}\delta_{\mathbf q',\mathbf 0} f_{\nu',\mathbf k'} f_{\sigma ,\mathbf p'} f_{\nu,\mathbf k}  \nonumber \\
	& - \delta_{\sigma,\nu'} \delta_{\sigma,\nu}  \delta_{\mathbf k,\mathbf p'} \delta_{\mathbf k+\mathbf q , \mathbf k' - \mathbf q'} f_{\nu,\mathbf k + \mathbf q} f_{\nu ' , \mathbf k' - \mathbf q} f_{\nu,\mathbf k}.
\end{align}
The six terms therein correspond to the six possible pairings of one creation and one annihilation operators. Doing the same for the remaining terms and replacing the final solution into Eq.~\eqref{A6} permits to relate $a^{\nu,\nu'}_{\mathbf k,\mathbf k',\mathbf q}$ with carrier distribution functions, thereby closing the system. Then, Eq.~\eqref{C3} becomes 
\begin{align}
\mathcal C_{\nu,\mathbf k}  =  \frac{2\pi}{\hbar} \sum_{\nu',\mathbf k',\mathbf q} |V^{\nu,\nu'}_\mathbf{q}|^2\delta\big( \chi^{\nu,\nu'}_{\mathbf k,\mathbf k',\mathbf q}\big) \Big[  f_{\nu,\mathbf k + \mathbf q} f_{\nu',\mathbf k' - \mathbf q}(1-f_{\nu',\mathbf k'} ) (1-f_{\nu,\mathbf k})-f_{\nu,\mathbf k} f_{\nu',\mathbf k'}(1-    f_{\nu,\mathbf k + \mathbf q} )(1-f_{\nu',\mathbf k' - \mathbf q}) \Big],
\end{align}
as desired.
\end{widetext}

\section{Born-Markov approximation for phonon coupling}\label{Ap_Phonon}
Consider a single type of carrier coupled to a phononic field. The Hamiltonian is decomposed as $\hat H = \hat H_0 + \hat V$, with
 \begin{align}
 	\hat H_0 &= \sum_{\mathbf k} E_{\mathbf k} \hat c^\dagger_{\mathbf k} \hat c_{\mathbf k} + \sum_{\mathbf q} \xi \hat b^\dagger_{\mathbf q} \hat b_{\mathbf q} 
 \end{align}
 and
 \begin{align}
 	\hat V &=  \sum_{\mathbf k,\mathbf q} \lambda_{\mathbf q} (\hat b_{\mathbf q} - \hat b^\dagger_{\mathbf q}) \hat c_{\mathbf k+\mathbf q}^\dagger \hat  c_{\mathbf k}.
\end{align}
The total Hamiltonian $\hat H$ acts on a space spanned by the many-body basis $\{\ket{\mathbf n} = \ket{\mathbf m}_{c} \otimes \ket{\mathbf g}_{b}\}$, where $\mathbf m = (m_1,m_2,...)$ is a vector of fermionic occupations, {\it i.e.},  $m_j\in \{0,1\}$, and $\mathbf g = (g_1,g_2,...)$ the photonic counterpart, with $g_j \in \mathbb{Z}^{+}$.  \par 
Let us now assume that the phonon system can be considered a static reservoir in thermal equilibrium at all times. The density matrix for the reduced carrier system is defined as $\hat \rho_S(t) = \text{tr} \, \hat \rho(t)$ where $\hat \rho(t)$ is the total density matrix and 'tr' denotes the trace with respect to phonon states. While interactions may induce correlations, for a sufficiently large reservoir such correlations decay over a characteristic time $\tau_c \ll \Delta t$, allowing the factorized approximation $\hat\rho(t+\Delta t) \simeq \hat \rho_S(t+\Delta t) \otimes \hat \rho_E(0)$. Here, $\hat \rho_E(0) = Z^{-1} \exp(-\beta \sum_{\mathbf q} \hbar\omega_{LO} \hat b^\dagger_{\mathbf q} \hat b_{\mathbf q})$ is the thermal state of the reservoir. \par 
Iterating the Dyson-Born series for the total density matrix up to second order in $\lambda$ and taking the trace with respect to the reservoir yields the leading contribution to the master equation for the reduced system due to the phonon bath:
\begin{widetext}
\begin{align}
	\frac{\partial \hat \rho_S(t)}{\partial t} \Bigg|_{\lambda}&= -\frac{1}{ \hbar^2\Delta t} \int_{t}^{t+\Delta t} dt' \int_{t}^{t'}dt'' \Big( \big\langle \hat V_i(t')\hat V_i(t'') \big\rangle_{\!E} \, \hat \rho_S(t) + \hat \rho_S(t)\big\langle \hat V_i(t')\hat V_i(t'') \big\rangle_{\!E} \nonumber  \\
	& - \big\langle \hat V_i(t')\hat \rho_S(t)\hat V_i(t'') \big\rangle_{\!E}- \big\langle \hat V_i(t'')\hat \rho_S(t)\hat V_i(t') \big\rangle_{\!E} \Big) .\label{BMapp}
\end{align}	
Above, $\hat V_i(t) = e^{i\hat H_0 t} \hat V e^{-i\hat H_0 t}$ is the potential in the interaction picture and $\langle \hat O \rangle_{\! E} \equiv \text{tr}\Big[ \rho_E(0)\hat O\Big]$ denotes the expectation value with respect to the reservoir. Using time-translation invariance, $ \big\langle \hat V_i(t+\tau)\hat V_i(t) \big\rangle_{\!E} = \big\langle \hat V_i(\tau)\hat V_i(0) \big\rangle_{\!E}$, and noting that $\big\langle \hat V_i(\tau)\hat V_i(0) \big\rangle_{\!E}$ is nonvanishing only for $\tau \lesssim \tau_e$, Eq.\eqref{BMapp} can be approximated by
\begin{align}
	\frac{\partial \hat \rho_S(t)}{\partial t} \Bigg|_{\lambda}& \simeq  -\frac{1}{\hbar^2} \int_{0}^{\infty}d\tau  \Big( \big\langle \hat V_i(\tau)\hat V_i(0) \big\rangle_{\!E} \, \hat \rho_S(t) + \hat \rho_S(t)\big\langle \hat V_i(-\tau)\hat V_i(0) \big\rangle_{\!E}- \big\langle \hat V_i(\tau)\hat \rho_S(t)\hat V_i(0) \big\rangle_{\!E}- \big\langle \hat V_i(-\tau)\hat \rho_S(t)\hat V_i(0) \big\rangle_{\!E} \Big).\label{BMapp2}
\end{align}	
\end{widetext}
The thermal state of the bath provides $\big\langle\hat b^{\dagger}_{\mathbf q}\hat b_{\mathbf q'}\big\rangle_{\!E}  = \delta_{\mathbf q, \mathbf q'} \mathcal N$ and $\big\langle\hat b_{\mathbf q}\hat b^\dagger_{\mathbf q'}\big\rangle_{\!E}  = \delta_{\mathbf q, \mathbf q'}(1+ \mathcal N)$, with $\mathcal N \equiv \mathcal N(\hbar \omega_{LO})$ the Bose-Einstein function evaluated at the phonon energy. The correlation function becomes

\begin{align}
&\big\langle \hat V_i(\tau)\hat V_i(0) \big\rangle_{\!E}  = -\frac{1}{\hbar^2}\sum_{\mathbf k,\mathbf k',\mathbf q} |\lambda_{\mathbf q}|^2 e^{-i\tau (E_{\mathbf k+\mathbf q} - E_{\mathbf k})/\hbar} \nonumber \\
& \times \hat c_{\mathbf k+\mathbf q}^\dagger \hat  c_{\mathbf k}\hat c_{\mathbf k'+\mathbf q}^\dagger \hat  c_{\mathbf k'} \Big[ e^{-i\omega_{LO}\tau}(1+\mathcal N) + e^{i\omega_{LO}\tau} \mathcal N\Big].
\end{align}
Inserting this expression into Eq.~\eqref{BMapp2} and performing the time integrations leads to the final expression of Lindblad form, 
\begin{align}
	\frac{\partial \hat \rho_S(t)}{\partial t} \Bigg|_{\lambda} = \sum_{\mathbf k,\mathbf k',\mathbf q} C_{\mathbf k,\mathbf k',\mathbf q} \Big[ \big\{\hat c_{\mathbf k+\mathbf q}^\dagger \hat  c_{\mathbf k}\hat c_{\mathbf k'+\mathbf q}^\dagger \hat  c_{\mathbf k'},\hat \rho_S(t)\big\} \nonumber \\
	- 2 \hat c_{\mathbf k+\mathbf q}^\dagger \hat  c_{\mathbf k}\hat \rho_S(t) \hat c_{\mathbf k'+\mathbf q}^\dagger \hat  c_{\mathbf k'}\Big],
\end{align}	
where we defined the coefficients
\begin{align}
	 C_{\mathbf k,\mathbf k',\mathbf q} &= \frac{2\pi}{\hbar}|\lambda_{\mathbf q}|^2 \Big[\delta(\hbar \omega_{LO}+ E_{\mathbf k+\mathbf q} - E_{\mathbf k}) (1+\mathcal N) \nonumber \\
	  &+\delta(E_{\mathbf k+\mathbf q} - E_{\mathbf k} -\hbar \omega_{LO}+ ) \mathcal N\Big].
\end{align}
The generalization to include more than one type of carrier is straightforward. 

\bibliographystyle{apsrev4-2}
\bibliography{references}

\begin{thebibliography}{51}%
\makeatletter
\providecommand \@ifxundefined [1]{%
 \@ifx{#1\undefined}
}%
\providecommand \@ifnum [1]{%
 \ifnum #1\expandafter \@firstoftwo
 \else \expandafter \@secondoftwo
 \fi
}%
\providecommand \@ifx [1]{%
 \ifx #1\expandafter \@firstoftwo
 \else \expandafter \@secondoftwo
 \fi
}%
\providecommand \natexlab [1]{#1}%
\providecommand \enquote  [1]{``#1''}%
\providecommand \bibnamefont  [1]{#1}%
\providecommand \bibfnamefont [1]{#1}%
\providecommand \citenamefont [1]{#1}%
\providecommand \href@noop [0]{\@secondoftwo}%
\providecommand \href [0]{\begingroup \@sanitize@url \@href}%
\providecommand \@href[1]{\@@startlink{#1}\@@href}%
\providecommand \@@href[1]{\endgroup#1\@@endlink}%
\providecommand \@sanitize@url [0]{\catcode `\\12\catcode `\$12\catcode
  `\&12\catcode `\#12\catcode `\^12\catcode `\_12\catcode `\%12\relax}%
\providecommand \@@startlink[1]{}%
\providecommand \@@endlink[0]{}%
\providecommand \url  [0]{\begingroup\@sanitize@url \@url }%
\providecommand \@url [1]{\endgroup\@href {#1}{\urlprefix }}%
\providecommand \urlprefix  [0]{URL }%
\providecommand \Eprint [0]{\href }%
\providecommand \doibase [0]{https://doi.org/}%
\providecommand \selectlanguage [0]{\@gobble}%
\providecommand \bibinfo  [0]{\@secondoftwo}%
\providecommand \bibfield  [0]{\@secondoftwo}%
\providecommand \translation [1]{[#1]}%
\providecommand \BibitemOpen [0]{}%
\providecommand \bibitemStop [0]{}%
\providecommand \bibitemNoStop [0]{.\EOS\space}%
\providecommand \EOS [0]{\spacefactor3000\relax}%
\providecommand \BibitemShut  [1]{\csname bibitem#1\endcsname}%
\let\auto@bib@innerbib\@empty
\bibitem [{\citenamefont {Klaers}\ \emph
  {et~al.}(2010{\natexlab{a}})\citenamefont {Klaers}, \citenamefont {Schmitt},
  \citenamefont {Vewinger},\ and\ \citenamefont {Weitz}}]{Klaers2010_2}%
  \BibitemOpen
  \bibfield  {author} {\bibinfo {author} {\bibfnamefont {J.}~\bibnamefont
  {Klaers}}, \bibinfo {author} {\bibfnamefont {J.}~\bibnamefont {Schmitt}},
  \bibinfo {author} {\bibfnamefont {F.}~\bibnamefont {Vewinger}},\ and\
  \bibinfo {author} {\bibfnamefont {M.}~\bibnamefont {Weitz}},\ }\href
  {https://doi.org/10.1038/nature09567} {\bibfield  {journal} {\bibinfo
  {journal} {Nature}\ }\textbf {\bibinfo {volume} {468}},\ \bibinfo {pages}
  {545} (\bibinfo {year} {2010}{\natexlab{a}})}\BibitemShut {NoStop}%
\bibitem [{\citenamefont {Schofield}\ \emph {et~al.}(2024)\citenamefont
  {Schofield}, \citenamefont {Fu}, \citenamefont {Clarke}, \citenamefont
  {Farrer}, \citenamefont {Trapalis}, \citenamefont {Dhar}, \citenamefont
  {Mukherjee}, \citenamefont {Severs~Millard}, \citenamefont {Heffernan},
  \citenamefont {Mintert}, \citenamefont {Nyman},\ and\ \citenamefont
  {Oulton}}]{Schofield2024}%
  \BibitemOpen
  \bibfield  {author} {\bibinfo {author} {\bibfnamefont {R.~C.}\ \bibnamefont
  {Schofield}}, \bibinfo {author} {\bibfnamefont {M.}~\bibnamefont {Fu}},
  \bibinfo {author} {\bibfnamefont {E.}~\bibnamefont {Clarke}}, \bibinfo
  {author} {\bibfnamefont {I.}~\bibnamefont {Farrer}}, \bibinfo {author}
  {\bibfnamefont {A.}~\bibnamefont {Trapalis}}, \bibinfo {author}
  {\bibfnamefont {H.~S.}\ \bibnamefont {Dhar}}, \bibinfo {author}
  {\bibfnamefont {R.}~\bibnamefont {Mukherjee}}, \bibinfo {author}
  {\bibfnamefont {T.}~\bibnamefont {Severs~Millard}}, \bibinfo {author}
  {\bibfnamefont {J.}~\bibnamefont {Heffernan}}, \bibinfo {author}
  {\bibfnamefont {F.}~\bibnamefont {Mintert}}, \bibinfo {author} {\bibfnamefont
  {R.~A.}\ \bibnamefont {Nyman}},\ and\ \bibinfo {author} {\bibfnamefont
  {R.~F.}\ \bibnamefont {Oulton}},\ }\href
  {https://doi.org/10.1038/s41566-024-01491-2} {\bibfield  {journal} {\bibinfo
  {journal} {Nature Photonics}\ }\textbf {\bibinfo {volume} {18}},\ \bibinfo
  {pages} {1083} (\bibinfo {year} {2024})}\BibitemShut {NoStop}%
\bibitem [{\citenamefont {Keeling}\ and\ \citenamefont
  {Kirton}(2016)}]{PhysRevA.93.013829}%
  \BibitemOpen
  \bibfield  {author} {\bibinfo {author} {\bibfnamefont {J.}~\bibnamefont
  {Keeling}}\ and\ \bibinfo {author} {\bibfnamefont {P.}~\bibnamefont
  {Kirton}},\ }\href {https://doi.org/10.1103/PhysRevA.93.013829} {\bibfield
  {journal} {\bibinfo  {journal} {Phys. Rev. A}\ }\textbf {\bibinfo {volume}
  {93}},\ \bibinfo {pages} {013829} (\bibinfo {year} {2016})}\BibitemShut
  {NoStop}%
\bibitem [{\citenamefont {Klaers}\ \emph {et~al.}(2012)\citenamefont {Klaers},
  \citenamefont {Schmitt}, \citenamefont {Damm}, \citenamefont {Vewinger},\
  and\ \citenamefont {Weitz}}]{PhysRevLett.108.160403}%
  \BibitemOpen
  \bibfield  {author} {\bibinfo {author} {\bibfnamefont {J.}~\bibnamefont
  {Klaers}}, \bibinfo {author} {\bibfnamefont {J.}~\bibnamefont {Schmitt}},
  \bibinfo {author} {\bibfnamefont {T.}~\bibnamefont {Damm}}, \bibinfo {author}
  {\bibfnamefont {F.}~\bibnamefont {Vewinger}},\ and\ \bibinfo {author}
  {\bibfnamefont {M.}~\bibnamefont {Weitz}},\ }\href
  {https://doi.org/10.1103/PhysRevLett.108.160403} {\bibfield  {journal}
  {\bibinfo  {journal} {Phys. Rev. Lett.}\ }\textbf {\bibinfo {volume} {108}},\
  \bibinfo {pages} {160403} (\bibinfo {year} {2012})}\BibitemShut {NoStop}%
\bibitem [{\citenamefont {Klaers}\ \emph
  {et~al.}(2010{\natexlab{b}})\citenamefont {Klaers}, \citenamefont
  {Vewinger},\ and\ \citenamefont {Weitz}}]{Klaers2010}%
  \BibitemOpen
  \bibfield  {author} {\bibinfo {author} {\bibfnamefont {J.}~\bibnamefont
  {Klaers}}, \bibinfo {author} {\bibfnamefont {F.}~\bibnamefont {Vewinger}},\
  and\ \bibinfo {author} {\bibfnamefont {M.}~\bibnamefont {Weitz}},\ }\href
  {https://doi.org/10.1038/nphys1680} {\bibfield  {journal} {\bibinfo
  {journal} {Nature Physics}\ }\textbf {\bibinfo {volume} {6}},\ \bibinfo
  {pages} {512} (\bibinfo {year} {2010}{\natexlab{b}})}\BibitemShut {NoStop}%
\bibitem [{\citenamefont {\"Ozt\"urk}\ \emph {et~al.}(2023)\citenamefont
  {\"Ozt\"urk}, \citenamefont {Vewinger}, \citenamefont {Weitz},\ and\
  \citenamefont {Schmitt}}]{PhysRevLett.130.033602}%
  \BibitemOpen
  \bibfield  {author} {\bibinfo {author} {\bibfnamefont {F.~E.}\ \bibnamefont
  {\"Ozt\"urk}}, \bibinfo {author} {\bibfnamefont {F.}~\bibnamefont
  {Vewinger}}, \bibinfo {author} {\bibfnamefont {M.}~\bibnamefont {Weitz}},\
  and\ \bibinfo {author} {\bibfnamefont {J.}~\bibnamefont {Schmitt}},\ }\href
  {https://doi.org/10.1103/PhysRevLett.130.033602} {\bibfield  {journal}
  {\bibinfo  {journal} {Phys. Rev. Lett.}\ }\textbf {\bibinfo {volume} {130}},\
  \bibinfo {pages} {033602} (\bibinfo {year} {2023})}\BibitemShut {NoStop}%
\bibitem [{\citenamefont {Vlaho}\ and\ \citenamefont
  {Eckardt}(2021)}]{PhysRevA.104.063709}%
  \BibitemOpen
  \bibfield  {author} {\bibinfo {author} {\bibfnamefont {M.}~\bibnamefont
  {Vlaho}}\ and\ \bibinfo {author} {\bibfnamefont {A.}~\bibnamefont
  {Eckardt}},\ }\href {https://doi.org/10.1103/PhysRevA.104.063709} {\bibfield
  {journal} {\bibinfo  {journal} {Phys. Rev. A}\ }\textbf {\bibinfo {volume}
  {104}},\ \bibinfo {pages} {063709} (\bibinfo {year} {2021})}\BibitemShut
  {NoStop}%
\bibitem [{\citenamefont {Kirton}\ and\ \citenamefont
  {Keeling}(2015)}]{PhysRevA.91.033826}%
  \BibitemOpen
  \bibfield  {author} {\bibinfo {author} {\bibfnamefont {P.}~\bibnamefont
  {Kirton}}\ and\ \bibinfo {author} {\bibfnamefont {J.}~\bibnamefont
  {Keeling}},\ }\href {https://doi.org/10.1103/PhysRevA.91.033826} {\bibfield
  {journal} {\bibinfo  {journal} {Phys. Rev. A}\ }\textbf {\bibinfo {volume}
  {91}},\ \bibinfo {pages} {033826} (\bibinfo {year} {2015})}\BibitemShut
  {NoStop}%
\bibitem [{\citenamefont {Kasprzak}\ \emph {et~al.}(2006)\citenamefont
  {Kasprzak}, \citenamefont {Richard}, \citenamefont {Kundermann},
  \citenamefont {Baas}, \citenamefont {Jeambrun}, \citenamefont {Keeling},
  \citenamefont {Marchetti}, \citenamefont {Szyma{\'n}ska}, \citenamefont
  {Andr{\'e}}, \citenamefont {Staehli} \emph {et~al.}}]{kasprzak2006bose}%
  \BibitemOpen
  \bibfield  {author} {\bibinfo {author} {\bibfnamefont {J.}~\bibnamefont
  {Kasprzak}}, \bibinfo {author} {\bibfnamefont {M.}~\bibnamefont {Richard}},
  \bibinfo {author} {\bibfnamefont {S.}~\bibnamefont {Kundermann}}, \bibinfo
  {author} {\bibfnamefont {A.}~\bibnamefont {Baas}}, \bibinfo {author}
  {\bibfnamefont {P.}~\bibnamefont {Jeambrun}}, \bibinfo {author}
  {\bibfnamefont {J.~M.~J.}\ \bibnamefont {Keeling}}, \bibinfo {author}
  {\bibfnamefont {F.}~\bibnamefont {Marchetti}}, \bibinfo {author}
  {\bibfnamefont {M.}~\bibnamefont {Szyma{\'n}ska}}, \bibinfo {author}
  {\bibfnamefont {R.}~\bibnamefont {Andr{\'e}}}, \bibinfo {author}
  {\bibfnamefont {J.~a.}\ \bibnamefont {Staehli}}, \emph {et~al.},\ }\href@noop
  {} {\bibfield  {journal} {\bibinfo  {journal} {Nature}\ }\textbf {\bibinfo
  {volume} {443}},\ \bibinfo {pages} {409} (\bibinfo {year}
  {2006})}\BibitemShut {NoStop}%
\bibitem [{\citenamefont {Sun}\ \emph {et~al.}(2017)\citenamefont {Sun},
  \citenamefont {Wen}, \citenamefont {Yoon}, \citenamefont {Liu}, \citenamefont
  {Steger}, \citenamefont {Pfeiffer}, \citenamefont {West}, \citenamefont
  {Snoke},\ and\ \citenamefont {Nelson}}]{PhysRevLett.118.016602}%
  \BibitemOpen
  \bibfield  {author} {\bibinfo {author} {\bibfnamefont {Y.}~\bibnamefont
  {Sun}}, \bibinfo {author} {\bibfnamefont {P.}~\bibnamefont {Wen}}, \bibinfo
  {author} {\bibfnamefont {Y.}~\bibnamefont {Yoon}}, \bibinfo {author}
  {\bibfnamefont {G.}~\bibnamefont {Liu}}, \bibinfo {author} {\bibfnamefont
  {M.}~\bibnamefont {Steger}}, \bibinfo {author} {\bibfnamefont {L.~N.}\
  \bibnamefont {Pfeiffer}}, \bibinfo {author} {\bibfnamefont {K.}~\bibnamefont
  {West}}, \bibinfo {author} {\bibfnamefont {D.~W.}\ \bibnamefont {Snoke}},\
  and\ \bibinfo {author} {\bibfnamefont {K.~A.}\ \bibnamefont {Nelson}},\
  }\href {https://doi.org/10.1103/PhysRevLett.118.016602} {\bibfield  {journal}
  {\bibinfo  {journal} {Phys. Rev. Lett.}\ }\textbf {\bibinfo {volume} {118}},\
  \bibinfo {pages} {016602} (\bibinfo {year} {2017})}\BibitemShut {NoStop}%
\bibitem [{\citenamefont {Moilanen}\ \emph {et~al.}(2021)\citenamefont
  {Moilanen}, \citenamefont {Daskalakis}, \citenamefont {Taskinen},\ and\
  \citenamefont {T\"orm\"a}}]{PhysRevLett.127.255301}%
  \BibitemOpen
  \bibfield  {author} {\bibinfo {author} {\bibfnamefont {A.~J.}\ \bibnamefont
  {Moilanen}}, \bibinfo {author} {\bibfnamefont {K.~S.}\ \bibnamefont
  {Daskalakis}}, \bibinfo {author} {\bibfnamefont {J.~M.}\ \bibnamefont
  {Taskinen}},\ and\ \bibinfo {author} {\bibfnamefont {P.}~\bibnamefont
  {T\"orm\"a}},\ }\href {https://doi.org/10.1103/PhysRevLett.127.255301}
  {\bibfield  {journal} {\bibinfo  {journal} {Phys. Rev. Lett.}\ }\textbf
  {\bibinfo {volume} {127}},\ \bibinfo {pages} {255301} (\bibinfo {year}
  {2021})}\BibitemShut {NoStop}%
\bibitem [{\citenamefont {Hakala}\ \emph {et~al.}(2018)\citenamefont {Hakala},
  \citenamefont {Moilanen}, \citenamefont {V{\"a}kev{\"a}inen}, \citenamefont
  {Guo}, \citenamefont {Martikainen}, \citenamefont {Daskalakis}, \citenamefont
  {Rekola}, \citenamefont {Julku},\ and\ \citenamefont
  {T{\"o}rm{\"a}}}]{hakala2018bose}%
  \BibitemOpen
  \bibfield  {author} {\bibinfo {author} {\bibfnamefont {T.~K.}\ \bibnamefont
  {Hakala}}, \bibinfo {author} {\bibfnamefont {A.~J.}\ \bibnamefont
  {Moilanen}}, \bibinfo {author} {\bibfnamefont {A.~I.}\ \bibnamefont
  {V{\"a}kev{\"a}inen}}, \bibinfo {author} {\bibfnamefont {R.}~\bibnamefont
  {Guo}}, \bibinfo {author} {\bibfnamefont {J.-P.}\ \bibnamefont
  {Martikainen}}, \bibinfo {author} {\bibfnamefont {K.~S.}\ \bibnamefont
  {Daskalakis}}, \bibinfo {author} {\bibfnamefont {H.~T.}\ \bibnamefont
  {Rekola}}, \bibinfo {author} {\bibfnamefont {A.}~\bibnamefont {Julku}},\ and\
  \bibinfo {author} {\bibfnamefont {P.}~\bibnamefont {T{\"o}rm{\"a}}},\
  }\href@noop {} {\bibfield  {journal} {\bibinfo  {journal} {Nature Physics}\
  }\textbf {\bibinfo {volume} {14}},\ \bibinfo {pages} {739} (\bibinfo {year}
  {2018})}\BibitemShut {NoStop}%
\bibitem [{\citenamefont {Schmitt}\ \emph {et~al.}(2014)\citenamefont
  {Schmitt}, \citenamefont {Damm}, \citenamefont {Dung}, \citenamefont
  {Vewinger}, \citenamefont {Klaers},\ and\ \citenamefont
  {Weitz}}]{PhysRevLett.112.030401}%
  \BibitemOpen
  \bibfield  {author} {\bibinfo {author} {\bibfnamefont {J.}~\bibnamefont
  {Schmitt}}, \bibinfo {author} {\bibfnamefont {T.}~\bibnamefont {Damm}},
  \bibinfo {author} {\bibfnamefont {D.}~\bibnamefont {Dung}}, \bibinfo {author}
  {\bibfnamefont {F.}~\bibnamefont {Vewinger}}, \bibinfo {author}
  {\bibfnamefont {J.}~\bibnamefont {Klaers}},\ and\ \bibinfo {author}
  {\bibfnamefont {M.}~\bibnamefont {Weitz}},\ }\href
  {https://doi.org/10.1103/PhysRevLett.112.030401} {\bibfield  {journal}
  {\bibinfo  {journal} {Phys. Rev. Lett.}\ }\textbf {\bibinfo {volume} {112}},\
  \bibinfo {pages} {030401} (\bibinfo {year} {2014})}\BibitemShut {NoStop}%
\bibitem [{\citenamefont {Marelic}\ and\ \citenamefont
  {Nyman}(2015)}]{PhysRevA.91.033813}%
  \BibitemOpen
  \bibfield  {author} {\bibinfo {author} {\bibfnamefont {J.}~\bibnamefont
  {Marelic}}\ and\ \bibinfo {author} {\bibfnamefont {R.~A.}\ \bibnamefont
  {Nyman}},\ }\href {https://doi.org/10.1103/PhysRevA.91.033813} {\bibfield
  {journal} {\bibinfo  {journal} {Phys. Rev. A}\ }\textbf {\bibinfo {volume}
  {91}},\ \bibinfo {pages} {033813} (\bibinfo {year} {2015})}\BibitemShut
  {NoStop}%
\bibitem [{\citenamefont {Kirton}\ and\ \citenamefont
  {Keeling}(2018)}]{kirton2018superradiant}%
  \BibitemOpen
  \bibfield  {author} {\bibinfo {author} {\bibfnamefont {P.}~\bibnamefont
  {Kirton}}\ and\ \bibinfo {author} {\bibfnamefont {J.}~\bibnamefont
  {Keeling}},\ }\href@noop {} {\bibfield  {journal} {\bibinfo  {journal} {New
  Journal of Physics}\ }\textbf {\bibinfo {volume} {20}},\ \bibinfo {pages}
  {015009} (\bibinfo {year} {2018})}\BibitemShut {NoStop}%
\bibitem [{\citenamefont {Rodrigues}\ \emph {et~al.}(2021)\citenamefont
  {Rodrigues}, \citenamefont {Dhar}, \citenamefont {Walker}, \citenamefont
  {Smith}, \citenamefont {Oulton}, \citenamefont {Mintert},\ and\ \citenamefont
  {Nyman}}]{rodrigues2021learning}%
  \BibitemOpen
  \bibfield  {author} {\bibinfo {author} {\bibfnamefont {J.~D.}\ \bibnamefont
  {Rodrigues}}, \bibinfo {author} {\bibfnamefont {H.~S.}\ \bibnamefont {Dhar}},
  \bibinfo {author} {\bibfnamefont {B.~T.}\ \bibnamefont {Walker}}, \bibinfo
  {author} {\bibfnamefont {J.~M.}\ \bibnamefont {Smith}}, \bibinfo {author}
  {\bibfnamefont {R.~F.}\ \bibnamefont {Oulton}}, \bibinfo {author}
  {\bibfnamefont {F.}~\bibnamefont {Mintert}},\ and\ \bibinfo {author}
  {\bibfnamefont {R.~A.}\ \bibnamefont {Nyman}},\ }\href@noop {} {\bibfield
  {journal} {\bibinfo  {journal} {Physical Review Letters}\ }\textbf {\bibinfo
  {volume} {126}},\ \bibinfo {pages} {150602} (\bibinfo {year}
  {2021})}\BibitemShut {NoStop}%
\bibitem [{\citenamefont {Damm}\ \emph {et~al.}(2017)\citenamefont {Damm},
  \citenamefont {Dung}, \citenamefont {Vewinger}, \citenamefont {Weitz},\ and\
  \citenamefont {Schmitt}}]{damm2017first}%
  \BibitemOpen
  \bibfield  {author} {\bibinfo {author} {\bibfnamefont {T.}~\bibnamefont
  {Damm}}, \bibinfo {author} {\bibfnamefont {D.}~\bibnamefont {Dung}}, \bibinfo
  {author} {\bibfnamefont {F.}~\bibnamefont {Vewinger}}, \bibinfo {author}
  {\bibfnamefont {M.}~\bibnamefont {Weitz}},\ and\ \bibinfo {author}
  {\bibfnamefont {J.}~\bibnamefont {Schmitt}},\ }\href@noop {} {\bibfield
  {journal} {\bibinfo  {journal} {Nature communications}\ }\textbf {\bibinfo
  {volume} {8}},\ \bibinfo {pages} {158} (\bibinfo {year} {2017})}\BibitemShut
  {NoStop}%
\bibitem [{\citenamefont {Marelic}\ \emph {et~al.}(2016)\citenamefont
  {Marelic}, \citenamefont {Walker},\ and\ \citenamefont
  {Nyman}}]{marelic2016phase}%
  \BibitemOpen
  \bibfield  {author} {\bibinfo {author} {\bibfnamefont {J.}~\bibnamefont
  {Marelic}}, \bibinfo {author} {\bibfnamefont {B.~T.}\ \bibnamefont
  {Walker}},\ and\ \bibinfo {author} {\bibfnamefont {R.~A.}\ \bibnamefont
  {Nyman}},\ }\href@noop {} {\bibfield  {journal} {\bibinfo  {journal}
  {Physical Review A}\ }\textbf {\bibinfo {volume} {94}},\ \bibinfo {pages}
  {063812} (\bibinfo {year} {2016})}\BibitemShut {NoStop}%
\bibitem [{\citenamefont {Khitrova}\ \emph {et~al.}(1999)\citenamefont
  {Khitrova}, \citenamefont {Gibbs}, \citenamefont {Jahnke}, \citenamefont
  {Kira},\ and\ \citenamefont {Koch}}]{RevModPhys.71.1591}%
  \BibitemOpen
  \bibfield  {author} {\bibinfo {author} {\bibfnamefont {G.}~\bibnamefont
  {Khitrova}}, \bibinfo {author} {\bibfnamefont {H.~M.}\ \bibnamefont {Gibbs}},
  \bibinfo {author} {\bibfnamefont {F.}~\bibnamefont {Jahnke}}, \bibinfo
  {author} {\bibfnamefont {M.}~\bibnamefont {Kira}},\ and\ \bibinfo {author}
  {\bibfnamefont {S.~W.}\ \bibnamefont {Koch}},\ }\href
  {https://doi.org/10.1103/RevModPhys.71.1591} {\bibfield  {journal} {\bibinfo
  {journal} {Rev. Mod. Phys.}\ }\textbf {\bibinfo {volume} {71}},\ \bibinfo
  {pages} {1591} (\bibinfo {year} {1999})}\BibitemShut {NoStop}%
\bibitem [{\citenamefont {Amo}\ \emph {et~al.}(2009)\citenamefont {Amo},
  \citenamefont {Sanvitto}, \citenamefont {Laussy}, \citenamefont {Ballarini},
  \citenamefont {Valle}, \citenamefont {Martin}, \citenamefont {Lemaitre},
  \citenamefont {Bloch}, \citenamefont {Krizhanovskii}, \citenamefont
  {Skolnick} \emph {et~al.}}]{amo2009collective}%
  \BibitemOpen
  \bibfield  {author} {\bibinfo {author} {\bibfnamefont {A.}~\bibnamefont
  {Amo}}, \bibinfo {author} {\bibfnamefont {D.}~\bibnamefont {Sanvitto}},
  \bibinfo {author} {\bibfnamefont {F.}~\bibnamefont {Laussy}}, \bibinfo
  {author} {\bibfnamefont {D.}~\bibnamefont {Ballarini}}, \bibinfo {author}
  {\bibfnamefont {E.~d.}\ \bibnamefont {Valle}}, \bibinfo {author}
  {\bibfnamefont {M.}~\bibnamefont {Martin}}, \bibinfo {author} {\bibfnamefont
  {A.}~\bibnamefont {Lemaitre}}, \bibinfo {author} {\bibfnamefont
  {J.}~\bibnamefont {Bloch}}, \bibinfo {author} {\bibfnamefont
  {D.}~\bibnamefont {Krizhanovskii}}, \bibinfo {author} {\bibfnamefont
  {M.}~\bibnamefont {Skolnick}}, \emph {et~al.},\ }\href@noop {} {\bibfield
  {journal} {\bibinfo  {journal} {Nature}\ }\textbf {\bibinfo {volume} {457}},\
  \bibinfo {pages} {291} (\bibinfo {year} {2009})}\BibitemShut {NoStop}%
\bibitem [{\citenamefont {Egorov}\ \emph {et~al.}(2009)\citenamefont {Egorov},
  \citenamefont {Skryabin}, \citenamefont {Yulin},\ and\ \citenamefont
  {Lederer}}]{egorov2009bright}%
  \BibitemOpen
  \bibfield  {author} {\bibinfo {author} {\bibfnamefont {O.}~\bibnamefont
  {Egorov}}, \bibinfo {author} {\bibfnamefont {D.~V.}\ \bibnamefont
  {Skryabin}}, \bibinfo {author} {\bibfnamefont {A.}~\bibnamefont {Yulin}},\
  and\ \bibinfo {author} {\bibfnamefont {F.}~\bibnamefont {Lederer}},\
  }\href@noop {} {\bibfield  {journal} {\bibinfo  {journal} {Physical review
  letters}\ }\textbf {\bibinfo {volume} {102}},\ \bibinfo {pages} {153904}
  (\bibinfo {year} {2009})}\BibitemShut {NoStop}%
\bibitem [{\citenamefont {Utsunomiya}\ \emph {et~al.}(2008)\citenamefont
  {Utsunomiya}, \citenamefont {Tian}, \citenamefont {Roumpos}, \citenamefont
  {Lai}, \citenamefont {Kumada}, \citenamefont {Fujisawa}, \citenamefont
  {Kuwata-Gonokami}, \citenamefont {L{\"o}ffler}, \citenamefont {H{\"o}fling},
  \citenamefont {Forchel} \emph {et~al.}}]{utsunomiya2008observation}%
  \BibitemOpen
  \bibfield  {author} {\bibinfo {author} {\bibfnamefont {S.}~\bibnamefont
  {Utsunomiya}}, \bibinfo {author} {\bibfnamefont {L.}~\bibnamefont {Tian}},
  \bibinfo {author} {\bibfnamefont {G.}~\bibnamefont {Roumpos}}, \bibinfo
  {author} {\bibfnamefont {C.}~\bibnamefont {Lai}}, \bibinfo {author}
  {\bibfnamefont {N.}~\bibnamefont {Kumada}}, \bibinfo {author} {\bibfnamefont
  {T.}~\bibnamefont {Fujisawa}}, \bibinfo {author} {\bibfnamefont
  {M.}~\bibnamefont {Kuwata-Gonokami}}, \bibinfo {author} {\bibfnamefont
  {A.}~\bibnamefont {L{\"o}ffler}}, \bibinfo {author} {\bibfnamefont
  {S.}~\bibnamefont {H{\"o}fling}}, \bibinfo {author} {\bibfnamefont
  {A.}~\bibnamefont {Forchel}}, \emph {et~al.},\ }\href@noop {} {\bibfield
  {journal} {\bibinfo  {journal} {Nature Physics}\ }\textbf {\bibinfo {volume}
  {4}},\ \bibinfo {pages} {700} (\bibinfo {year} {2008})}\BibitemShut {NoStop}%
\bibitem [{\citenamefont {Fontaine}\ \emph {et~al.}(2018)\citenamefont
  {Fontaine}, \citenamefont {Bienaim\'e}, \citenamefont {Pigeon}, \citenamefont
  {Giacobino}, \citenamefont {Bramati},\ and\ \citenamefont
  {Glorieux}}]{PhysRevLett.121.183604}%
  \BibitemOpen
  \bibfield  {author} {\bibinfo {author} {\bibfnamefont {Q.}~\bibnamefont
  {Fontaine}}, \bibinfo {author} {\bibfnamefont {T.}~\bibnamefont
  {Bienaim\'e}}, \bibinfo {author} {\bibfnamefont {S.}~\bibnamefont {Pigeon}},
  \bibinfo {author} {\bibfnamefont {E.}~\bibnamefont {Giacobino}}, \bibinfo
  {author} {\bibfnamefont {A.}~\bibnamefont {Bramati}},\ and\ \bibinfo {author}
  {\bibfnamefont {Q.}~\bibnamefont {Glorieux}},\ }\href
  {https://doi.org/10.1103/PhysRevLett.121.183604} {\bibfield  {journal}
  {\bibinfo  {journal} {Phys. Rev. Lett.}\ }\textbf {\bibinfo {volume} {121}},\
  \bibinfo {pages} {183604} (\bibinfo {year} {2018})}\BibitemShut {NoStop}%
\bibitem [{\citenamefont {Larr\'e}\ and\ \citenamefont
  {Carusotto}(2015)}]{PhysRevA.92.043802}%
  \BibitemOpen
  \bibfield  {author} {\bibinfo {author} {\bibfnamefont {P.-E.}\ \bibnamefont
  {Larr\'e}}\ and\ \bibinfo {author} {\bibfnamefont {I.}~\bibnamefont
  {Carusotto}},\ }\href {https://doi.org/10.1103/PhysRevA.92.043802} {\bibfield
   {journal} {\bibinfo  {journal} {Phys. Rev. A}\ }\textbf {\bibinfo {volume}
  {92}},\ \bibinfo {pages} {043802} (\bibinfo {year} {2015})}\BibitemShut
  {NoStop}%
\bibitem [{\citenamefont {Alyatkin}\ \emph {et~al.}(2021)\citenamefont
  {Alyatkin}, \citenamefont {Sigurdsson}, \citenamefont {Askitopoulos},
  \citenamefont {T{\"o}pfer},\ and\ \citenamefont
  {Lagoudakis}}]{alyatkin2021quantum}%
  \BibitemOpen
  \bibfield  {author} {\bibinfo {author} {\bibfnamefont {S.}~\bibnamefont
  {Alyatkin}}, \bibinfo {author} {\bibfnamefont {H.}~\bibnamefont
  {Sigurdsson}}, \bibinfo {author} {\bibfnamefont {A.}~\bibnamefont
  {Askitopoulos}}, \bibinfo {author} {\bibfnamefont {J.~D.}\ \bibnamefont
  {T{\"o}pfer}},\ and\ \bibinfo {author} {\bibfnamefont {P.~G.}\ \bibnamefont
  {Lagoudakis}},\ }\href@noop {} {\bibfield  {journal} {\bibinfo  {journal}
  {Nature Communications}\ }\textbf {\bibinfo {volume} {12}},\ \bibinfo {pages}
  {5571} (\bibinfo {year} {2021})}\BibitemShut {NoStop}%
\bibitem [{\citenamefont {Carusotto}\ and\ \citenamefont
  {Ciuti}(2013)}]{RevModPhys.85.299}%
  \BibitemOpen
  \bibfield  {author} {\bibinfo {author} {\bibfnamefont {I.}~\bibnamefont
  {Carusotto}}\ and\ \bibinfo {author} {\bibfnamefont {C.}~\bibnamefont
  {Ciuti}},\ }\href {https://doi.org/10.1103/RevModPhys.85.299} {\bibfield
  {journal} {\bibinfo  {journal} {Rev. Mod. Phys.}\ }\textbf {\bibinfo {volume}
  {85}},\ \bibinfo {pages} {299} (\bibinfo {year} {2013})}\BibitemShut
  {NoStop}%
\bibitem [{\citenamefont {Kirton}\ and\ \citenamefont
  {Keeling}(2013)}]{PhysRevLett.111.100404}%
  \BibitemOpen
  \bibfield  {author} {\bibinfo {author} {\bibfnamefont {P.}~\bibnamefont
  {Kirton}}\ and\ \bibinfo {author} {\bibfnamefont {J.}~\bibnamefont
  {Keeling}},\ }\href {https://doi.org/10.1103/PhysRevLett.111.100404}
  {\bibfield  {journal} {\bibinfo  {journal} {Phys. Rev. Lett.}\ }\textbf
  {\bibinfo {volume} {111}},\ \bibinfo {pages} {100404} (\bibinfo {year}
  {2013})}\BibitemShut {NoStop}%
\bibitem [{\citenamefont {Loirette-Pelous}\ and\ \citenamefont
  {Greffet}(2023)}]{loirette2023photon}%
  \BibitemOpen
  \bibfield  {author} {\bibinfo {author} {\bibfnamefont {A.}~\bibnamefont
  {Loirette-Pelous}}\ and\ \bibinfo {author} {\bibfnamefont {J.-J.}\
  \bibnamefont {Greffet}},\ }\href@noop {} {\bibfield  {journal} {\bibinfo
  {journal} {Laser \& Photonics Reviews}\ }\textbf {\bibinfo {volume} {17}},\
  \bibinfo {pages} {2300366} (\bibinfo {year} {2023})}\BibitemShut {NoStop}%
\bibitem [{\citenamefont {Pieczarka}\ \emph {et~al.}(2024)\citenamefont
  {Pieczarka}, \citenamefont {G{\k{e}}bski}, \citenamefont {Piasecka},
  \citenamefont {Lott}, \citenamefont {Pelster}, \citenamefont {Wasiak},\ and\
  \citenamefont {Czyszanowski}}]{Pieczarka2024}%
  \BibitemOpen
  \bibfield  {author} {\bibinfo {author} {\bibfnamefont {M.}~\bibnamefont
  {Pieczarka}}, \bibinfo {author} {\bibfnamefont {M.}~\bibnamefont
  {G{\k{e}}bski}}, \bibinfo {author} {\bibfnamefont {A.~N.}\ \bibnamefont
  {Piasecka}}, \bibinfo {author} {\bibfnamefont {J.~A.}\ \bibnamefont {Lott}},
  \bibinfo {author} {\bibfnamefont {A.}~\bibnamefont {Pelster}}, \bibinfo
  {author} {\bibfnamefont {M.}~\bibnamefont {Wasiak}},\ and\ \bibinfo {author}
  {\bibfnamefont {T.}~\bibnamefont {Czyszanowski}},\ }\href
  {https://doi.org/10.1038/s41566-024-01478-z} {\bibfield  {journal} {\bibinfo
  {journal} {Nature Photonics}\ }\textbf {\bibinfo {volume} {18}},\ \bibinfo
  {pages} {1090} (\bibinfo {year} {2024})}\BibitemShut {NoStop}%
\bibitem [{\citenamefont {Sob'yanin}(2012)}]{Sobyanin2012}%
  \BibitemOpen
  \bibfield  {author} {\bibinfo {author} {\bibfnamefont {D.~N.}\ \bibnamefont
  {Sob'yanin}},\ }\href {https://doi.org/10.1103/PhysRevE.85.061120} {\bibfield
   {journal} {\bibinfo  {journal} {Phys. Rev. E}\ }\textbf {\bibinfo {volume}
  {85}},\ \bibinfo {pages} {061120} (\bibinfo {year} {2012})}\BibitemShut
  {NoStop}%
\bibitem [{\citenamefont {Grimaldi}(2008)}]{PhysRevB.77.024306}%
  \BibitemOpen
  \bibfield  {author} {\bibinfo {author} {\bibfnamefont {C.}~\bibnamefont
  {Grimaldi}},\ }\href {https://doi.org/10.1103/PhysRevB.77.024306} {\bibfield
  {journal} {\bibinfo  {journal} {Phys. Rev. B}\ }\textbf {\bibinfo {volume}
  {77}},\ \bibinfo {pages} {024306} (\bibinfo {year} {2008})}\BibitemShut
  {NoStop}%
\bibitem [{\citenamefont {Breuer}\ and\ \citenamefont
  {Petruccione}(2002)}]{breuer2002theory}%
  \BibitemOpen
  \bibfield  {author} {\bibinfo {author} {\bibfnamefont {H.}~\bibnamefont
  {Breuer}}\ and\ \bibinfo {author} {\bibfnamefont {F.}~\bibnamefont
  {Petruccione}},\ }\href {https://books.google.co.uk/books?id=0Yx5VzaMYm8C}
  {\emph {\bibinfo {title} {The Theory of Open Quantum Systems}}}\ (\bibinfo
  {publisher} {Oxford University Press},\ \bibinfo {year} {2002})\BibitemShut
  {NoStop}%
\bibitem [{\citenamefont {Gardiner}(1991)}]{gardiner1991quantum}%
  \BibitemOpen
  \bibfield  {author} {\bibinfo {author} {\bibfnamefont {C.}~\bibnamefont
  {Gardiner}},\ }\href {https://books.google.co.uk/books?id=eFAbAQAAIAAJ}
  {\emph {\bibinfo {title} {Quantum Noise}}},\ Springer series in synergetics\
  (\bibinfo  {publisher} {Springer-Verlag},\ \bibinfo {year}
  {1991})\BibitemShut {NoStop}%
\bibitem [{\citenamefont {Fricke}(1996)}]{fricke1996transport}%
  \BibitemOpen
  \bibfield  {author} {\bibinfo {author} {\bibfnamefont {J.}~\bibnamefont
  {Fricke}},\ }\href@noop {} {\bibfield  {journal} {\bibinfo  {journal} {Annals
  of physics}\ }\textbf {\bibinfo {volume} {252}},\ \bibinfo {pages} {479}
  (\bibinfo {year} {1996})}\BibitemShut {NoStop}%
\bibitem [{\citenamefont {{Kira}}\ and\ \citenamefont
  {{Koch}}(2006)}]{2006PQE}%
  \BibitemOpen
  \bibfield  {author} {\bibinfo {author} {\bibfnamefont {M.}~\bibnamefont
  {{Kira}}}\ and\ \bibinfo {author} {\bibfnamefont {S.~W.}\ \bibnamefont
  {{Koch}}},\ }\href {https://doi.org/10.1016/j.pquantelec.2006.12.002}
  {\bibfield  {journal} {\bibinfo  {journal} {Progress in Quantum Electronics}\
  }\textbf {\bibinfo {volume} {30}},\ \bibinfo {pages} {155} (\bibinfo {year}
  {2006})}\BibitemShut {NoStop}%
\bibitem [{Note1()}]{Note1}%
  \BibitemOpen
  \bibinfo {note} {Equation \protect \textup {\hbox {\mathsurround \z@ \protect
  \normalfont (\ignorespaces \ref {cDcDcc}\unskip \@@italiccorr )}} defines the
  fluctuations through $\delta u_{\protect \mathbf k} \equiv u_{\protect
  \mathbf k} - |p_{\protect \mathbf k }|^2 - f_{e,\protect \mathbf k} f_{h,-
  \protect \mathbf k}$, which can be interpreted as the departure of
  $u_{\protect \mathbf k}$ from its classical value.}\BibitemShut {Stop}%
\bibitem [{\citenamefont {Kira}\ and\ \citenamefont
  {Koch}(2011)}]{Kira_Koch_2011}%
  \BibitemOpen
  \bibfield  {author} {\bibinfo {author} {\bibfnamefont {M.}~\bibnamefont
  {Kira}}\ and\ \bibinfo {author} {\bibfnamefont {S.~W.}\ \bibnamefont
  {Koch}},\ }\href@noop {} {\emph {\bibinfo {title} {Semiconductor Quantum
  Optics}}}\ (\bibinfo  {publisher} {Cambridge University Press},\ \bibinfo
  {year} {2011})\BibitemShut {NoStop}%
\bibitem [{\citenamefont {Vasko}\ and\ \citenamefont
  {Raichev}(2005)}]{vasko2005quantum}%
  \BibitemOpen
  \bibfield  {author} {\bibinfo {author} {\bibfnamefont {F.}~\bibnamefont
  {Vasko}}\ and\ \bibinfo {author} {\bibfnamefont {O.}~\bibnamefont
  {Raichev}},\ }\href {https://books.google.co.uk/books?id=kFg3kQWKt2MC} {\emph
  {\bibinfo {title} {Quantum Kinetic Theory and Applications: Electrons,
  Photons, Phonons}}}\ (\bibinfo  {publisher} {Springer},\ \bibinfo {year}
  {2005})\BibitemShut {NoStop}%
\bibitem [{\citenamefont {Richter}\ \emph {et~al.}(2009)\citenamefont
  {Richter}, \citenamefont {Carmele}, \citenamefont {Butscher}, \citenamefont
  {Bücking}, \citenamefont {Milde}, \citenamefont {Kratzer}, \citenamefont
  {Scheffler},\ and\ \citenamefont {Knorr}}]{10.1063/1.3117236}%
  \BibitemOpen
  \bibfield  {author} {\bibinfo {author} {\bibfnamefont {M.}~\bibnamefont
  {Richter}}, \bibinfo {author} {\bibfnamefont {A.}~\bibnamefont {Carmele}},
  \bibinfo {author} {\bibfnamefont {S.}~\bibnamefont {Butscher}}, \bibinfo
  {author} {\bibfnamefont {N.}~\bibnamefont {Bücking}}, \bibinfo {author}
  {\bibfnamefont {F.}~\bibnamefont {Milde}}, \bibinfo {author} {\bibfnamefont
  {P.}~\bibnamefont {Kratzer}}, \bibinfo {author} {\bibfnamefont
  {M.}~\bibnamefont {Scheffler}},\ and\ \bibinfo {author} {\bibfnamefont
  {A.}~\bibnamefont {Knorr}},\ }\href {https://doi.org/10.1063/1.3117236}
  {\bibfield  {journal} {\bibinfo  {journal} {Journal of Applied Physics}\
  }\textbf {\bibinfo {volume} {105}},\ \bibinfo {pages} {122409} (\bibinfo
  {year} {2009})},\ \Eprint
  {https://arxiv.org/abs/https://pubs.aip.org/aip/jap/article-pdf/doi/10.1063/1.3117236/13906634/122409\_1\_online.pdf}
  {https://pubs.aip.org/aip/jap/article-pdf/doi/10.1063/1.3117236/13906634/122409\_1\_online.pdf}
  \BibitemShut {NoStop}%
\bibitem [{\citenamefont {Hesten}\ \emph {et~al.}(2018)\citenamefont {Hesten},
  \citenamefont {Nyman},\ and\ \citenamefont
  {Mintert}}]{PhysRevLett.120.040601}%
  \BibitemOpen
  \bibfield  {author} {\bibinfo {author} {\bibfnamefont {H.~J.}\ \bibnamefont
  {Hesten}}, \bibinfo {author} {\bibfnamefont {R.~A.}\ \bibnamefont {Nyman}},\
  and\ \bibinfo {author} {\bibfnamefont {F.}~\bibnamefont {Mintert}},\ }\href
  {https://doi.org/10.1103/PhysRevLett.120.040601} {\bibfield  {journal}
  {\bibinfo  {journal} {Phys. Rev. Lett.}\ }\textbf {\bibinfo {volume} {120}},\
  \bibinfo {pages} {040601} (\bibinfo {year} {2018})}\BibitemShut {NoStop}%
\bibitem [{\citenamefont {Rozas}\ \emph {et~al.}(2018)\citenamefont {Rozas},
  \citenamefont {Mart{\'\i}n}, \citenamefont {Tejedor}, \citenamefont
  {Vi{\~n}a}, \citenamefont {Deligeorgis}, \citenamefont {Hatzopoulos},\ and\
  \citenamefont {Savvidis}}]{rozas2018temperature}%
  \BibitemOpen
  \bibfield  {author} {\bibinfo {author} {\bibfnamefont {E.}~\bibnamefont
  {Rozas}}, \bibinfo {author} {\bibfnamefont {M.}~\bibnamefont {Mart{\'\i}n}},
  \bibinfo {author} {\bibfnamefont {C.}~\bibnamefont {Tejedor}}, \bibinfo
  {author} {\bibfnamefont {L.}~\bibnamefont {Vi{\~n}a}}, \bibinfo {author}
  {\bibfnamefont {G.}~\bibnamefont {Deligeorgis}}, \bibinfo {author}
  {\bibfnamefont {Z.}~\bibnamefont {Hatzopoulos}},\ and\ \bibinfo {author}
  {\bibfnamefont {P.}~\bibnamefont {Savvidis}},\ }\href@noop {} {\bibfield
  {journal} {\bibinfo  {journal} {Physical Review B}\ }\textbf {\bibinfo
  {volume} {97}},\ \bibinfo {pages} {075442} (\bibinfo {year}
  {2018})}\BibitemShut {NoStop}%
\bibitem [{\citenamefont {Schneider}\ \emph {et~al.}(2013)\citenamefont
  {Schneider}, \citenamefont {Rahimi-Iman}, \citenamefont {Kim}, \citenamefont
  {Fischer}, \citenamefont {Savenko}, \citenamefont {Amthor}, \citenamefont
  {Lermer}, \citenamefont {Wolf}, \citenamefont {Worschech}, \citenamefont
  {Kulakovskii} \emph {et~al.}}]{schneider2013electrically}%
  \BibitemOpen
  \bibfield  {author} {\bibinfo {author} {\bibfnamefont {C.}~\bibnamefont
  {Schneider}}, \bibinfo {author} {\bibfnamefont {A.}~\bibnamefont
  {Rahimi-Iman}}, \bibinfo {author} {\bibfnamefont {N.~Y.}\ \bibnamefont
  {Kim}}, \bibinfo {author} {\bibfnamefont {J.}~\bibnamefont {Fischer}},
  \bibinfo {author} {\bibfnamefont {I.~G.}\ \bibnamefont {Savenko}}, \bibinfo
  {author} {\bibfnamefont {M.}~\bibnamefont {Amthor}}, \bibinfo {author}
  {\bibfnamefont {M.}~\bibnamefont {Lermer}}, \bibinfo {author} {\bibfnamefont
  {A.}~\bibnamefont {Wolf}}, \bibinfo {author} {\bibfnamefont {L.}~\bibnamefont
  {Worschech}}, \bibinfo {author} {\bibfnamefont {V.~D.}\ \bibnamefont
  {Kulakovskii}}, \emph {et~al.},\ }\href@noop {} {\bibfield  {journal}
  {\bibinfo  {journal} {Nature}\ }\textbf {\bibinfo {volume} {497}},\ \bibinfo
  {pages} {348} (\bibinfo {year} {2013})}\BibitemShut {NoStop}%
\bibitem [{\citenamefont {Bhattacharya}\ \emph {et~al.}(2012)\citenamefont
  {Bhattacharya}, \citenamefont {Pal},\ and\ \citenamefont
  {Bansal}}]{10.1063/1.4721495}%
  \BibitemOpen
  \bibfield  {author} {\bibinfo {author} {\bibfnamefont {R.}~\bibnamefont
  {Bhattacharya}}, \bibinfo {author} {\bibfnamefont {B.}~\bibnamefont {Pal}},\
  and\ \bibinfo {author} {\bibfnamefont {B.}~\bibnamefont {Bansal}},\ }\href
  {https://doi.org/10.1063/1.4721495} {\bibfield  {journal} {\bibinfo
  {journal} {Applied Physics Letters}\ }\textbf {\bibinfo {volume} {100}},\
  \bibinfo {pages} {222103} (\bibinfo {year} {2012})},\ \Eprint
  {https://arxiv.org/abs/https://pubs.aip.org/aip/apl/article-pdf/doi/10.1063/1.4721495/14249149/222103\_1\_online.pdf}
  {https://pubs.aip.org/aip/apl/article-pdf/doi/10.1063/1.4721495/14249149/222103\_1\_online.pdf}
  \BibitemShut {NoStop}%
\bibitem [{\citenamefont {Katzer}\ \emph {et~al.}(2023)\citenamefont {Katzer},
  \citenamefont {Selig}, \citenamefont {Christiansen}, \citenamefont
  {Ballottin}, \citenamefont {Christianen},\ and\ \citenamefont
  {Knorr}}]{PhysRevLett.131.146201}%
  \BibitemOpen
  \bibfield  {author} {\bibinfo {author} {\bibfnamefont {M.}~\bibnamefont
  {Katzer}}, \bibinfo {author} {\bibfnamefont {M.}~\bibnamefont {Selig}},
  \bibinfo {author} {\bibfnamefont {D.}~\bibnamefont {Christiansen}}, \bibinfo
  {author} {\bibfnamefont {M.~V.}\ \bibnamefont {Ballottin}}, \bibinfo {author}
  {\bibfnamefont {P.~C.~M.}\ \bibnamefont {Christianen}},\ and\ \bibinfo
  {author} {\bibfnamefont {A.}~\bibnamefont {Knorr}},\ }\href
  {https://doi.org/10.1103/PhysRevLett.131.146201} {\bibfield  {journal}
  {\bibinfo  {journal} {Phys. Rev. Lett.}\ }\textbf {\bibinfo {volume} {131}},\
  \bibinfo {pages} {146201} (\bibinfo {year} {2023})}\BibitemShut {NoStop}%
\bibitem [{\citenamefont {van Roosbroeck}\ and\ \citenamefont
  {Shockley}(1954)}]{PhysRev.94.1558}%
  \BibitemOpen
  \bibfield  {author} {\bibinfo {author} {\bibfnamefont {W.}~\bibnamefont {van
  Roosbroeck}}\ and\ \bibinfo {author} {\bibfnamefont {W.}~\bibnamefont
  {Shockley}},\ }\href {https://doi.org/10.1103/PhysRev.94.1558} {\bibfield
  {journal} {\bibinfo  {journal} {Phys. Rev.}\ }\textbf {\bibinfo {volume}
  {94}},\ \bibinfo {pages} {1558} (\bibinfo {year} {1954})}\BibitemShut
  {NoStop}%
\bibitem [{\citenamefont {Ren}\ \emph {et~al.}(2022)\citenamefont {Ren},
  \citenamefont {Li}, \citenamefont {Liu}, \citenamefont {Chen}, \citenamefont
  {Chen}, \citenamefont {Peng},\ and\ \citenamefont {Liu}}]{yuhao_2022}%
  \BibitemOpen
  \bibfield  {author} {\bibinfo {author} {\bibfnamefont {Y.}~\bibnamefont
  {Ren}}, \bibinfo {author} {\bibfnamefont {P.}~\bibnamefont {Li}}, \bibinfo
  {author} {\bibfnamefont {Z.}~\bibnamefont {Liu}}, \bibinfo {author}
  {\bibfnamefont {Z.}~\bibnamefont {Chen}}, \bibinfo {author} {\bibfnamefont
  {Y.-L.}\ \bibnamefont {Chen}}, \bibinfo {author} {\bibfnamefont
  {C.}~\bibnamefont {Peng}},\ and\ \bibinfo {author} {\bibfnamefont
  {J.}~\bibnamefont {Liu}},\ }\href {https://doi.org/10.1126/sciadv.ade8817}
  {\bibfield  {journal} {\bibinfo  {journal} {Science Advances}\ }\textbf
  {\bibinfo {volume} {8}},\ \bibinfo {pages} {eade8817} (\bibinfo {year}
  {2022})},\ \Eprint
  {https://arxiv.org/abs/https://www.science.org/doi/pdf/10.1126/sciadv.ade8817}
  {https://www.science.org/doi/pdf/10.1126/sciadv.ade8817} \BibitemShut
  {NoStop}%
\bibitem [{\citenamefont {Colombelli}\ and\ \citenamefont
  {Manceau}(2015)}]{colombelli_2015}%
  \BibitemOpen
  \bibfield  {author} {\bibinfo {author} {\bibfnamefont {R.}~\bibnamefont
  {Colombelli}}\ and\ \bibinfo {author} {\bibfnamefont {J.-M.}\ \bibnamefont
  {Manceau}},\ }\href {https://doi.org/10.1103/PhysRevX.5.011031} {\bibfield
  {journal} {\bibinfo  {journal} {Phys. Rev. X}\ }\textbf {\bibinfo {volume}
  {5}},\ \bibinfo {pages} {011031} (\bibinfo {year} {2015})}\BibitemShut
  {NoStop}%
\bibitem [{\citenamefont {Lagr\'ee}\ \emph {et~al.}(2024)\citenamefont
  {Lagr\'ee}, \citenamefont {Jeannin}, \citenamefont {Quinchard}, \citenamefont
  {Pes}, \citenamefont {Evirgen}, \citenamefont {Delga}, \citenamefont
  {Trinit\'e},\ and\ \citenamefont {Colombelli}}]{lagree_2024}%
  \BibitemOpen
  \bibfield  {author} {\bibinfo {author} {\bibfnamefont {M.}~\bibnamefont
  {Lagr\'ee}}, \bibinfo {author} {\bibfnamefont {M.}~\bibnamefont {Jeannin}},
  \bibinfo {author} {\bibfnamefont {G.}~\bibnamefont {Quinchard}}, \bibinfo
  {author} {\bibfnamefont {S.}~\bibnamefont {Pes}}, \bibinfo {author}
  {\bibfnamefont {A.}~\bibnamefont {Evirgen}}, \bibinfo {author} {\bibfnamefont
  {A.}~\bibnamefont {Delga}}, \bibinfo {author} {\bibfnamefont
  {V.}~\bibnamefont {Trinit\'e}},\ and\ \bibinfo {author} {\bibfnamefont
  {R.}~\bibnamefont {Colombelli}},\ }\href
  {https://doi.org/10.1103/PhysRevApplied.21.034002} {\bibfield  {journal}
  {\bibinfo  {journal} {Phys. Rev. Appl.}\ }\textbf {\bibinfo {volume} {21}},\
  \bibinfo {pages} {034002} (\bibinfo {year} {2024})}\BibitemShut {NoStop}%
\bibitem [{\citenamefont {Vretenar}\ \emph {et~al.}(2021)\citenamefont
  {Vretenar}, \citenamefont {Kassenberg}, \citenamefont {Bissesar},
  \citenamefont {Toebes},\ and\ \citenamefont {Klaers}}]{vrenetar_2021}%
  \BibitemOpen
  \bibfield  {author} {\bibinfo {author} {\bibfnamefont {M.}~\bibnamefont
  {Vretenar}}, \bibinfo {author} {\bibfnamefont {B.}~\bibnamefont
  {Kassenberg}}, \bibinfo {author} {\bibfnamefont {S.}~\bibnamefont
  {Bissesar}}, \bibinfo {author} {\bibfnamefont {C.}~\bibnamefont {Toebes}},\
  and\ \bibinfo {author} {\bibfnamefont {J.}~\bibnamefont {Klaers}},\ }\href
  {https://doi.org/10.1103/PhysRevResearch.3.023167} {\bibfield  {journal}
  {\bibinfo  {journal} {Phys. Rev. Res.}\ }\textbf {\bibinfo {volume} {3}},\
  \bibinfo {pages} {023167} (\bibinfo {year} {2021})}\BibitemShut {NoStop}%
\bibitem [{\citenamefont {Reiserer}(2022)}]{reiserer_2022}%
  \BibitemOpen
  \bibfield  {author} {\bibinfo {author} {\bibfnamefont {A.}~\bibnamefont
  {Reiserer}},\ }\href {https://doi.org/10.1103/RevModPhys.94.041003}
  {\bibfield  {journal} {\bibinfo  {journal} {Rev. Mod. Phys.}\ }\textbf
  {\bibinfo {volume} {94}},\ \bibinfo {pages} {041003} (\bibinfo {year}
  {2022})}\BibitemShut {NoStop}%
\bibitem [{\citenamefont {Haug}\ and\ \citenamefont
  {Koch}(2004)}]{haug2004quantum}%
  \BibitemOpen
  \bibfield  {author} {\bibinfo {author} {\bibfnamefont {H.}~\bibnamefont
  {Haug}}\ and\ \bibinfo {author} {\bibfnamefont {S.}~\bibnamefont {Koch}},\
  }\href {https://books.google.co.uk/books?id=-UoG0Hx0w04C} {\emph {\bibinfo
  {title} {Quantum Theory of the Optical and Electronic Properties of
  Semiconductors}}},\ G - Reference,Information and Interdisciplinary Subjects
  Series\ (\bibinfo  {publisher} {World Scientific},\ \bibinfo {year}
  {2004})\BibitemShut {NoStop}%
\end{thebibliography}%

\end{document}